\documentclass[12pt,a4paper]{article} 

\usepackage{amsfonts}
\usepackage{amstext}
\usepackage{amsmath}
\usepackage{amssymb}
\usepackage{bm,empheq}

\usepackage{graphicx}
\usepackage{color}
\usepackage{here}

\numberwithin{equation}{section}
\numberwithin{figure}{section}
\numberwithin{table}{section}

\newcommand{\rf}[1]{(\ref{#1})}
\newcommand{\beq}{\begin{equation}}
\newcommand{\eeq}{\end{equation}}
\newcommand{\bea}{\begin{eqnarray}}
\newcommand{\eea}{\end{eqnarray}}

\renewcommand{\l}{\lambda}

\newcommand{\ep}{\varepsilon}

\newcommand{\cG}{{\cal G}}
\newcommand{\cH}{{\cal H}}

\newcommand{\calF}{\mathcal{F}{\negdbltinyspace}}

\newcommand{\calG}{\cG}
\newcommand{\calH}{{\negdbltinyspace}\mathcal{H}{\negdbltinyspace}}
\newcommand{\calJ}{\mathcal{J}{\negdbltinyspace}}
\newcommand{\calPhi}{\phi}
\newcommand{\calPsi}{\psi}
\newcommand{\calOmega}{\Omega{\halftinyspace}}
\newcommand{\calomega}{\omega}

\newcommand{\halftinyspace}{\hspace{0.0278em}} 
\newcommand{\tinyspace}{\hspace{0.0556em}}
\newcommand{\trehalftinyspace}{\hspace{0.0834em}}
\newcommand{\dbltinyspace}{\hspace{0.1112em}}
\newcommand{\trpltinyspace}{\hspace{0.1668em}}
\newcommand{\qdrpltinyspace}{\hspace{0.2224em}}
\newcommand{\neghalftinyspace}{\hspace{-0.0278em}}
\newcommand{\negtinyspace}{\hspace{-0.0556em}}
\newcommand{\negtrehalftinyspace}{\hspace{-0.0834em}}
\newcommand{\negdbltinyspace}{\hspace{-0.1112em}}
\newcommand{\negfemhalftinyspace}{\hspace{-0.1390em}}
\newcommand{\negtrpltinyspace}{\hspace{-0.1668em}}
\newcommand{\negqdrpltinyspace}{\hspace{-0.2224em}}
\newcommand{\negoctpltinyspace}{\hspace{-0.4448em}}

\newcommand{\define}{\leftdefine}
\newcommand{\leftdefine}{:=}

\newcommand{\bra}[1]{\langle #1 |}

\newcommand{\ket}[1]{| #1 \rangle}

\newcommand{\vac}{\bra{{\rm vac}}}
\newcommand{\cuum}{\ket{{\rm vac}}}
\newcommand{\expect}[1]{\langle #1 \tinyspace\rangle}

\newcommand{\combi}[2]{\left( \!\! \begin{array}{c} 
	\raise0.5ex\hbox{$#1$} \\ \lower0.5ex\hbox{$#2$} \\ 
	\end{array} \!\! \right)}

\newcommand{\commutator}[2]{[\, #1 {\dbltinyspace}, #2 \,]}

\newcommand{\cc}{\mu} 

\font\twelvemsbm = msbm10 scaled\magstep1
\font\tenmsbm = msbm10
\font\eightmsbm = msbm8
\font\sixmsbm = msbm6
\newcommand{\Dbl}[1]{\leavevmode\raise-.10ex\hbox{\twelvemsbm #1}}
\newcommand{\dbl}[1]{\leavevmode\raise-.00ex\hbox{\tenmsbm #1}}
\newcommand{\dblsmall}[1]{\leavevmode\raise-.05ex\hbox{\eightmsbm #1}}
\newcommand{\dbltiny}[1]{\leavevmode\raise-.05ex\hbox{\sixmsbm #1}}

\newcommand{\pder}[1]{\frac{\partial}{\partial #1}}

\newcommand{\Hop}{H}

\newcount\figX
\newcount\figY
\newcount\figXl
\newcount\figXr
\newcount\figXll
\newcount\figXrr
\newcount\figXlll
\newcount\figXrrr
\newcommand{\minipagewidth}{\linewidth}
\newcommand{\figwidth}{\unitlength}

\newcommand{\pictureY}{141.4}

\newcommand{\figuresize}[3]{
  \figX=#2
  \multiply \figX by -1
  \advance  \figX by 100
  \divide   \figX by 2
  \figY=0
  \renewcommand{\minipagewidth}{#1\linewidth}
  \setlength\unitlength{#1\linewidth}
  \setlength\unitlength{0.01\unitlength}
  \renewcommand{\figwidth}{#2\unitlength}
  \renewcommand{\pictureY}{#3}
}

\newcommand{\figureshift}[2]{
  \figXl=#1
  \advance  \figXl by \figX
  \figXr=#1
  \multiply \figXr by -1
  \advance  \figXr by \figX
  \figXll=#1
  \multiply \figXll by 2
  \advance  \figXll by \figX
  \figXrr=#1
  \multiply \figXrr by -2
  \advance  \figXrr by \figX
  \figXlll=#1
  \multiply \figXlll by 3
  \advance  \figXlll by \figX
  \figXrrr=#1
  \multiply \figXrrr by -3
  \advance  \figXrrr by \figX
  \advance  \figY by #2
}

\newcommand{\tildePsidag}{\tilde\Psi^\dagger}
\newcommand{\tildePhidag}{\tilde\Phi^\dagger}

\newcommand{\lambdaXXX}{\lambda}

\newcommand{\OmegaXXX}{\Omega}

\newcommand{\AmpFXXX}{\tilde{F}}
\newcommand{\Const}{C}
\newcommand{\ConstXXX}{C}

\newcommand{\T}{T}

\newcommand{\NN}{N}

\newcommand{\E}{{\rm e}}

\newcommand{\G}{G}

\newcommand{\X}{\bm{x}}

\def\theequation{\arabic{section}.\arabic{equation}}
\def\thefigure{\arabic{section}.\arabic{figure}}

\allowdisplaybreaks

\begin{document}
\topmargin 0pt
\oddsidemargin 5mm                              
\headheight 0pt
\headsep 0pt
\topskip 9mm

\thispagestyle{empty}

\begin{center}
  \vspace{24pt}
  {\large \bf
Dynamical Triangulations for 2D Pure Gravity\\ and Topological Recursion
    }

  \vspace{24pt}

  {\sl Hiroyuki Fuji}

  \vspace{6pt}

{\small
  Center for Mathematical and Data Sciences\\ and Department of Mathematics, Kobe University\\ 
   Rokko, Kobe 657-8501, Japan
}
\vspace{6pt}

  \vspace{12pt}

  {\sl Masahide Manabe}

  \vspace{6pt}

{\small
  Liberal Arts \& Data Science Unit\\
  Tottori University\\
  4-101 Koyama-cho Minami, Tottori, 680-8550, Japan
}

  \vspace{12pt}

  and

  \vspace{12pt}

  {\sl Yoshiyuki Watabiki}

  \vspace{6pt}

{\small
  Department of Physics, 
  Institute of Science Tokyo\\
  Oh-okayama 2-12-1, Meguro-ku, Tokyo 152-8551, Japan
}

\end{center}
\vspace{0pt}

\vfill

\begin{center}
  {\bf Abstract}
\end{center}

\vspace{2pt}

\noindent
We show that, in two-dimensional Euclidean quantum gravity without matter fields, the Schwinger-Dyson equations derived within the Hamiltonian framework of non-critical string field theory can be reformulated in terms of the Chekhov-Eynard-Orantin topological recursion, and we explicitly compute the associated low-order amplitudes. In particular, we establish this reformulation for two discrete models---the basic type and the strip type---as well as for the continuum limit of dynamical triangulations.\vfill

\vspace{36pt}



\newpage

\setcounter{page}{1}

\tableofcontents

\newpage


\section{Introduction}

The fundamental theory currently known to us combines 
General Relativity and the Standard Model based on the 
${\rm SU}(3)\times{\rm SU}(2)\times{\rm U}(1)$ Yang-Mills theory. 
The former describes gravity, while the latter explains 
the strong, weak, and electromagnetic forces.
However, General Relativity remains unquantized, 
and its quantization is an essential challenge. 
Quantizing General Relativity in four-dimensional spacetime 
has proven exceedingly difficult. 
While theories like string theory, 
often referred to as superstring theory, 
are considered strong candidates for quantum gravity, 
they have yet to address phenomena in extremely small-scale regimes, 
such as those before the Big Bang, leaving them far from conclusive. 
On the other hand, although not in four-dimensional spacetime, 
the quantization of General Relativity in two-dimensional (2D) Euclidean space, 
known as 2D quantum gravity, has been successfully achieved. 
Notable approaches to this include 
Liouville gravity models \cite{LG:Polyakov,LG:KPZ}, 
matrix models \cite{MM:BIPZ,MM:DS,MM:GM,MM:BK,MM:DVV,MM:FKN}, 
and 
Dynamical Triangulations (DT) \cite{DT:David,DT:ADF,DT:KKM}.
In this paper, 
we calculate the amplitudes of 
pure DT---a type of DTs without matter fields---using 
string field theory and topological recursion,
both of which enable a non-perturbative approach.

The development of the string field theory for pure DT has progressed as follows:
Initially, 
the concept of time was introduced on the 2D surface of pure DT 
in \cite{preSFT:AM,preSFT:KKSW,preSFT:KKMW},
where time was identified with the geodesic distance.
In \cite{preSFT:AM}, the geodesic distance on DT was first defined.
The scaling properties of the geodesic distance 
were subsequently observed in \cite{preSFT:KKSW},
and the continuum limit of the geodesic distance 
was successfully obtained in \cite{preSFT:KKMW}.
In the discrete setting, 
the model includes not only propagator-like contributions
but also an infinite tower of interaction vertices, 
such as four-point, five-point, and higher.
It was shown that in the continuum limit, 
all these higher-order interactions vanish,
leaving only tadpole terms and three-string interaction terms.
This disappearance of the propagator reflects 
the fractal nature of the 2D surface.
In \cite{SFT:IK}, 
the string field theory of pure DT was formulated 
in order to realize the fractal structure studied in \cite{preSFT:KKMW}.
The authors demonstrated that 
their string field theory is equivalent to 
the matrix model without matter fields.
The relation between the string field theory of DT and the matrix model 
was also discussed in \cite{SFT:JR}. 
However, the Hamiltonian in this work 
is not well-defined because it is only defined in the continuum limit,
requiring a well-known regularization for practical calculations.
A well-defined formulation of  the string field theory 
was achieved in \cite{SFT:Watabiki}
by introducing the ``peeling decomposition'' at the discretized level.
In \cite{SFT:AW}, the string field theory was constructed 
at the continuous level in well-defined form 
by introducing the $W^{(3)}$ operator of conformal field theory.
In this paper, we focus on two models of pure DT in \cite{SFT:Watabiki}, 
one is the model constructed only by triangulations, 
and the other is the model constructed not only from triangulations but also from strips. 
We refer to the former as ``DT (basic type)'' and the latter as ``DT (strip type)'', 
where both models, in the continuum limit, are shown to yield the same pure DT referred to as ``DT (continuous level)''.

On the other hand, the topological recursion, formulated by Eynard and Orantin in \cite{Eynard:2007kz}, recursively defines a set of multi-differentials $\omega_{N}^{(h)}(z_1,\ldots,z_N)$ on $\Sigma^{N}$, 
labeled by two integers $h \ge 0$ and $N \ge 1$, 
from spectral curve data $(\Sigma; x,y,B)$, where $x=x(z), y=y(z)$ $[z \in \Sigma]$ are meromorphic functions on a Riemann surface $\Sigma$, 
and $B=B(z_1, z_2)$ denotes a bi-differential on $\Sigma^{2}$. 
The topological recursion has its origin in the loop equations for matrix models \cite{Alexandrov:2003pj,Eynard:2004mh,Chekhov:2006vd}, and is shown to be applicable to many examples beyond the scope of matrix models (see, e.g., \cite{Eynard:2016yaa, Bouchard:2024fih} for a review).
In matrix models, the multi-differential $\omega_{N}^{(h)}(z_1,\ldots,z_N)$ calculates the genus $h$ part of the $N$-point amplitude of resolvents, and 
$x=x(z), y=y(z)$ are provided by the disk amplitude, while $B=B(z_1, z_2)$ is provided by the cylinder amplitude.
For the three models discussed in this paper, the spectral curve data are
$$
\Sigma=\mathbb{P}^1\,,\qquad
B(z_1, z_2)=\frac{dz_1dz_2}{\left(z_1-z_2\right)^2}\,,
$$
and
\begin{align*}
&
\textrm{DT (basic type)}:\quad
x^2y^2=
\frac12 \left(x-\frac{c}{\kappa}\right)^2\left(x-\frac{4\kappa}{c^2}\right)\,,
\quad \left[c\left(1-c^2\right)=8 \kappa^2\right],
\\
&\hspace{8.5em}
x(z)=\frac{\kappa}{c^2}\left(2 + z + \frac{1}{z}\right),
\\
&
\textrm{DT (strip type)}:\quad
y^2=\frac{\kappa^2}{4}\left(x-\frac{2-a-b}{2\kappa}\right)^2\left(x-\frac{a}{\kappa}\right)\left(x-\frac{b}{\kappa}\right),
\\
&\hspace{8.5em}
\left[2\left(a+b\right)\left(2-a-b\right)=\left(a-b\right)^2,\quad
\left(a-b\right)^2\left(1-a-b\right)=16 \kappa^2
\right],
\\
&\hspace{8.5em}
x(z)=\frac{a+b}{2\kappa}
+\frac{b-a}{4\kappa} \left( z + \frac{1}{z} \right),
\\
&
\textrm{DT (continuous level)}:\quad
y^2=\left(x-\frac{\sqrt{\cc}}{2}\right)^2\left(x+\sqrt{\cc}\right),
\\
&\hspace{11em}
x(z)=z^2-\sqrt{\cc}\,,
\end{align*}
where $\kappa$ and $\cc$ are cosmological constants at the discrete and continuous levels, respectively. 
Note that the choice of the variable $z \in \mathbb{P}^1$ for each model is not unique, and in DT (continuous level) we instead use the notation $\xi=x$, $\eta=z$.
We show that the multi-differentials $\omega_{N}^{(h)}(z_1,\ldots,z_N)$ determined by the topological recursion for the above spectral curves give the amplitudes of each model of pure DT.%
\footnote{
Rescaling the variables $x\to \mu^{1/2}x/2$ and $y\to \mu^{3/4}y/\sqrt{2}$ in the spectral curve for the DT (continuous level), one obtains the defining equation 
\begin{align}
y^2=(x-1)^2(x+2),
\nonumber
\end{align}
as derived in Section 5 of \cite{Eynard:2016yaa}. 
This form of the spectral curve can also be obtained by taking an appropriate scaling limit of the matrix model with a cubic potential.
}

Here, we emphasize the novel contributions of this work.
The discrete DT model of the strip type was introduced to reproduce the Schwinger-Dyson equation for the matrix model with the cubic potential \cite{preSFT:KKMW,SFT:Watabiki}.
Therefore, by construction, its reformulation via topological recursion can be regarded as anticipated, although the direct derivation has not been explicitly presented.
In contrast, the discrete DT model of the basic type was originally proposed as a random lattice model based on equilateral triangulations, and its realization in terms of a matrix model remains unclear.
Therefore, it is not evident whether its reformulation via topological recursion can be achieved in this case.

Furthermore, the novel results for the DT model at the continuous level can be summarized as follows.
In \cite{SFT:IK}, the Virasoro constraint was derived from the Schwinger-Dyson equation%
\footnote{
The Schwinger-Dyson equation is essentially obtained via the continuum limit of the matrix model in \cite{SFT:IK}. 
In \cite{SFT:Watabiki}, the Hamiltonian framework for 2D pure gravity was formulated at the discretized level, and the continuum limit was properly taken without invoking the matrix model.
Based on this well-defined Hamiltonian formulation of the string field theory, we derive the explicit form of the Schwinger-Dyson equation for the  $N$-point amplitude directly in this work.
} 
for the DT model at the continuous level and was shown to agree with the constraint obtained from the continuum limit of the matrix model \cite{MM:DVV,MM:FKN}. 
Essentially, the topological recursion is equivalent to the Virasoro constraint, and it is natural to expect that this Schwinger-Dyson equation can also be reformulated in terms of the topological recursion. 
However, a direct derivation of the topological recursion for the DT model at the continuous level has not been presented, and establishing this connection is another novel result of this work.%
\footnote{
In the accompanying paper \cite{FMW2}, we further investigate the string field theory for multicritical continuum DT models and propose an analogue for multicritical CDT models. In a subsequent work \cite{FMW3}, we extend the Hamiltonian formalism of the string field theory to a broader class of models beyond DT, reconstructing the Hamiltonian  from the spectral curve data.
}

The structure of this paper is as follows.
Section \ref{sec:DTND} and \ref{sec:DTSTD} focus, respectively,  on the basic type and the strip type of pure DT.
We provide a detailed summary of these models, addressing aspects that were left unclear in previous studies \cite{SFT:Watabiki,SFT:AW}.
We define the amplitudes for each type 
and calculate them using the peeling decomposition.
We show that the Schwinger-Dyson equations for the amplitudes lead to the topological recursion.
In Section \ref{sec:DTcontinuouslevel}, we first review the fact that the continuum limit of the pure DTs in Section \ref{sec:DTND} and \ref{sec:DTSTD} yields the same continuous level pure DT. 
We then show that the Schwinger-Dyson equation for the amplitudes of the continuous level pure DT also leads to the topological recursion.
Finally, we conclude this paper in Section \ref{sec:conclusions}.
Appendix \ref{app:CalculationConnectivity} summarizes formulas for connected amplitudes, and Appendix \ref{app:list_amplitudes} provides a list of connected amplitudes of pure DTs calculated by the topological recursion.
In Appendix \ref{app:1DQG},  we consider a toy model of 1D pure quantum gravity.
Through this simple model, we clarify the essential meaning of the continuum limit.
In Appendix \ref{app:SD_MM}, we discuss
the equivalence between the Schwinger-Dyson equations for DT (strip-type) and the cubic matrix model.


\section{Dynamical Triangulation (Basic Type)}
\label{sec:DTND}

\subsection{Fundamental properties}
\label{sec:PropertiesDiscretNaiveDT}

In this section,
we define and compute the amplitudes of pure DT of the basic type.
Triangulations in this model 
divide an orientable two-dimensional surface 
into equilateral triangles of the same size, 
allowing only triangle-based gluings---no additional structures 
such as matter fields or higher-order polygons are introduced.

\begin{figure}[t] 
\begin{center}
\includegraphics[width=11cm,pagebox=cropbox,clip]{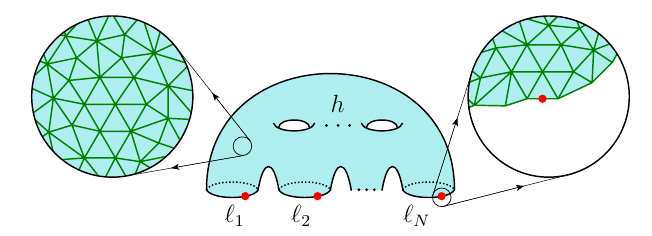}
  \caption{A (basic type) triangulated 2D surface 
with $\NN$ boundaries and $h$ handles.
Red points on the boundaries indicate marked points.
The figures in two large circles show 
the detailed structures of the triangulated 2D surface.
}
  \label{fig:DTamplitude}
\end{center}
\end{figure}

We denote by ${\cal T}_\NN^{(h)}(\ell_1, \ldots, \ell_\NN; N_2)$ 
the set of all triangulated, oriented, and connected surfaces in pure DT 
that have $\NN$ boundary loops of lengths 
$\ell_1, \ldots, \ell_\NN$ $[\in \Dbl{N}]$, 
$h$ handles, and that are composed of $N_2$ equilateral triangles of the same size.
The boundary of the two-dimensional surface 
consists of one-dimensional loops,
each formed by several edges of triangles.
To fix the rotational symmetry of each loop,
one edge per loop is marked at its midpoint.
In this triangulated surface, 
all triangles are connected via shared edges.
Triangles that touch only at a single vertex are not regarded as connected.
For example, if a part of an annular surface (a ring-shaped region)
is attached to another part solely at a single vertex,
these parts are treated as disconnected.
Thus, the connectedness of the entire surface 
is determined solely by whether the triangles are joined edge to edge.
Triangulations are regarded as identical if they can be matched 
by gluing along edges in the same pattern, up to rotation.
However, since all triangles are oriented, 
configurations that match only after a flip 
(i.e., orientation reversal) are not regarded as identical.
Fig.~\ref{fig:DTamplitude} shows a typical example 
of a triangulated 2D surface 
with $\NN$ boundaries and $h$ handles. 
We refer to this as ``DT (basic type)''.
Unlike DT (strip type), 
which will be described in Section \ref{sec:DTSTD},
DT (basic type) does not exhibit strips at its boundaries.

The amplitude of pure DT, 
$F_\NN^{{\tinyspace}{\rm conn}}(\ell_1,\ldots,\ell_\NN;\G)$, 
is defined by 
\begin{equation}\label{DTamplitude}
F_\NN^{{\tinyspace}{\rm conn}}(\ell_1,\ldots,\ell_\NN;\G{\tinyspace})
\,:=\,
\sum_{h=0}^\infty\,
\sum_{N_2=1}^\infty\,
\sum_{S {\tinyspace}\in{\trehalftinyspace}
      {\cal T}_\NN^{(h)}(\ell_1,\ldots,\ell_\NN, N_2)}
\G^{\;\!h+\NN-1} \kappa^{N_2}
\,,
\end{equation}
where the third summation is taken over 
all possible 
$S {\negtrpltinyspace}\in{\negtrpltinyspace}
 {\cal T}_\NN^{(h)}(\ell_1, \ldots, \ell_\NN; N_2)$. 
$\G$ is the parameter that counts the number of handles, 
and 
$\kappa$ is the discrete cosmological constant, 
which specifies the number of triangles in the triangulated 2D surface.
The dependence on $\kappa$ is omitted in the expression of 
$F_\NN^{{\tinyspace}{\rm conn}}(\ell_1,\ldots,\ell_\NN;G{\tinyspace})$ 
for simplicity.

\subsubsection{Decomposition of triangulated surface}
\label{sec:DecompositionOfDT}

\begin{figure}[t] 
\begin{center}
\includegraphics[width=13cm,pagebox=cropbox,clip]{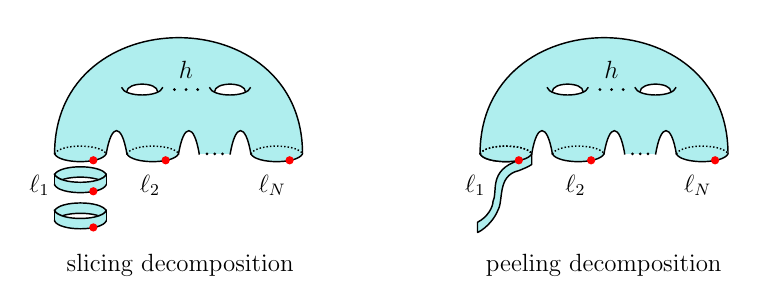}
  \caption{2D surface 
with $N$ boundaries and $h$ handles 
by single-slicing decomposition and single-peeling decomposition.
}
  \label{fig:DTamplitudeSingleSlicingPeeling}
\end{center}
\end{figure}

To calculate the amplitude (\ref{DTamplitude}),
one must count the number of triangulated surfaces 
with $\NN$ boundaries, $h$ handles, and $N_2$ triangles.
This can be done by first choosing 
one of the $\NN$ boundaries---called 
the selected boundary---and 
removing all triangles adjacent to it. 
After removing these triangles,
the boundary is updated with a new edge,
and a new marked point is placed
at the midpoint of the edge closest to the removed region.
At each subsequent step, a boundary is chosen 
from those that newly appear, 
and the removal process is repeated. 
This iterative procedure, 
known as the ``slicing decomposition'' \cite{preSFT:KKMW}, 
is illustrated in the left-hand figure in 
Fig.~\ref{fig:DTamplitudeSingleSlicingPeeling}.

Another decomposition method compares two amplitudes that differ by a single triangle.
The operation of selecting one of the $\NN$ boundaries 
is the same as in the slicing decomposition discussed earlier.
The comparison procedure proceeds by iteratively removing one triangle at a time.
At each step, 
the triangle to be removed is the one facing the boundary and carrying a marked point.
After removing this triangle, 
a new edge on the boundary is chosen,
and its midpoint is marked. 
This process continues sequentially,
in a manner analogous to peeling the skin off an apple.
The marked point on the boundary loop is introduced 
to break rotational symmetry;
in the peeling decomposition, however, 
it also specifies the location of the next peeling step.
Thus, it plays a dual role in the process.
This iterative removal procedure, 
known as the ``peeling decomposition'' \cite{SFT:Watabiki}, 
is illustrated in the right-hand figure of 
Fig.~\ref{fig:DTamplitudeSingleSlicingPeeling}.

\begin{figure}[t] 
\begin{center}
\includegraphics[width=14cm,pagebox=cropbox,clip]{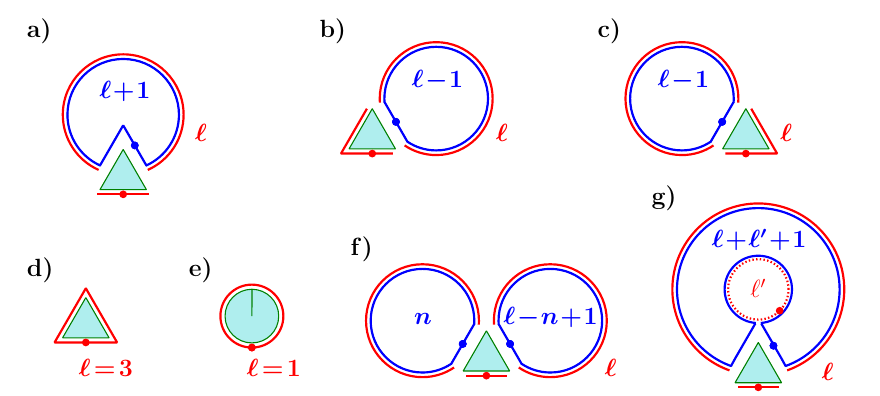}
  \caption{Decompositions by removing a triangle.
The solid red line represents the initial boundary of length $\ell$,
while the solid blue line represents the boundary 
after the removal of a triangle in each figure.
The dashed red line represents the adjacent boundary of length $\ell'$ 
that merges with the initial boundary. 
Light blue shaded triangles represent those that are removed.
}
  \label{fig:TriangleDecomposition}
\end{center}
\end{figure}

There are seven possible ways to remove a triangle with a marked point.
These operations are illustrated in 
Fig.~\ref{fig:TriangleDecomposition}.
In these figures, the red marked point represents 
the position of the marked point on the boundary before the removal of a triangle, 
while the blue marked point indicates its position after the removal.
The changes in the boundary length caused by each operation are as follows:
\begin{itemize}
\item
In Fig.~\ref{fig:TriangleDecomposition} {\bf a)} , 
a boundary of length $\ell$ becomes a boundary of length $\ell+1$. 
\item
In Figs.~\ref{fig:TriangleDecomposition} {\bf b)} and {\bf c)}, 
a boundary of length $\ell$ is reduced to a boundary of length $\ell-1$. 
\item
In Figs.~\ref{fig:TriangleDecomposition} {\bf d)} and {\bf e)}, 
the entire triangulation disappears.
\item
In Fig.~\ref{fig:TriangleDecomposition} {\bf f)}, 
a boundary of length $\ell$ splits into 
two boundaries of lengths $n$ and $\ell-n+1$.
\item
In Fig.~\ref{fig:TriangleDecomposition} {\bf g)}, 
a boundary of length $\ell$ merges with 
an adjacent boundary of length $\ell'$, 
forming a single boundary of length $\ell+\ell'+1$.
\end{itemize}
From the perspective of the string field theory,
Figs.~\ref{fig:TriangleDecomposition} {\bf a)}--{\bf c)}
represent propagators.
Figs.~\ref{fig:TriangleDecomposition} {\bf d)} and {\bf e)}
illustrate the disappearance of a single string.
In Fig.~\ref{fig:TriangleDecomposition} {\bf f)},
one string splits into two strings, while 
in Fig.~\ref{fig:TriangleDecomposition} {\bf g)},
two strings merge into a single string.

\subsubsection{String field theory of dynamical triangulation}
\label{sec:SFTOfDT}

In the string field theory \cite{SFT:Watabiki,SFT:WatabikiReview}, 
we introduce the creation operator 
$\Psi^\dagger(\ell)$
and the annihilation operator 
$\Psi(\ell)$
for each boundary of length 
$\ell$ [\,$\ell {\negtrpltinyspace}\in{\negtrpltinyspace} \Dbl{N}$\,], 
which satisfy the commutation relations, 
\begin{equation}\label{CommutationRelationLength}
\commutator{\Psi(\ell)}{\Psi^\dagger(\ell')}
=
\delta_{\ell,\ell'}
\,,
\qquad
\commutator{\Psi^\dagger(\ell)}{\Psi^\dagger(\ell')}
=
0
\,,
\qquad
\commutator{\Psi(\ell)}{\Psi(\ell')}
=
0
\,.
\end{equation}
The vacuum states 
$\vac$ and $\cuum$ 
satisfy the conditions 
\begin{equation}\label{vacuumCondition_phi}
\expect{{\rm vac}|{\rm vac}} = 1
\,,
\qquad
\vac \Psi^\dagger(\ell) = 0
\,,
\qquad
\Psi(\ell) \cuum = 0
\,.
\end{equation}
Then, the state consisting of $N$ string fields with lengths 
$\ell_1$, \ldots, $\ell_N$ is expressed as 
\begin{equation}\label{StringStateN}
\Psi^\dagger(\ell_1)
\ldots
\Psi^\dagger(\ell_N)
\cuum
\,.
\end{equation}

Using a time $T$ and a Hamiltonian $\Hop$, 
the time evolution of the state \rf{StringStateN} 
is expressed as 
\begin{equation}\label{StringStateNtime}
\E^{-T \Hop} {\tinyspace}
\Psi^\dagger(\ell_1)
\ldots
\Psi^\dagger(\ell_N)
\cuum
\,.
\end{equation}
In the DT case, the time variable 
$T$ corresponds to the geodesic distance or its discrete analogue, 
rather than reflecting any causal structure.
A key property of DT geometries is that 
any point can be reached by successively moving to neighboring triangles, 
provided that the triangulated space is connected.
This implies that there is no privileged origin 
of space---no specific point acts as the birth point of geometry.
The presence of causality would impose directionality 
on such local moves and obstruct this universal reachability.
Therefore, 
the notion of a ``big bang''---where space would emerge 
from a unique origin---is incompatible with DT, 
as illustrated in Fig.~6 of \cite{SFT:AW} 
and Fig.~1 of \cite{SFT:WatabikiReview}, 
and discussed in detail therein.
Equivalently, 
allowing a big bang would lead to ambiguity in the time evolution 
from a given configuration, resulting in overcounting.
Accordingly, the Hamiltonian $\Hop$ satisfies 
the so-called ``no big-bang condition'':
\begin{equation}\label{NoBigBangCondition}
\Hop \!\> \cuum
\,=\, 0
\,.
\end{equation}
We note here that 
this condition will play a crucial role 
in the construction of more general theories. 
Since this paper focuses only on Hamiltonians 
that inherently satisfy this condition, 
further discussion is beyond the scope of this study.

\begin{figure}[t] 
\begin{center}
\includegraphics[width=15cm,pagebox=cropbox,clip]{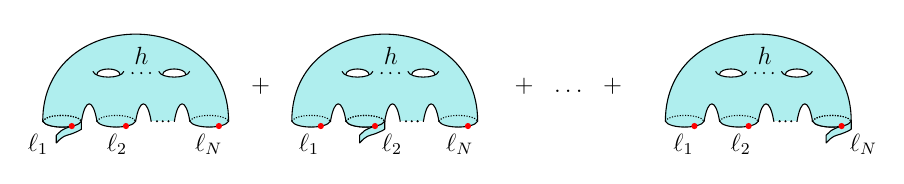}
  \caption{2D surface with $N$ boundaries and $h$ handles 
by multi-peeling decomposition.
This decomposition is implemented 
within the framework of the string field theory.
}
  \label{fig:DTamplitudeMultiPeeling}
\end{center}
\end{figure}

We now turn to a feature specific to field theory. 
The decompositions shown in 
Fig.~\ref{fig:DTamplitudeSingleSlicingPeeling} 
do not work well in the string field theory 
because all creation operators act on the same vacuum and 
are indistinguishable with respect to the decomposition scheme.
For the peeling decomposition to work effectively in the string field theory, 
the single-peeling decomposition must be modified to 
a multi-peeling decomposition,
in which peeling operations are performed simultaneously on all boundaries.
The multi-peeling decomposition is shown 
in Fig.~\ref{fig:DTamplitudeMultiPeeling}.
The same applies to the slicing decomposition.

In the slicing decomposition,
removing all triangles from a boundary of length $\ell$
can be regarded as a single step of decomposition
in terms of geodesic distance.
This ``step'' corresponds to a unit of time evolution in field theory.
Accordingly,
in the peeling decomposition,
removing a single triangle from a boundary of length $\ell$
is considered a $1/\ell$-step decomposition.%
\footnote{%
The former decomposition was introduced in \cite{preSFT:KKMW},
and the latter in \cite{SFT:Watabiki}.
}
This concept lies at the core of the ``peeling decomposition''.
Consequently, 
the Hamiltonian that implements the decompositions shown in
Fig.~\ref{fig:TriangleDecomposition} satisfies 
\begin{equation}\label{OneOverLstepMovement}
\Psi^{{\rm [Decomp]}\dagger}(\ell)
-
\Psi^\dagger(\ell)
\,=\,
-{\dbltinyspace}\frac{1}{\ell}{\dbltinyspace}
\commutator{\Hop}{\Psi^\dagger(\ell)}
\,.
\end{equation}
%
Within the DT framework, 
because the slicing decomposition is technically 
much more difficult than the peeling decomposition, 
we henceforth focus exclusively on the latter.

In the string field theory, 
the amplitude \eqref{DTamplitude} is obtained as the connected part of the disconnected amplitude
\begin{equation}\label{DTamplitudeFieldTheory}
F_N(\ell_1,\ldots,\ell_N;G{\tinyspace})
=
\lim_{T \to \infty}
\vac \E^{-T \Hop} {\tinyspace}
     \Psi^\dagger(\ell_1) \ldots \Psi^\dagger(\ell_N)
\cuum
\,,
\end{equation}
where 
$\Hop$ is 
the Hamiltonian that implements the decompositions shown in
Fig.~\ref{fig:TriangleDecomposition}. 
Here, we introduce the Laplace transform of the amplitudes%
\footnote{%
The standard discrete Laplace transform is
\begin{equation}\label{DTamplitudeLaplaceTransf}
\tilde{F}_N^{[{\rm LT}]}(x_1,\ldots,x_N;G{\tinyspace})
\,:=\,
\sum_{\ell_1=1}^\infty \ldots \sum_{\ell_N=1}^\infty
x_1^{\ell_1}
\ldots
x_N^{\ell_N}
F_N(\ell_1,\ldots,\ell_N;G{\tinyspace})
\,.
\nonumber
\end{equation}
However, we use (\ref{DTamplitudeMatrixModel}) 
rather than this definition,
since (\ref{DTamplitudeMatrixModel}) 
is the standard expression in the matrix-model literature.
}
\begin{equation}\label{DTamplitudeMatrixModel}
\tilde{F}_N(x_1,\ldots,x_N;G{\tinyspace})
\,:=\,
\sum_{\ell_1=1}^\infty \ldots \sum_{\ell_N=1}^\infty
x_1^{- \ell_1 -1}
\ldots
x_N^{- \ell_N -1}
F_N(\ell_1,\ldots,\ell_N;G{\tinyspace})
\,,
\end{equation}
for the purpose of simplifying the subsequent mathematical analysis. 
Substituting \rf{DTamplitudeFieldTheory} 
into \rf{DTamplitudeMatrixModel}, 
we obtain 
\begin{align}\label{DTamplitudeSFT}
\tilde{F}_N(x_1,\ldots,x_N;G{\tinyspace})
&=
\lim_{T \to \infty}
\vac \E^{-T \Hop} {\tinyspace}
     \tilde\Psi^\dagger(x_1) \ldots \tilde\Psi^\dagger(x_N)
\cuum
\,,
\end{align}
where the Laplace-transformed operators are defined by
\begin{equation}\label{DiscreteLaplaceTransfWaveFun}
\tilde\Psi^\dagger(x)
\,\define\,
\sum_{\ell=1}^\infty x^{-\ell-1} \Psi^\dagger(\ell)
\,,
\qquad
\tilde\Psi(y)
\,\define\,
\sum_{\ell=1}^\infty y^{-\ell-1} \Psi(\ell)
\,,
\end{equation}
and the commutation relations \rf{CommutationRelationLength} imply
\begin{equation}\label{CommutationRelationConj}
\commutator{\tilde\Psi(y)}{\Psi^\dagger(x)}
=
\frac{1 / (y x)^2}{1 - 1/(y x)}
\,,
\qquad
\commutator{\tilde\Psi^\dagger(x)}{\tilde\Psi^\dagger(x')}
=
0
\,,
\qquad
\commutator{\tilde\Psi(y)}{\tilde\Psi(y')}
=
0
\,.
\end{equation}

The decompositions in 
Fig.~\ref{fig:TriangleDecomposition} 
change the creation operator 
$\Psi^\dagger(\ell)$ 
as follows:
\begin{equation}\label{BasicTypeWaveFun}
\Psi^\dagger(\ell)
\,\to\,
\left\{
  \begin{array}{ll}
   \displaystyle
    \kappa {\tinyspace}
    \Psi^\dagger(\ell{\negdbltinyspace}+{\negdbltinyspace}1)
    &
    \hbox{[\,$\ell {\negdbltinyspace}\ge{\negdbltinyspace} 1$\,]}
   \\
   \displaystyle
    \kappa{\tinyspace}
    \Psi^\dagger(\ell{\negdbltinyspace}-{\negdbltinyspace}1)
    &
    \hbox{[\,$\ell {\negdbltinyspace}\ge{\negdbltinyspace} 2$\,]}
   \\
   \displaystyle
    \kappa{\tinyspace}
    \Psi^\dagger(\ell{\negdbltinyspace}-{\negdbltinyspace}1)
    &
    \hbox{[\,$\ell {\negdbltinyspace}\ge{\negdbltinyspace} 2$\,]}
   \\
   \displaystyle
    \kappa
    &
    \hbox{[\,$\ell {\negdbltinyspace}={\negdbltinyspace} 3$\,]}
   \\
   \displaystyle
    \kappa
    &
    \hbox{[\,$\ell {\negdbltinyspace}={\negdbltinyspace} 1$\,]}
   \\
   \displaystyle
    \kappa{\tinyspace}
    \sum_{n=1}^{\ell}{\negtrpltinyspace}
      \Psi^\dagger(n)
      \Psi^\dagger(\ell{\negdbltinyspace}-{\negdbltinyspace}n{\negdbltinyspace}+{\negdbltinyspace}1)
    &
    \hbox{[\,$\ell {\negdbltinyspace}\ge{\negdbltinyspace} 1$\,]}
   \\
   \displaystyle
    \kappa
    \sum_{\ell'=1}^{\infty}{\negtinyspace}
      \Psi^\dagger(\ell{\negdbltinyspace}+{\negdbltinyspace}\ell'{\negdbltinyspace}+{\negdbltinyspace}1)
      \ell'{\halftinyspace}
      \Psi(\ell')
    \hspace{30pt}
    &
    \hbox{[\,$\ell {\negdbltinyspace}\ge{\negdbltinyspace} 1$\,]}
  \end{array}
\right.
\,.
\end{equation}
Each line of \rf{BasicTypeWaveFun} 
corresponds to a specific case in 
Fig.~\ref{fig:TriangleDecomposition}.
In Fig.~\ref{fig:TriangleDecomposition} {\bf f)}, 
the boundary splits into 
two boundaries of lengths $n$ and $\ell - n + 1$; 
in Fig.~\ref{fig:TriangleDecomposition} {\bf g)}, 
the boundary merges with another boundary of length $\ell'$. 
Note that the last case in \rf{BasicTypeWaveFun}, 
which corresponds to Fig.~\ref{fig:TriangleDecomposition} {\bf g)},  
accounts for $\ell'$ distinct configurations 
for the possible positions 
of the marked point on the merged boundary.
Therefore, 
the creation operator $\Psi^\dagger(\ell)$ 
is transformed by the removal of a triangle, as follows: 
\begin{eqnarray}\label{PeelingDecomposition}
\Psi^\dagger(\ell)
\!\!&\to&\!\!
\Psi^{{\rm [Decomp]}\dagger}(\ell) =
\kappa \bigg(
    \Psi^\dagger(\ell{\negdbltinyspace}+{\negdbltinyspace}1)
+
    2{\tinyspace}\theta_{\ell,2}{\tinyspace}
    \Psi^\dagger(\ell{\negdbltinyspace}-{\negdbltinyspace}1)
+
    \delta_{\ell,3}
+
    \delta_{\ell,1}
\nonumber\\&&\!\!\phantom{%
\Psi^{{\rm [Decomp]}\dagger}(\ell) =
\kappa \bigg(
}%
+
    \sum_{n=1}^{\ell}{\negtrpltinyspace}
      \Psi^\dagger(n)
      \Psi^\dagger(\ell{\negdbltinyspace}-{\negdbltinyspace}n{\negdbltinyspace}+{\negdbltinyspace}1)
+
    \sum_{\ell'=1}^{\infty}{\negtinyspace}
      \Psi^\dagger(\ell{\negdbltinyspace}+{\negdbltinyspace}\ell'{\negdbltinyspace}+{\negdbltinyspace}1)
      \ell'{\halftinyspace}
      \Psi(\ell')
\bigg)
,
\qquad
\end{eqnarray}
where 
$\theta_{\ell,k} {\negtrpltinyspace}={\negtrpltinyspace} 1$ if 
$\ell {\negtrpltinyspace}\ge{\negtrpltinyspace} k$ 
and 
$\theta_{\ell,k} {\negtrpltinyspace}={\negtrpltinyspace} 0$ if 
$\ell {\negtrpltinyspace}<{\negtrpltinyspace} k$. 
Using the peeling decomposition 
\rf{PeelingDecomposition} and \rf{OneOverLstepMovement},
one finds that 
the Hamiltonian satisfying the ``no big-bang condition'' \rf{NoBigBangCondition} takes the form
\begin{eqnarray}\label{BasicTypeHamiltonianLength}
\Hop
\!\!&=&\!\!
\sum_{\ell=1}^\infty{\negtinyspace}
  \Psi^\dagger(\ell)
  {\tinyspace}\ell{\tinyspace}\Psi(\ell)
-
  \kappa
  \sum_{\ell=1}^\infty{\negtinyspace}
    \Psi^\dagger(\ell{\negdbltinyspace}+{\negdbltinyspace}1)
  {\tinyspace}\ell{\tinyspace}\Psi(\ell)
-
  2{\tinyspace}\kappa
  \sum_{\ell=2}^\infty{\negtinyspace}
    \Psi^\dagger(\ell{\negdbltinyspace}-{\negdbltinyspace}1)
  {\tinyspace}\ell{\tinyspace}\Psi(\ell)
\nonumber\\&&\!\!
-{\dbltinyspace}
  3{\tinyspace}\kappa{\tinyspace} \Psi(3)
-
  \kappa{\tinyspace} \Psi(1)
\nonumber\\&&\!\!
-{\dbltinyspace}
  \kappa
  \sum_{\ell=1}^\infty \sum_{n=1}^\ell
    \Psi^\dagger(n)
    \Psi^\dagger(\ell{\negdbltinyspace}-{\negdbltinyspace}n
                     {\negdbltinyspace}+{\negdbltinyspace}1)
  {\tinyspace}\ell{\tinyspace}\Psi(\ell)
\nonumber\\&&\!\!
-{\dbltinyspace}
 \G{\tinyspace}\kappa
  \sum_{\ell=1}^\infty \sum_{\ell'=1}^\infty
    \Psi^\dagger(\ell{\negdbltinyspace}+{\negdbltinyspace}\ell'
                     {\negdbltinyspace}+{\negdbltinyspace}1)
    {\tinyspace}\ell'{\tinyspace}\Psi(\ell')
    {\tinyspace}\ell{\tinyspace}\Psi(\ell)
\,,
\end{eqnarray}
where the parameter $G$ is introduced to count the number of handles 
according to \rf{DTamplitude}.

\begin{figure}[t] 
\begin{center}
\includegraphics[width=16cm,pagebox=cropbox,clip]{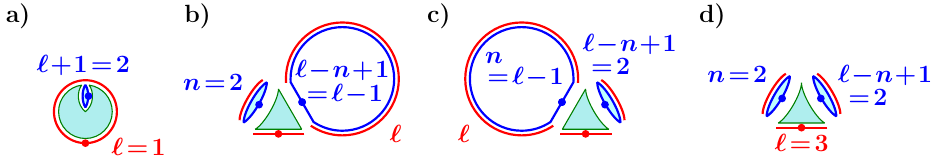}
  \caption{Special cases of the decompositions 
in Figs.~\ref{fig:TriangleDecomposition} {\bf a)} and {\bf f)} 
featuring 2-gons.
The solid red line represents the initial boundary of length $\ell$, 
while the solid blue line represents the boundary after the removal of a single triangle in each figure.
Polygons outlined in blue and filled in light blue denote 2-gons.
}
  \label{fig:TriangleDecompositionTwoGon}
\end{center}
\end{figure}

Note that in DT (basic type), 
adding $\delta_{\ell,2}$ 
to $\Psi^\dagger(\ell)$ reduces 
Figs.~\ref{fig:TriangleDecomposition} {\bf a)} and {\bf e)} to 
Fig.~\ref{fig:TriangleDecomposition} {\bf a)}, 
and 
Figs.~\ref{fig:TriangleDecomposition} {\bf b)}--{\bf d)} and {\bf f)} to 
Fig.~\ref{fig:TriangleDecomposition} {\bf f)}. 
In other words, the decomposition figures are reduced to only 
three cases, 
Figs.~\ref{fig:TriangleDecomposition} {\bf a)}, {\bf f)} and {\bf g)}. 
The geometric meaning of the term  
$\delta_{\ell,2}$ in 
$\Psi^\dagger(\ell) {\negtinyspace}+{\negtinyspace} \delta_{\ell,2}$ 
is that it generates a single 2-gon 
(a 2D surface with no handles, zero area, and a single boundary of length 2)
when $\ell {\negtrpltinyspace}={\negtrpltinyspace} 2$, 
and nothing 
when $\ell {\negtrpltinyspace}\neq{\negtrpltinyspace} 2$. 
When a triangle with a marked edge is removed, 
a 2-gon appears along each of its unmarked edges 
that lies on the boundary of the triangulated surface.
That is, if one or both of the remaining edges are on the boundary, 
a corresponding number of 2-gons will be created along those boundary edges.
Fig.~\ref{fig:TriangleDecompositionTwoGon} {\bf a)},
corresponding to 
Fig.~\ref{fig:TriangleDecomposition} {\bf e)},
is a specific variation of 
Fig.~\ref{fig:TriangleDecomposition} {\bf a)}, 
and
Figs.~\ref{fig:TriangleDecompositionTwoGon} {\bf b)}--{\bf d)},
corresponding to 
Figs.~\ref{fig:TriangleDecomposition} {\bf b)}--{\bf d)},
are specific variations of 
Fig.~\ref{fig:TriangleDecomposition} {\bf f)}.
If a 2-gon shares only a vertex with another 2-gon or with a triangle, 
they are treated as disconnected, just as in the case 
where two triangles share only a single vertex.
Consequently, the Hamiltonian \rf{BasicTypeHamiltonianLength} 
simplifies to 
\begin{eqnarray}\label{BasicTypeAnotherHamiltonianLength}
\Hop
\!\!&=&\!\!
\sum_{\ell=1}^\infty{\negtinyspace}
  \big(
    \Psi^\dagger(\ell) + \delta_{\ell,2}
  \big)
  {\tinyspace}\ell{\tinyspace}\Psi(\ell)
-
  2{\tinyspace}\Psi(2)
-
  \kappa
  \sum_{\ell=1}^\infty{\negtinyspace}
    \big(
      \Psi^\dagger(\ell{\negdbltinyspace}+{\negdbltinyspace}1)
      + \delta_{\ell+1,2}
    \big)
  {\tinyspace}\ell{\tinyspace}\Psi(\ell)
\nonumber\\&&\!\!
-{\dbltinyspace}
  \kappa
  \sum_{\ell=1}^\infty \sum_{n=1}^\ell
    \big(
      \Psi^\dagger(n)
      + \delta_{n,2}
    \big)
    \big(
      \Psi^\dagger(\ell{\negdbltinyspace}-{\negdbltinyspace}n
                       {\negdbltinyspace}+{\negdbltinyspace}1)
      + \delta_{\ell-n+1,2}
    \big)
  {\tinyspace}\ell{\tinyspace}\Psi(\ell)
\nonumber\\&&\!\!
-{\dbltinyspace}
 \G{\tinyspace}\kappa
  \sum_{\ell=1}^\infty \sum_{\ell'=1}^\infty
    \big(
      \Psi^\dagger(\ell{\negdbltinyspace}+{\negdbltinyspace}\ell'
                       {\negdbltinyspace}+{\negdbltinyspace}1)
      + \delta_{\ell+\ell'+1,2}
    \big)
    {\tinyspace}\ell'{\tinyspace}\Psi(\ell')
    {\tinyspace}\ell{\tinyspace}\Psi(\ell)
\,.
\end{eqnarray}
The $\delta_{\ell,2}$ appearing in 
$\Psi^\dagger(\ell) {\negtinyspace}+{\negtinyspace} \delta_{\ell,2}$ 
in the first term of \rf{BasicTypeAnotherHamiltonianLength} 
is canceled by the second term, $-2{\tinyspace}\Psi(2)$.
The  $\delta$-term in the third term corresponds to 
Fig.~\ref{fig:TriangleDecompositionTwoGon} {\bf a)} 
and matches the fifth term of \rf{BasicTypeHamiltonianLength}.
The $\delta$-term in the fourth term corresponds to 
Figs.~\ref{fig:TriangleDecompositionTwoGon} {\bf b)}--{\bf d)}, 
and matches the third and fourth terms of \rf{BasicTypeHamiltonianLength}, 
respectively.
The final $\delta$-term is identically zero 
and is included only as a formal expression.

The Hamiltonians 
\rf{BasicTypeHamiltonianLength} 
and 
\rf{BasicTypeAnotherHamiltonianLength} 
are equivalent.
Using the Laplace-transformed operators \rf{DiscreteLaplaceTransfWaveFun}, 
the Hamiltonian \rf{BasicTypeAnotherHamiltonianLength} takes the form 
\begin{eqnarray}
\Hop
\!\!&=&\!\!
\underset{z=0}{\rm Res}{\trpltinyspace}
 \frac{1}{z}
\bigg[
  \Big\{{\negqdrpltinyspace}-{\negdbltinyspace}
    \frac{1}{z^2}
    +
	\frac{
      V'{\negtinyspace}(z)}{z}
    {\negtinyspace}\Big({\negtinyspace}
      z \tildePsidag(z) + \frac{1}{z^2}
    \Big)
    -
    \kappa{\halftinyspace}z
    \Big({\negtinyspace}
      z \tildePsidag(z) + \frac{1}{z^2}
    \Big)^{{\negdbltinyspace}2}
  {\tinyspace}\Big\}
  \Big\{{\negtinyspace}
    z \pder{z}
    \Big( \frac{1}{z} \tilde\Psi \Big(\frac{1}{z}\Big)\!\Big)
  \!\Big\}
\nonumber\\&&\!\!\phantom{%
\underset{z=0}{\rm Res}{\trpltinyspace}
 \frac{1}{z}
\bigg[
}-
  \G {\tinyspace} \kappa{\halftinyspace}z
  {\negtinyspace}\Big({\negtinyspace}
    z \tildePsidag(z) + \frac{1}{z^2}
  \Big){\negdbltinyspace}
  \Big\{{\negtinyspace}
    z \pder{z}
    \Big( \frac{1}{z} \tilde\Psi \Big(\frac{1}{z}\Big)\!\Big)
  \!\Big\}^{{\negtinyspace}2}
{\trpltinyspace}\bigg]
\label{BasicTypeHamiltonianConjLength}
\,,
\end{eqnarray}
where $\underset{z=0}{\rm Res}$ denotes taking the residue 
at $z{\negdbltinyspace}={\negdbltinyspace}0$ 
in the complex $z$-plane. 
The derivative of the potential 
\begin{equation}\label{MxModelPotentialDer}
V'{\negtinyspace}(z)  \,\define\,  
z - \kappa{\halftinyspace}z^2\,,
\end{equation}
of the cubic matrix model appears 
in the kinetic term of the Hamiltonian \rf{BasicTypeHamiltonianConjLength}. 
The factor $\kappa{\halftinyspace}z$ 
in \rf{BasicTypeHamiltonianConjLength}, 
which appears not only in $V'(z)/z$ 
but also in front of two three-loop interactions, 
represents the operation of removing one triangle.
It is noteworthy that
the creation operator $\tilde\Psi^\dagger(z)$ 
always appears in \rf{BasicTypeHamiltonianConjLength} in the combination
$z \tilde\Psi^\dagger(z) + 1/z^2$.
This indicates that 
a single 2-gon is effectively being added 
to the disk amplitude. 
However, the tadpole term $-1/z^2$,
which appears as the first term in \rf{BasicTypeHamiltonianConjLength},
was originally introduced to cancel the 2-gon state 
represented by 
$\Psi^\dagger(2)$.

\subsubsection{Schwinger-Dyson equation}

Since the peeling decomposition reduces the number of triangles one by one,
all triangles are eventually removed after a finite number of steps.
Therefore, the amplitude defined by \rf{DTamplitude} is finite 
and satisfies the so-called Schwinger-Dyson (SD) equation,%
\footnote{%
In matrix models, 
this equation is commonly called the ``loop equation'',
but in this paper, 
we refer to it as the ``Schwinger-Dyson equation''
in accordance with the conventions of standard quantum field theory.
}
\begin{equation}\label{SDeqDT}
\lim_{T \to \infty} \pder{{\tinyspace}T}
\vac \E^{-T \Hop} {\tinyspace}
     \tilde\Psi^\dagger(x_1) \ldots \tilde\Psi^\dagger(x_\NN)
\cuum
\,=\,
0
\,.
\end{equation}

Let us consider the SD equation in the case of $\NN=1$. 
To this end, we employ the well-known technique of 
completing the square for the creation operators by applying the shift 
\begin{equation}\label{WaveFunShiftDef}
\tilde\Phi^\dagger(x)
\,\define\,
\tilde\Psi^\dagger(x)
-
\lambda(x)
\,,
\end{equation}
and then, the Hamiltonian 
\rf{BasicTypeHamiltonianConjLength} becomes 
\begin{eqnarray}\label{BasicTypeHamiltonianConjLengthShift}
\Hop
\!\!&=&\!\!
\underset{z=0}{\rm Res}{\trpltinyspace}
 \frac{\kappa}{z}
\bigg[
  z^3
  \big\{
    \OmegaXXX(z)
    -
      \tilde\Phi^\dagger(z)^{2}
  {\tinyspace}\big\}
  \Big\{{\negtinyspace}
    z \pder{z}
    \Big( \frac{1}{z} \tilde\Psi\Big(\frac{1}{z}\Big)\!\Big)
  \!\Big\}
\nonumber\\&&\!\!\phantom{%
\underset{z=0}{\rm Res}{\trpltinyspace}
 \frac{1}{z}
\bigg[
}-
  \G {\tinyspace}
  z^2
  \Big(
    \tilde\Phi^\dagger(z) +
    \frac{V'{\negtinyspace}(z)}
         {2{\halftinyspace}\kappa{\halftinyspace}z^3}
  {\tinyspace}\Big)
  \Big\{{\negtinyspace}
    z \pder{z}
    \Big( \frac{1}{z} \tilde\Psi\Big(\frac{1}{z}\Big)\!\Big)
  \!\Big\}^{{\negtinyspace}2}
{\dbltinyspace}\bigg]
\,,
\end{eqnarray}
where 
\begin{align}
\lambdaXXX(x)
&\define
  \frac{V'{\negtinyspace}(x)}
       {2{\halftinyspace}\kappa{\halftinyspace}x^3}
-
  \frac{1}{x^3}\,,
\label{BasicTypeLambdaDef}
\\
\OmegaXXX(x)
&\define
 \lambda(x)^2
 - \frac{1}{x^4}
 - \frac{1}{x^6}
\,.
\label{BasicTypeOmegaDef}
\end{align}

For $\NN=1$, 
the commutation relation between the Hamiltonian and
$\tilde\Psi^\dagger(x)$ is 
\begin{eqnarray}\label{BasicTypeHamiltonianCommutationRelationN1}
&&
\commutator{\Hop}
           {\tilde\Psi^\dagger(x)}
\cuum
\,=\,
-\,
\kappa {\dbltinyspace}
\pder{x}
  \Big(
    x^3
    \big\{
      \OmegaXXX(x)
      -
      \tilde\Phi^\dagger(x)^{2}
    \big\}
  {\negdbltinyspace}\Big)
\cuum
\,.
\end{eqnarray}
Then, 
the SD equation \rf{SDeqDT} with the ``no big-bang condition'' \rf{NoBigBangCondition} becomes%
\footnote{%
Details to be noted in this calculation are provided 
in Appendix \ref{app:CalculationConnectivity}.
\label{fn:AppConnectivity}
}
\begin{eqnarray}\label{BasicTypeSDeqN1}
0
\!\!&=&\!\!
\frac{1}{\kappa}
\lim_{T \to \infty}
\vac \E^{-T \Hop} {\dbltinyspace}
\Hop {\dbltinyspace} \tilde\Psi^\dagger(x) \cuum
\nonumber\\&=&\!\!
\pder{x}
  \Big(
    x^3
    \big\{
      \OmegaXXX(x)
      -
      \tilde{F}_2^{{\tinyspace}{\rm conn}}(x,x;G)
      -
      \tilde{F}_1^{{\tinyspace}{\rm conn}}(x;G)^2
    {\tinyspace}\big\}
  {\negdbltinyspace}\Big)
\,,
\end{eqnarray}
where 
\begin{eqnarray}
 \AmpFXXX_1^{{\tinyspace}{\rm conn}}(x;G)
 \!\!&\define&\!\!
\tilde{F}_1(x;G)
-
\lambdaXXX(x)
\,=\,
\lim_{T \to \infty}
\vac \E^{-T \Hop} {\tinyspace}
     \tilde\Phi^\dagger(x)
\cuum
\,,
\label{DefFcon1}
\\
 \AmpFXXX_2^{{\tinyspace}{\rm conn}}(x_1,x_2;G)
\!\!&\define&\!\!
 \AmpFXXX_2(x_1,x_2;G)
-
\tilde{F}_1(x_1;G)
\tilde{F}_1(x_2;G)
\,.
\label{DefFcon2}
\end{eqnarray}
Note that
$\tilde{F}_1^{{\tinyspace}{\rm conn}}(x;G)$ 
lacks a geometric interpretation and is therefore
{\it a formal} amplitude
of a 2D surface with one boundary.
In contrast,
$\tilde{F}_1(x;G)$ 
and 
$\tilde{F}_2^{{\tinyspace}{\rm conn}}(x_1, x_2;G)$ 
are the amplitudes of a connected 2D surface 
with one boundary and with two boundaries, respectively.

For $\NN=2$, 
the commutation relation between the Hamiltonian and 
$\tilde\Psi^\dagger(x_1) \tilde\Psi^\dagger(x_2)$ 
is 
\begin{eqnarray}\label{BasicTypeHamiltonianCommutationRelationN2}
&&
\commutator{\Hop}
           {\tilde\Psi^\dagger(x_1) \tilde\Psi^\dagger(x_2)}
\cuum
\nonumber\\
&&=\,
-\,
\kappa {\dbltinyspace}
\pder{x_1}
  \Big(
    x_1^3
    \big\{
      \OmegaXXX(x_1)
      -
      \tilde\Phi^\dagger(x_1)^{2}
    \big\}
  {\negdbltinyspace}\Big)
{\tinyspace}
\tilde\Psi^\dagger(x_2)
\cuum
\nonumber\\
&&\phantom{=\,}%
-
\kappa {\dbltinyspace}
\pder{x_2}
  \Big(
    x_2^3
    \big\{
      \OmegaXXX(x_2)
      -
      \tilde\Phi^\dagger(x_2)^{2}
    \big\}
  {\negdbltinyspace}\Big)
{\tinyspace}
\tilde\Psi^\dagger(x_1)
\cuum
\nonumber\\
&&\phantom{=\,}%
-
2{\tinyspace}\kappa{\tinyspace}\G{\dbltinyspace}
\pder{x_1}
\pder{x_2}
{\tinyspace}
  \frac{
        x_1^3 {\tinyspace}
        \big(
          \tilde\Phi^\dagger(x_1) +
          \frac{V'{\negtinyspace}(x_1)}
               {2{\tinyspace}\kappa{\halftinyspace}x_1^3}
        \big)
        -
        x_2^3 {\tinyspace}
        \big(
          \tilde\Phi^\dagger(x_2) +
          \frac{V'{\negtinyspace}(x_2)}
               {2{\tinyspace}\kappa{\halftinyspace}x_2^3}
        \big)
       }
       {x_1 {\negtinyspace}-{\negtinyspace} x_2}
\,.
\end{eqnarray}
Then, 
the SD equation \rf{SDeqDT} with the ``no big-bang condition'' \rf{NoBigBangCondition} 
becomes${}^{\ref{fn:AppConnectivity}}$
\begin{eqnarray}\label{BasicTypeSDeqN2}
0
\!\!&=&\!\!
\frac{1}{\kappa}
\lim_{T \to \infty}
\vac \E^{-T \Hop} {\dbltinyspace}
  \Hop {\dbltinyspace} \tilde\Psi^\dagger(x_1) \tilde\Psi^\dagger(x_2)
\cuum^{{\negtinyspace}{\rm conn}}
\nonumber\\&=&\!\!
\pder{x_1}
 \Big(
  x_1^3 {\tinyspace}
  \big\{
    \AmpFXXX_3^{{\tinyspace}{\rm conn}}(x_1,x_1,x_2;G)
    {\tinyspace}+{\tinyspace}
    2{\tinyspace}
    \AmpFXXX_1^{{\tinyspace}{\rm conn}}(x_1;G)
    \AmpFXXX_2^{{\tinyspace}{\rm conn}}(x_1,x_2;G)
  \big\}
 {\negtinyspace}\Big)
\nonumber\\&&\!\!
+{\trpltinyspace}
\pder{x_2}
 \Big(
  x_2^3 {\tinyspace}
  \big\{
    \AmpFXXX_3^{{\tinyspace}{\rm conn}}(x_1,x_2,x_2;G)
    {\tinyspace}+{\tinyspace}
    2{\tinyspace}
      \AmpFXXX_1^{{\tinyspace}{\rm conn}}(x_2;G)
      \AmpFXXX_2^{{\tinyspace}{\rm conn}}(x_1,x_2;G)
  \big\}
 {\negtinyspace}\Big)
\nonumber\\&&\!\!
+{\trpltinyspace}
2{\tinyspace}\G{\dbltinyspace}
\pder{x_1}
\pder{x_2}
{\tinyspace}
  \frac{
        x_1^3 {\tinyspace}
        \AmpFXXX_1^{{\tinyspace}{\rm conn}}(x_1;G)
        -
        x_2^3 {\tinyspace}
        \AmpFXXX_1^{{\tinyspace}{\rm conn}}(x_2;G)
       }
       {x_1 {\negtinyspace}-{\negtinyspace} x_2}
\,,
\end{eqnarray}
where 
the notation ``conn'' of $\vac \ldots \cuum^{{\negtinyspace}{\rm conn}}$ 
signifies that only connected 2D surfaces are considered. 
$\AmpFXXX_1^{{\tinyspace}{\rm conn}}(x;G)$ 
and 
$\AmpFXXX_2^{{\tinyspace}{\rm conn}}(x_1,x_2;G)$ 
are defined by 
\rf{DefFcon1} and \rf{DefFcon2}, respectively, 
and 
$\AmpFXXX_3^{{\tinyspace}{\rm conn}}(x_1,x_2,x_3;G)$ 
is defined by 
\begin{eqnarray}
 \AmpFXXX_3^{{\tinyspace}{\rm conn}}(x_1,x_2,x_3;G)
\!\!&\define&\!\!
 \AmpFXXX_3(x_1,x_2,x_3;G)
\nonumber\\&&\!\!
-{\trpltinyspace}
\tilde{F}_1(x_1;G)
\tilde{F}_2^{{\tinyspace}{\rm conn}}(x_2,x_3;G)
\nonumber\\&&\!\!
-{\trpltinyspace}
\tilde{F}_1(x_2;G)
\tilde{F}_2^{{\tinyspace}{\rm conn}}(x_3,x_1;G)
\nonumber\\&&\!\!
-{\trpltinyspace}
\tilde{F}_1(x_3;G)
\tilde{F}_2^{{\tinyspace}{\rm conn}}(x_1,x_2;G)
\nonumber\\&&\!\!
-{\trpltinyspace}
\tilde{F}_1(x_1;G)
\tilde{F}_1(x_2;G)
\tilde{F}_1(x_3;G)
\,.
\label{DefFcon3}
\end{eqnarray}

The SD equation \rf{BasicTypeSDeqN2} 
is derived from the ``multi-peeling decomposition''. 
Its first and second lines correspond to removing a single triangle 
from two boundaries with $x_1$ and $x_2$, respectively, 
while the last line represents merging the two boundaries into one. 
By transitioning to the ``single-peeling decomposition'' 
(the right-hand figure of Fig.~\ref{fig:DTamplitudeSingleSlicingPeeling}), 
we obtain the modified SD equation, 
\begin{eqnarray}\label{BasicTypeSDeqN2singlePeeling}
0
\!\!&=&\!\!
\pder{x_1}
 \Big(
  x_1^3 {\tinyspace}
  \big\{
    \AmpFXXX_3^{{\tinyspace}{\rm conn}}(x_1,x_1,x_2;G)
    {\tinyspace}+{\tinyspace}
    2{\tinyspace}
    \AmpFXXX_1^{{\tinyspace}{\rm conn}}(x_1;G)
    \AmpFXXX_2^{{\tinyspace}{\rm conn}}(x_1,x_2;G)
  \big\}
 {\negdbltinyspace}\Big)
\nonumber\\&&\!\!
+{\trpltinyspace}
\G{\dbltinyspace}
\pder{x_1}
\pder{x_2}
{\tinyspace}
  \frac{
        x_1^3 {\tinyspace}
        \AmpFXXX_1^{{\tinyspace}{\rm conn}}(x_1;G)
        -
        x_2^3 {\tinyspace}
        \AmpFXXX_1^{{\tinyspace}{\rm conn}}(x_2;G)
       }
       {x_1 {\negtinyspace}-{\negtinyspace} x_2}
\,.
\end{eqnarray}

For $\NN \ge 3$, 
using with the ``no big-bang condition'' \rf{NoBigBangCondition}, 
the SD equation \rf{SDeqDT} becomes${}^{\ref{fn:AppConnectivity}}$
\begin{eqnarray}\label{BasicTypeSDeqGeneralNN}
0
\!\!&=&\!\!
\frac{1}{\kappa}
\lim_{T \to \infty}
\vac \E^{-T \Hop} {\dbltinyspace}
\Hop {\dbltinyspace} \prod_{k=1}^\NN{\negtinyspace} \tildePsidag(x_k)
 \cuum^{{\negtinyspace}{\rm conn}}
\nonumber\\&=&\!\!
\sum_{i=1}^\NN {\negtinyspace}
  \pder{x_i}
    \bigg({\negtinyspace}
     x_i^3 \Big\{
      \tilde{F}_{N+1}^{{\tinyspace}{\rm conn}}(x_i,x_i,\X_{I \backslash \{i\}};G)
      + 
\sum_{I_1 \cup I_2=I \backslash \{i\}}
      \tilde{F}_{|I_1|+1}^{{\tinyspace}{\rm conn}}(x_i,\bm{x}_{I_1};G)
      \tilde{F}_{|I_2|+1}^{{\tinyspace}{\rm conn}}(x_i,\bm{x}_{I_2};G)
     \Big\}
\nonumber\\&&\!\!
\phantom{%
\sum_{i=1}^\NN {\negtinyspace}
  \pder{x_i}
    \bigg({\negtinyspace}
}{\negqdrpltinyspace}%
+
\G {\negdbltinyspace}
\sum_{{\scriptstyle j=1} \atop (j \neq i)}^{\NN}
{\negtrpltinyspace}
  \pder{x_j}
     \frac{
           x_i^3 {\tinyspace}
           \tilde{F}_{N-1}^{{\tinyspace}{\rm conn}}(\X_{I \backslash \{j\}};G)
           -
           x_j^3 {\tinyspace}
           \tilde{F}_{N-1}^{{\tinyspace}{\rm conn}}(\X_{I \backslash \{i\}};G)
          }
          {x_i - x_j}
    \bigg)
\,,
\end{eqnarray}
where $I=\{1, \ldots, \NN\}$, $\X_I=\{x_1, \ldots, x_\NN\}$, and
$I_1=\{i_1,\ldots,i_{|I_1|}\}$, $I_2=\{i_{{|I_1|}+1},\ldots,i_{N-1}\}$ 
are disjoint subsets of $I \backslash \{i\}$, $\bm{x}_{I_1}=\{x_{i_1}, \ldots, x_{i_{|I_1|}}\}$, $\bm{x}_{I_2}=\{x_{i_{|I_1|+1}}, \ldots, x_{i_{N-1}}\}$.
The shifted amplitude $\AmpFXXX_1^{{\tinyspace}{\rm conn}}(x{\halftinyspace};G)$ 
is defined by 
\rf{DefFcon1}, and the other 
$\AmpFXXX_\NN^{{\tinyspace}{\rm conn}}(x_1,\ldots,x_\NN;G)$ 
[\,$\NN \ge 2$\,]
are defined as 
\begin{equation}\label{DefFconN}
 \AmpFXXX_\NN^{{\tinyspace}{\rm conn}}(x_1,\ldots,x_\NN;G)
 \,\define\,
 \hbox{connected part of}\ 
 \AmpFXXX_\NN(x_1,\ldots,x_\NN;G)
\qquad\quad
\hbox{[\,$\NN \ge 2$\,]}
\,,
\end{equation}
in the same manner as in \rf{DefFcon2} and \rf{DefFcon3}.

For DT (basic type) with general $\NN$, 
the SD equations \rf{BasicTypeSDeqGeneralNN} 
follow the same pattern as in the case $N\!=\!2$, 
being derived from the ``multi-peeling decomposition''. 
Applying the ``single-peeling decomposition'',
we similarly obtain, including the case $N=1$,
\begin{eqnarray}\label{BasicTypeSDeqGeneralNNsinglePeeling}
0
\!\!&=&\!\!
  \pder{x_i}
    \bigg({\negtinyspace}
     x_i^3 \Big\{
      \tilde{F}_{N+1}^{{\tinyspace}{\rm conn}}(x_i,x_i,\X_{I \backslash \{i\}};G)
      + 
\sum_{I_1 \cup I_2=I \backslash \{i\}}
      \tilde{F}_{|I_1|+1}^{{\tinyspace}{\rm conn}}(x_i,\bm{x}_{I_1};G)
      \tilde{F}_{|I_2|+1}^{{\tinyspace}{\rm conn}}(x_i,\bm{x}_{I_2};G)
\Big\}
\nonumber\\&&\!\!
\phantom{%
  \pder{x_i}
    \bigg({\negtinyspace}
}{\negoctpltinyspace}%
- x_i^3\, \OmegaXXX(x_i)\, \delta_{N,1}
+
\G {\negdbltinyspace}
\sum_{{\scriptstyle j=1} \atop (j \neq i)}^{\NN}
{\negtrpltinyspace}
  \pder{x_j}
     \frac{
           x_i^3 {\tinyspace}
           \tilde{F}_{N-1}^{{\tinyspace}{\rm conn}}(\X_{I \backslash \{j\}};G)
           -
           x_j^3 {\tinyspace}
           \tilde{F}_{N-1}^{{\tinyspace}{\rm conn}}(\X_{I \backslash \{i\}};G)
          }
          {x_i - x_j}
    \bigg)
\,.
\end{eqnarray}
Here, expanding \rf{DTamplitude} in powers of $G$ (which counts the number of handles of 2D surfaces)
we write the connected amplitudes as
\begin{align}\label{AmplitudeGexpansionN}
\AmpFXXX_\NN^{{\tinyspace}{\rm conn}}(x_1,\ldots,x_\NN;G)
&=
\sum_{h=0}^\infty G^{\;\!h+\NN-1}
\AmpFXXX_{N}^{{\tinyspace}{\rm conn}(h)}{\negdbltinyspace}(x_1,\ldots,x_\NN)
\,.
\end{align}
For small $\NN$, we have 
\begin{align}
\AmpFXXX_1^{{\tinyspace}{\rm conn}}(x;G)
&=
\sum_{h=0}^\infty G^{\;\!h} 
\AmpFXXX_{1}^{{\tinyspace}{\rm conn}(h)}{\negdbltinyspace}(x)
\,,
\label{AmplitudeGexpansion1}
\\
\AmpFXXX_2^{{\tinyspace}{\rm conn}}(x_1,x_2;G)
&=
\sum_{h=0}^\infty G^{\;\!h+1} 
\AmpFXXX_{2}^{{\tinyspace}{\rm conn}(h)}{\negdbltinyspace}(x_1,x_2)
\,.
\label{AmplitudeGexpansion2}
\end{align}
$\AmpFXXX_{1}^{{\tinyspace}{\rm conn}(0)}{\negdbltinyspace}(x)$ 
and 
$\AmpFXXX_{2}^{{\tinyspace}{\rm conn}(0)}{\negdbltinyspace}(x_1,x_2)$ 
represent the amplitudes with 
disk topology and with the cylinder topology, respectively.

\subsection{Amplitudes and topological recursion}\label{subsec:TR_Basic}

\subsubsection{Disk amplitude}\label{subsubsec:Disk_Basic}

Using the expansion \rf{AmplitudeGexpansionN},
we extract the zeroth-order term in \rf{BasicTypeSDeqN1} with respect to $G$, which yields
\begin{eqnarray}\label{BasicTypeSDeqN1G0}
0
\!\!&=&\!\!
\pder{x}
  \Big({\tinyspace}
    x^3 
    \big\{{\negtinyspace}
      \big( \tilde{F}_1^{{\tinyspace}{\rm conn(0)}}(x)
      \big)^2
      -
      \OmegaXXX(x)
    \big\}
  {\neghalftinyspace}\Big)
\,.
\end{eqnarray}
Integrating \rf{BasicTypeSDeqN1G0} with respect to $x$, we obtain
\begin{equation}\label{BasicTypeDiskAmpTemp}
 \AmpFXXX_1^{{\tinyspace}{\rm conn}(0)}{\negdbltinyspace}(x)
\,=\,
\sqrt{\OmegaXXX(x) + \frac{\ConstXXX_1}{x^3}
}
\,,
\end{equation}
where $\ConstXXX_1$ denotes an integration constant that depends on $\kappa$.
Here, we assume that
$\AmpFXXX_1^{{\tinyspace}{\rm conn}(0)}{\negdbltinyspace}(x)$
has a single cut in the complex $x$-plane.%
\footnote{%
The physical meaning of this assumption 
is explained in detail in \cite{MM:BIPZ}.
} 
That is, 
\begin{equation}\label{BasicTypeDiskAmpAssumption}
 \AmpFXXX_1^{{\tinyspace}{\rm conn}(0)}{\negdbltinyspace}(x)
\,=\,
 \frac{1}{2{\tinyspace}x}
 \Big({\tinyspace}
   1 - \frac{c}{\kappa{\tinyspace}x}
 \Big)
 \sqrt{
    1
    -
    \frac{c{\dbltinyspace}'}{\kappa{\tinyspace}x}
 }
\,,
\end{equation}
where 
$c$ and $c{\dbltinyspace}'$ are functions of $\kappa$ 
to be determined below.
Comparing \rf{BasicTypeSDeqN1G0} and \rf{BasicTypeDiskAmpAssumption}, 
we obtain 
\begin{equation}\label{BasicTypeEdgeConditionC}
c{\dbltinyspace}'
{\tinyspace}={\tinyspace}
\frac{1 - c^2}{2{\tinyspace}c}
\,,
\qquad\quad
c {\dbltinyspace} ( 1 {\negtinyspace}-{\negtinyspace} c^2 )
{\tinyspace}={\tinyspace}
8 \kappa^2
\,,
\qquad\quad
C_1
{\tinyspace}={\tinyspace}
\frac{(1-c)(3{\tinyspace}c-1)}{8 \kappa c}
\,.
\end{equation}

Here, we discuss the valid ranges of the parameters $x$ and $\kappa$ for the disk amplitude $ \AmpFXXX_1^{{\tinyspace}{\rm conn}(0)}{\negdbltinyspace}(x)$.
To ensure consistency with the enumerative definitions given in \rf{DTamplitude} and \rf{DTamplitudeMatrixModel} for the DT amplitude, the expression \rf{BasicTypeDiskAmpAssumption} must admit a power series expansion of the form:
\begin{equation}\label{disk_expansion_basic}
\AmpFXXX_1^{(0)}{\negdbltinyspace}(x)
\,=\,
 \lambdaXXX(x)
 +
 \AmpFXXX_1^{{\tinyspace}{\rm conn}(0)}{\negdbltinyspace}(x)
\,=\,
\sum_{N_2\ge 1}\sum_{\ell_1\ge 1}\mathfrak{N}_{N_2, \ell_1}x^{-\ell-1}\kappa^{N_2}
\,,
\end{equation}
where $\mathfrak{N}_{N_2, \ell_1}$ denotes the number of triangulated disks with
$N_2$ triangles and boundary length $\ell_1$.
This expansion \rf{disk_expansion_basic} is valid only if the parameters $x$ and $\kappa$ lie within the domain of convergence of the series. 
Specifically, the expansion holds when
\begin{equation}\label{BasicRangexkappa}
0 \le x_{\rm c} \le x < \infty
\,,
\qquad\quad
0 \le \kappa \le \kappa_{\rm c}
\,,
\end{equation}
where $x_{\rm c}$ and $\kappa_{\rm c}$ are critical values 
determined by the radii of convergence around infinity and the origin, respectively. 
These critical values will be derived in the following analysis.

First, one finds three solutions for the parameter $c$ from the second equation in \rf{BasicTypeEdgeConditionC}, namely $c {\negdbltinyspace}={\negdbltinyspace} -1,0,1$ 
for $\kappa {\negdbltinyspace}={\negdbltinyspace} 0$.
Furthermore, the behavior of these solutions near $\kappa {\negdbltinyspace}={\negdbltinyspace} 0$ is given by
\begin{equation}
c {\tinyspace}\sim{\tinyspace} -1 - 4 \kappa^2\,,\;8 \kappa^2\,,\; 
1 - 4 \kappa^2\,,
\end{equation}
respectively.
To analyze the behavior of the disk amplitude $ \AmpFXXX_1^{{\tinyspace}{\rm conn}(0)}{\negdbltinyspace}(x)$ in \rf{BasicTypeDiskAmpAssumption}, 
we focus on its expansion around $x {\negdbltinyspace}={\negdbltinyspace} \infty$. 
The leading term of this expansion is
\begin{equation}\label{BasicTypeAmpExpansion}
 \AmpFXXX_1^{(0)}{\negdbltinyspace}(x)
\,=\,
 \frac{(3{\tinyspace}c - 1) {\tinyspace} \kappa}
 {c^2 {\tinyspace} (c + 1)}{\dbltinyspace}x^{-2}
+ {\cal O}(x^{-3})
\,.
\end{equation}
The coefficient of the leading term behaves as
\begin{equation} 
\frac{(3{\tinyspace}c - 1) {\tinyspace} \kappa}{c^2 {\tinyspace} (c + 1)}
 {\tinyspace}\sim{\tinyspace}
\frac{1}{\kappa}\,,\; -\frac{1}{64 \kappa^3}\,,\; \kappa\,,
\end{equation}
for the three respective solutions near $\kappa {\negdbltinyspace}={\negdbltinyspace} 0$. 
Since the expansion in \rf{disk_expansion_basic} must contain only positive powers of $\kappa$, 
we select the solution that behaves as $c {\negdbltinyspace}\sim{\negdbltinyspace} 1-4\kappa^2$ near 
$\kappa {\negdbltinyspace}={\negdbltinyspace} 0$.

The expansion \rf{disk_expansion_basic} remains valid for small values of $\kappa$
until the solution intersects another branch, which behaves as $c {\negdbltinyspace}\sim{\negdbltinyspace} 8\kappa^2$ near $\kappa {\negdbltinyspace}={\negdbltinyspace} 0$.
The expansion ceases to be applicable at the critical value 
$\kappa {\negdbltinyspace}={\negdbltinyspace} \kappa_{\rm c} {\negdbltinyspace}={\negdbltinyspace} 1/(2\cdot 3^{3/4})$, 
where the discriminant of the cubic polynomial 
$c {\dbltinyspace} ( 1 {\negtinyspace}-{\negtinyspace} c^2 ) {\negdbltinyspace}-{\negdbltinyspace} 8 \kappa^2$ in $c$, appearing in \rf{BasicTypeEdgeConditionC}, vanishes.
The critical value $c_{\rm c}$ corresponds to a multiple root of  $c {\dbltinyspace} ( 1 {\negtinyspace}-{\negtinyspace} c^2 ) {\negdbltinyspace}={\negdbltinyspace} 8 \kappa^2$, and is given by $c_{\rm c} {\negdbltinyspace}={\negdbltinyspace} 1/\sqrt{3}$.

Now we examine the validity of the expansion of \rf{BasicTypeDiskAmpAssumption} with respect to $x$ in the vicinity of infinity.
The expansion \rf{disk_expansion_basic} remains valid for $|\frac{c{\dbltinyspace}'}{\kappa{\tinyspace}x}|<1$, since the radius of convergence $u_{\rm c}$ of the Taylor expansion of the function $(1+u)^{1/2}$ near $u {\negdbltinyspace}={\negdbltinyspace} 0$ is $u_{\rm c} {\negdbltinyspace}={\negdbltinyspace} 1$. 
Consequently, the critical value $x_{\rm c}$ satisfies $c_{\rm c}{\dbltinyspace}'/(\kappa_{\rm c}{\tinyspace}x_{\rm c}) {\negdbltinyspace}={\negdbltinyspace} 1$
where  $c_{\rm c}{\dbltinyspace}' {\negdbltinyspace}={\negdbltinyspace} (1 - c_{\rm c}^2)/(2c_{\rm c})$, yielding $x_{\rm c} {\negdbltinyspace}={\negdbltinyspace} 2 \!\cdot\! 3^{1/4}$.

In summary, the critical values of the parameters $x,\kappa$, and $c$ are
\begin{equation}\label{BasicCriticalValues}
x_{\rm c} \,=\, 
2 \!\cdot\! 3^{1/4}
\,,
\qquad
\kappa_{\rm c} \,=\,
\frac{1}{2 \!\cdot\! 3^{3/4}}
\,,
\qquad
c_{\rm c} \,=\,
\frac{1}{\sqrt{3}}
\,.
\end{equation}
These values are universal for all amplitudes, even though they are derived from the disk amplitude within the framework of the SD equations. 
The continuum limit of this DT model is determined by the scaling behavior near these critical values, which will be discussed in Section \ref{sec:DTcontinuumLimit}.

Finally, we discuss a transformation of the parameter $x$.
The disk amplitude \rf{BasicTypeDiskAmpAssumption} 
can be written in the following form: 
\begin{equation}\label{BasicTypeDiskAmp}
 \AmpFXXX_1^{{\tinyspace}{\rm conn}(0)}{\negdbltinyspace}(x)
\,=\,
 \frac{1}{2{\tinyspace}x}
 \Big({\tinyspace}
   1 - \frac{c}{\kappa{\tinyspace}x}
 \Big)
 \sqrt{
    1
    -
    \frac{4{\tinyspace}\kappa}{c^2{\halftinyspace}x}
 }
\,=\,
M(x) \sqrt{\sigma(x)}
\,,
\end{equation}
where
\begin{align}\label{BasicTypeSpDef}
M(x) :=
\frac{1}{2x^3} \!\left(x-\frac{c}{\kappa}\right)\,,
\qquad
\sigma(x) :=
x \!\left(x-\frac{4\kappa}{c^2}\right)\,.
\end{align}
To derive the topological recursion from the Schwinger-Dyson equation, we introduce a variable $p \in \mathbb{P}^1$, called the Zhukovsky variable, defined by%
\footnote{%
In general, the Zhukovsky variable is introduced for a one-cut function with branch points $\alpha_1<\alpha_2$,
namely $\sqrt{\sigma(x)}=\sqrt{(x-\alpha_1)(x-\alpha_2)}$, by
$$
x(p)=\frac{\alpha_1+\alpha_2}{2} + \frac{\alpha_2-\alpha_1}{4}\left(p+p^{-1}\right),
$$
which gives $\sqrt{\sigma(x(p))}=(\alpha_2-\alpha_1)(p-p^{-1})/4$.
}
\begin{align}\label{BasicZ}
x(p):=
\frac{\kappa}{c^2}
\Big(
  2 + p + \frac{1}{p}
{\dbltinyspace}\Big)
\,,
\end{align}
and then the square root $\sqrt{\sigma(x)}$ in the disk amplitude yields
\begin{align}
\sqrt{\sigma(x(p))}
 =
\frac{\kappa}{c^2} \Big( p - \frac{1}{p} {\dbltinyspace}\Big)
\,.
\end{align}
Under this map, the branch point $x=0$ (resp. $4\kappa/c^2$) of 
the disk amplitude is mapped to $p=-1$ (resp. $1$), and 
the first sheet $\sqrt{\sigma(x)}$ (resp. the second sheet $-\sqrt{\sigma(x)}$) is mapped to the exterior of the unit disk $|p|\ge 1$ (resp. the interior, $|p|\le 1$).

\subsubsection{Cylinder amplitude}\label{subsubsec:Cylinder_Basic}
Using the expansion \rf{AmplitudeGexpansionN},
one can extract
the first-order term in $G$ from \rf{BasicTypeSDeqN2},
which yields
\begin{eqnarray}\label{BasicTypeSDeqN2G1}
0
\!\!&=&\!\!
  x_1^3 {\tinyspace}
    \AmpFXXX_1^{{\tinyspace}{\rm conn}(0)}{\negdbltinyspace}(x_1)
    \AmpFXXX_2^{{\tinyspace}{\rm conn}(0)}{\negdbltinyspace}(x_1,x_2)
\nonumber\\&&\!\!
+{\trpltinyspace}
\pder{x_2}
{\tinyspace}
  \frac{
        x_1^3 {\tinyspace}
        \AmpFXXX_1^{{\tinyspace}{\rm conn}(0)}{\negdbltinyspace}(x_1)
        -
        x_2^3 {\tinyspace}
        \AmpFXXX_1^{{\tinyspace}{\rm conn}(0)}{\negdbltinyspace}(x_2)
       }
       {2{\tinyspace} ( x_1 {\negtinyspace}-{\negtinyspace} x_2 )}
{\dbltinyspace}+{\dbltinyspace}
\ConstXXX_2(x_2)
\,,
\end{eqnarray}
where $\ConstXXX_2(x_2)$ is an integration constant with respect to $x_1$. 
Here, we assume that
$\AmpFXXX_2^{{\tinyspace}{\rm conn}(0)}{\negdbltinyspace}(x_1,x_2)$ has no poles at the zeros of $M(x)$ in \eqref{BasicTypeSpDef}.
This assumption fixes $\ConstXXX_2(x_2)$ as in \cite{Eynard:2004mh},
\begin{equation}
\ConstXXX_2(x_2)
\,=\,
-\,
\frac{1 - \frac{2 \kappa}{c^2 x_2}}
     {4 \sqrt{1 - \frac{4 \kappa}{c^2 x_2}}
     }
\,.
\end{equation}
Then, 
using the disk amplitude \rf{BasicTypeDiskAmp}, 
the cylinder amplitude becomes 
\begin{align}\label{BasicTypeCylinderAmp}
\AmpFXXX_2^{{\tinyspace}{\rm conn}(0)}{\negdbltinyspace}(x_1,x_2)
\,&=\,
 \frac{1}
      {2{\tinyspace} ( x_1 {\negtinyspace}-{\negtinyspace} x_2 )^2}
 \Bigg(
   \frac{
         1 - \frac{2{\tinyspace}\kappa}{c^2}
             ( \frac{1}{x_1} {\negtinyspace}+{\negtinyspace} \frac{1}{x_2} )
        }
        {
         \sqrt{1 - \frac{4{\tinyspace}\kappa}{c^2{\halftinyspace}x_1}}
         \sqrt{1 - \frac{4{\tinyspace}\kappa}{c^2{\halftinyspace}x_2}}
        }
   {\tinyspace}-{\tinyspace} 1
 \Bigg)
\nonumber\\
\,&=\,
\frac{1}
      {2{\tinyspace} ( x_1 {\negtinyspace}-{\negtinyspace} x_2 )^2}
\left(
\frac{x_1x_2-\frac{2\kappa}{c^2}\left(x_1+x_2\right)}
{\sqrt{\sigma(x_1)\sigma(x_2)}}-1\right),
\end{align}
and this can also be written by a bi-differential $B(p_{1},p_{2})$ as \cite{Eynard:2004mh},
\begin{align}\label{BasicBergman}
\AmpFXXX_2^{{\tinyspace}{\rm conn}(0)}{\negdbltinyspace}(x(p_1),x(p_2))\, dx(p_1)dx(p_2)
\,&=\,
\frac{dp_{1}dp_{2}}{\left(p_{1}-p_{2}\right)^2}
-\frac{dx(p_1)dx(p_2)}{\left(x(p_1)-x(p_2)\right)^2}
\nonumber\\
\,&=:\, B(p_{1},p_{2})-\frac{dx(p_1)dx(p_2)}{\left(x(p_1)-x(p_2)\right)^2}
\nonumber\\
\,&=\, \frac{dp_{1}dp_{2}}{\left(p_{1}p_{2}-1\right)^2}
\,=\, -B(p_{1},p_{2}^{-1})\,,
\end{align}
where $x=x(p)$ is defined by the map \eqref{BasicZ}.

\subsubsection{Topological recursion}

Integrating \rf{BasicTypeSDeqGeneralNNsinglePeeling} 
with respect to $x_i$, one finds 
\begin{eqnarray}\label{BasicTypeSDeqGeneralNNsinglePeelingInt}
0
\!\!&=&\!\!
     x_i^3 \Big\{
      \tilde{F}_{N+1}^{{\tinyspace}{\rm conn}}(x_i,x_i,\X_{I \backslash \{i\}})
      + 
\sum_{I_1 \cup I_2=I \backslash \{i\}}
      \tilde{F}_{|I_1|+1}^{{\tinyspace}{\rm conn}}(x_i,\bm{x}_{I_1};G)
      \tilde{F}_{|I_2|+1}^{{\tinyspace}{\rm conn}}(x_i,\bm{x}_{I_2};G)
- \OmegaXXX(x_i)\, \delta_{N,1}
     \Big\}
\nonumber\\&&\!\!
{\negqdrpltinyspace}%
+
\G {\negdbltinyspace}
\sum_{{\scriptstyle j=1} \atop (j \neq i)}^{\NN}
{\negtrpltinyspace}
  \pder{x_j}
     \frac{
           x_i^3 {\tinyspace}
           \tilde{F}_{N-1}^{{\tinyspace}{\rm conn}}(\X_{I \backslash \{j\}})
           -
           x_j^3 {\tinyspace}
           \tilde{F}_{N-1}^{{\tinyspace}{\rm conn}}(\X_{I \backslash \{i\}})
          }
          {x_i - x_j}
{\dbltinyspace}+{\dbltinyspace}
\ConstXXX_\NN(\X_{I \backslash \{i\}})
\,,
\end{eqnarray}
where $C_N(\X_{I \backslash \{i\}})$ is a function of $\X_{I \backslash \{i\}}$. 
Using the expansion \rf{AmplitudeGexpansionN} in powers of $G$, 
the equation \rf{BasicTypeSDeqGeneralNNsinglePeelingInt} with $i=1$ yields: 
\begin{align}
\label{BasicSD}
\AmpFXXX_{N}^{{\tinyspace}{\rm conn}(h)}{\negdbltinyspace}(\X_I)
&=
\frac{(-1)}{2\AmpFXXX_{1}^{{\tinyspace}{\rm conn}(0)}{\negdbltinyspace}(x_1)}
\Biggl[\AmpFXXX_{N+1}^{{\tinyspace}{\rm conn}(h-1)}{\negdbltinyspace}(x_1,x_1,\X_{I \backslash \{1\}})
\nonumber\\
&\hspace{7em}
+
\mathop{\sum_{h_1+h_2=h}}_{I_1 \cup I_2=\{2,\ldots,N\}}^{\textrm{no $(0,1)$}}
\AmpFXXX_{|I_1|+1}^{{\tinyspace}{\rm conn}(h_1)}{\negdbltinyspace}(x_1,\bm{x}_{I_1})
\AmpFXXX_{|I_2|+1}^{{\tinyspace}{\rm conn}(h_2)}{\negdbltinyspace}(x_1,\bm{x}_{I_2})
\nonumber\\
&\hspace{7em}
+
\sum_{i=2}^{\NN}
\frac{\AmpFXXX_{N-1}^{{\tinyspace}{\rm conn}(h)}{\negdbltinyspace}(\X_{I \backslash \{i\}})}
{(x_1-x_i)^2}\Biggr]
+R(x_1;\X_{I \backslash \{1\}})\,,
\\
R(x_1;\X_{I \backslash \{1\}})&:=
\frac{(-1)}{2 x_1^3 
\AmpFXXX_{1}^{{\tinyspace}{\rm conn}(0)}{\negdbltinyspace}(x_1)}
\Biggl[
-\sum_{i=2}^{\NN}
\pder{x_i}
\frac{x_i^3 \AmpFXXX_{N-1}^{{\tinyspace}{\rm conn}(h)}{\negdbltinyspace}(\X_{I \backslash \{1\}})}{x_1-x_i}
+ \ConstXXX_\NN^{(h)}(\X_{I \backslash \{1\}})
\Biggr]\,,
\label{rem_sd}
\end{align}
for $(h,N)\ne (0,1)$, 
where ``no $(0,1)$'' in the summation means that it excludes 
the disk amplitude 
$\AmpFXXX_{1}^{{\tinyspace}{\rm conn}(0)}{\negdbltinyspace}(x_1)$, 
and $\ConstXXX_\NN^{(h)}(\X_{I \backslash \{1\}})$ is 
a function of $\X_{I \backslash \{1\}}$. 
Assuming that the amplitudes 
$\AmpFXXX_{N}^{{\tinyspace}{\rm conn}(h)}{\negdbltinyspace}(\X_I)$  have 
no poles away from the branch cut $[0, 4\kappa/c^2]$ of the disk amplitude 
\rf{BasicTypeDiskAmp}, 
the equation \rf{BasicSD} is solved as in \cite{Eynard:2004mh} (see also \cite{Brini:2010fc}),
\begin{align}\label{sd_tr_d}
\AmpFXXX_{N}^{{\tinyspace}{\rm conn}(h)}{\negdbltinyspace}(\X_I)
&=
\frac{1}{2\pi \mathrm{i}}\oint_{x_0=x_1}
\frac{dx_0}{x_0-x_1}\,
\sqrt{\frac{\sigma(x_0)}{\sigma(x_1)}}\,
\AmpFXXX_{N}^{{\tinyspace}{\rm conn}(h)}{\negdbltinyspace}(x_0, \X_{I \backslash \{1\}})
\nonumber\\
&=
\frac{1}{2\pi \mathrm{i}}\oint_{[0, 4\kappa/c^2]}
\frac{dx_0}{x_1-x_0}\,
\sqrt{\frac{\sigma(x_0)}{\sigma(x_1)}}\,
\AmpFXXX_{N}^{{\tinyspace}{\rm conn}(h)}{\negdbltinyspace}(x_0, \X_{I \backslash \{1\}})
\nonumber\\
&=
\frac{(-1)}{2\pi \mathrm{i}}\oint_{[0, 4\kappa/c^2]}
\frac{dx_0\, dS_{p_0}(p_1)}{2\AmpFXXX_{1}^{{\tinyspace}{\rm conn}(0)}{\negdbltinyspace}(x_0)\, dx(p_1)}
\Biggl[
\AmpFXXX_{N+1}^{{\tinyspace}{\rm conn}(h-1)}{\negdbltinyspace}(x_0,x_0,\X_{I \backslash \{1\}})
\nonumber\\
&\hspace{8em}
+
\mathop{\sum_{h_1+h_2=h}}_{I_1 \cup I_2=\{2,\ldots,N\}}^{\textrm{no $(0,1)$}}
\mathfrak{\AmpFXXX}_{|I_1|+1}^{{\tinyspace}{\rm conn}(h_1)}{\negdbltinyspace}(x_0,\bm{x}_{I_1})\,
\mathfrak{\AmpFXXX}_{|I_2|+1}^{{\tinyspace}{\rm conn}(h_2)}{\negdbltinyspace}(x_0,\bm{x}_{I_2})
\Biggr],
\end{align}
where, in the second equality, the integration contour is deformed under the above assumption, and in the third equality we use the fact that 
$\sqrt{\sigma(x_0)}R(x_0; \X_{I \backslash \{1\}})$ has no poles along the branch cut in the variable $x_0$, and
\begin{align}
dS_{p_{0}}(p_1):&=
\frac{dx(p_1)}{x(p_1)-x(p_0)}\,
\sqrt{\frac{\sigma(x(p_0))}{\sigma(x(p_1))}}
\nonumber\\
&=
\frac{\left(p_0-p_0^{-1}\right) dp_1}{\left(p_1-p_0\right)\left(p_1-p_0^{-1}\right)}
\left(=
\int^{p_0}_{1/p_0} B(\cdot, p_1)
\right),
\label{differential_ds}
\\
\mathfrak{\AmpFXXX}_{|I|+1}^{{\tinyspace}{\rm conn}(h)}{\negdbltinyspace}(x_0,\bm{x}_I):&=
\AmpFXXX_{|I|+1}^{{\tinyspace}{\rm conn}(h)}{\negdbltinyspace}(x_0,\bm{x}_I)
+
\frac{\delta_{|I|,1}\, \delta_{h,0}}{2\left(x_0-x_i\right)^2}\,,\qquad 
i \in I\,.
\end{align}
Here the variables $p_i \in \mathbb{P}^1$ are introduced by the map \rf{BasicZ} as $x_i=x(p_i)$, and 
$B(\cdot, p_1)$ is the bi-differential in \rf{BasicBergman}.
We see that
$\mathfrak{\AmpFXXX}_{|I|+1}^{{\tinyspace}{\rm conn}(h)}{\negdbltinyspace}(x(p^{-1}), \bm{x}_I)=
-\mathfrak{\AmpFXXX}_{|I|+1}^{{\tinyspace}{\rm conn}(h)}{\negdbltinyspace}(x(p), \bm{x}_I)$, 
and that the integrand of the equation \rf{sd_tr_d} has no branch cuts. 
As a result, the equation \rf{sd_tr_d} yields
\begin{align}\label{top_rec_F_d}
\AmpFXXX_{N}^{{\tinyspace}{\rm conn}(h)}{\negdbltinyspace}(\X_I)
&=
\sum_{i=1,2}
\mathop{\mathrm{Res}}_{x_0=\alpha_i}
\frac{(-1)\, dx_0\, dS_{p_0}(p_{1})}{2\AmpFXXX_{1}^{{\tinyspace}{\rm conn}(0)}{\negdbltinyspace}(x_0)\, dx(p_1)}
\Biggl[
\AmpFXXX_{N+1}^{{\tinyspace}{\rm conn}(h-1)}{\negdbltinyspace}(x_0,x_0,\X_{I \backslash \{1\}})
\nonumber\\
&\hspace{8em}
+
\mathop{\sum_{h_1+h_2=h}}_{I_1 \cup I_2=\{2,\ldots,N\}}^{\textrm{no $(0,1)$}}
\mathfrak{\AmpFXXX}_{|I_1|+1}^{{\tinyspace}{\rm conn}(h_1)}{\negdbltinyspace}(x_0,\bm{x}_{I_1})\,
\mathfrak{\AmpFXXX}_{|I_2|+1}^{{\tinyspace}{\rm conn}(h_2)}{\negdbltinyspace}(x_0,\bm{x}_{I_2})
\Biggr]\,,
\end{align}
where $\alpha_1=0$ and $\alpha_2=4\kappa/c^2$. 
We now introduce the multi-differentials
\begin{align}
&
\omega_{2}^{(0)}(p_{1}, p_{2})=B(p_{1}, p_{2})
=\frac{dp_{1}dp_{2}}{\left(p_{1}-p_{2}\right)^2}\,,
\nonumber
\\
&
\omega_{N}^{(h)}(p_1,\ldots,p_N)=
\AmpFXXX_{N}^{{\tinyspace}{\rm conn}(h)}{\negdbltinyspace}(x(p_1),\ldots,x(p_N))\,
dx(p_{1}) \cdots dx(p_{N})
\ \ \textrm{for}\ \ (h,N)\ne (0,2)\,.
\label{F_mdiffW_d}
\end{align}
Then, the equation \rf{top_rec_F_d} can be written as \cite{Eynard:2004mh},
\begin{align}
\omega_{N}^{(h)}(\bm{p}_I)&=
\sum_{s=\pm 1}
\mathop{\mathrm{Res}}_{p_0=s}
K_{p_0}(p_1)
\Biggl[\omega_{N+1}^{(h-1)}(p_0,p_0^{-1},\bm{p}_{I \backslash \{1\}})
\nonumber
\\
&\hspace{8em}
+
\mathop{\sum_{h_1+h_2=h}}_{I_1 \cup I_2=\{2,\ldots,N\}}^{\textrm{no $(0,1)$}}
\omega_{|I_1|+1}^{(h_1)}(p_0, \bm{p}_{I_1})\,
\omega_{|I_2|+1}^{(h_2)}(p_0^{-1}, \bm{p}_{I_2})\Biggr]\,,
\label{top_rec_W_d}
\end{align}
where we used $\AmpFXXX_2^{{\tinyspace}{\rm conn}(0)}{\negdbltinyspace}(x(p_1),x(p_2))dx(p_1)dx(p_2)
=-B(p_{1}, p_{2}^{-1})$ from \eqref{BasicBergman}. 
Here $\bm{p}_I=\{p_1, \ldots, p_N\}$, and the recursion kernel
\begin{align}\label{top_rec_K}
K_{p_0}(p_1):=
\frac{dS_{p_0}(p_1)}{4{\tinyspace}\omega_1^{(0)}(p_0)}\,,
\end{align}
is defined.
Note that an extra factor of $2$ is introduced to the denominator of the topological recursion to account 
for the change of variables to $p_i$.

\begin{figure}[t] 
\begin{center}
\vspace{-5cm}
\hspace*{-1cm}
  \includegraphics[bb=0 0 595 842, width=130mm]{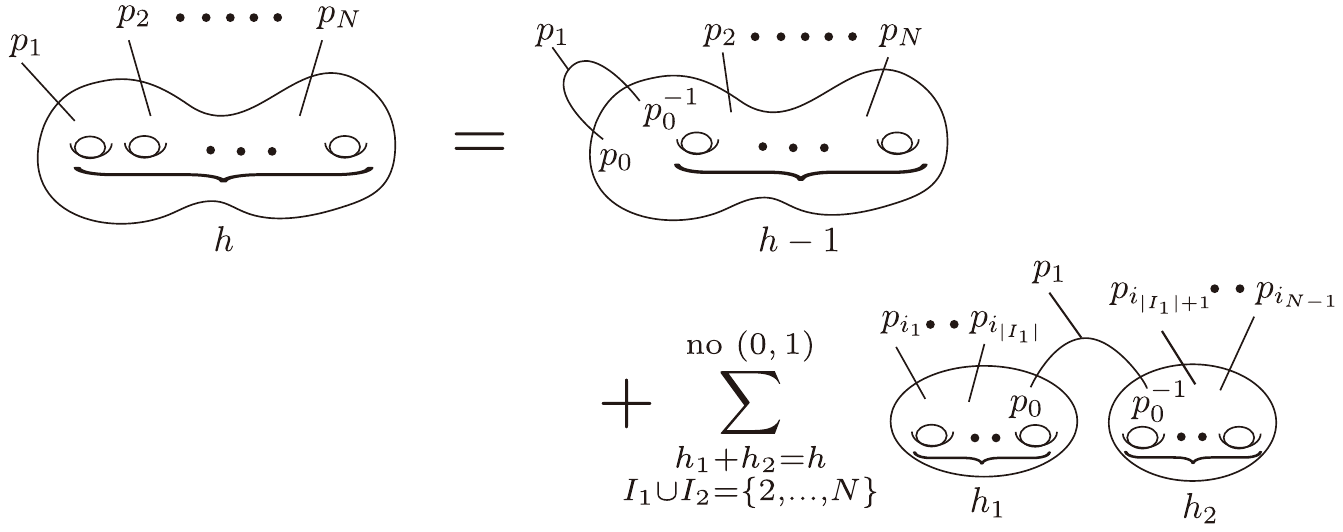}
 \vspace{-7.5cm}
\caption{Graphical presentation of the structure of the topological recursion \eqref{top_rec_W_d}.}
\label{fig:ceo_tr}
\end{center}
\end{figure}

The equation \rf{top_rec_W_d} is known as 
the Chekhov-Eynard-Orantin (CEO) topological recursion 
\cite{Chekhov:2006vd,Eynard:2007kz}.
Here we briefly comment on the recursive structure of the equation \rf{top_rec_W_d}.
Crucially, the presence of the term $R$ in \rf{BasicSD} makes it difficult to determine the amplitudes recursively from the SD equation, as it introduces an inhomogeneity and breaks the self-contained nature of the relations.
Owing to the absence of this term in \rf{top_rec_W_d}, we find homogeneous relations among the differentials $\omega_{N}^{(h)}$.
The recursive structure of the equation \rf{top_rec_W_d} can be seen in the graphical presentation in Fig.~\ref{fig:ceo_tr}, where a Riemann surface with $h$ handles and $N$ marked points
is depicted for the differential $\omega_{N}^{(h)}$.

We consider 
the sign-reversed Euler number, $2h+N-2$, of the Riemann surface with $h$ handles and $N$ marked points---shown on the left hand side of  Fig.~\ref{fig:ceo_tr}---as the grading number for the differential $\omega_{N}^{(h)}$.
The gradings of the differentials $\omega_{N+1}^{(h-1)}$, $\omega_{|I_1|+1}^{(h_1)}$, and $\omega_{|I_2|+1}^{(h_2)}$ that appear on the right hand side of the equation \rf{top_rec_W_d} are lower than that 
of $\omega_{N}^{(h)}$.
Thus, 
the equation 
\rf{top_rec_W_d} provides a recursion relation for $\omega_{N}^{(h)}$ with respect to this grading,\footnote{%
This property is the origin of the term ``topological recursion.''
}  
and the differentials $\omega_{N}^{(h)}$ for $2h+N \ge 3$ 
are determined iteratively from the initial data $\omega_{1}^{(0)}$ and $\omega_{2}^{(0)}$.

In the present case, the initial data is 
the spectral curve data $(\mathbb{P}^1; x,y,B)$ composed of the disk amplitude $y=\AmpFXXX_1^{{\tinyspace}{\rm conn}(0)}{\negdbltinyspace}(x)$ in \rf{BasicTypeDiskAmp},
\begin{align}
y=M(x) \sqrt{\sigma(x)}
=
\frac{1}{2x^3}\left(x-\gamma\right)\sqrt{x\left(x-\alpha\right)}\,,\ \
\alpha:=\frac{4\kappa}{c^2}\,,\ \
\gamma:=\frac{c}{\kappa}\,,
\end{align}
and the cylinder amplitude given by the bi-differential $B=B(p_{1},p_{2})$ in \eqref{F_mdiffW_d}.
Then, the topological recursion \eqref{top_rec_W_d} determines 
the differentials $\omega_{N}^{(h)}$ and the amplitudes 
$\AmpFXXX_N^{{\tinyspace}{\rm conn}(h)}$ 
for $2h+N \ge 3$, for example, as
\begin{align*}
&
\omega_{1}^{(1)}(p)=
\sum_{s= \pm 1}
\mathop{\mathrm{Res}}_{p_0=s} K_{p_0}(p)\, 
B(p_0,p_0^{-1})\,,
\\
&
\omega_{3}^{(0)}(p_{1},p_{2},p_{3})=
2\, \sum_{s= \pm 1}
\mathop{\mathrm{Res}}_{p_0=s} K_{p_0}(p_{1})\, 
B(p_0,p_{2})\, B(p_0^{-1},p_{3})\,,
\\
&
\omega_{2}^{(1)}(p_{1},p_{2})=
\sum_{s= \pm 1}
\mathop{\mathrm{Res}}_{p_0=s} K_{p_0}(p_{1})
\left[\omega_{3}^{(0)}(p_0,p_0^{-1},p_{2})
+ 2 B(p_0,p_{2})\, \omega_{1}^{(1)}(p_0^{-1})\right],
\\
&
\omega_{4}^{(0)}(p_{1},p_{2},p_{3},p_{4})=
2\, \sum_{s= \pm 1}
\mathop{\mathrm{Res}}_{p_0=s} K_{p_0}(p_{1})\, 
B(p_0,p_{2})\, \omega_{3}^{(0)}(p_0^{-1},p_{3},p_{4})\,,
\\
&
\omega_{1}^{(2)}(p)=
\sum_{s= \pm 1}
\mathop{\mathrm{Res}}_{p_0=s} K_{p_0}(p)
\left[\omega_{2}^{(1)}(p_0,p_0^{-1}) + \omega_{1}^{(1)}(p_0)\, \omega_{1}^{(1)}(p_0^{-1})
\right].
\end{align*}
Some computational results for amplitudes are listed in Appendix \ref{app:list_dt_basic}.
Here we note that the recursive structure implies that 
the amplitudes (or multi-differentials) for $2h+N \ge 3$ can be expressed in terms of the kernel differentials \cite{Bouchard:2007ys},
\begin{align}
\chi_i^{(n)}(x):=&\,
\mathop{\mathrm{Res}}_{x_0=\alpha_i}
\left(\frac{dS_{p_0}(p)}{\AmpFXXX_{1}^{{\tinyspace}{\rm conn}(0)}{\negdbltinyspace}(x_0)}\frac{dx_0}{(x_0-\alpha_i)^n}\right)
\nonumber\\
=&\,
\frac{dx}{(n-1)!\sqrt{\sigma(x)}}
\frac{\partial^{n-1}}{\partial x_0^{n-1}}\bigg|_{x_0=\alpha_i}
\frac{1}{M(x_0)\left(x-x_0\right)}\,,
\quad
i=1,2,\
n\ge 1\,,
\label{kernel_diff}
\end{align}
as
\begin{align}
\omega_{N}^{(h)}(p_1,\ldots,p_N)=
\sum_{i=1,2} \sum_{n_1,\ldots,n_N \ge 1}
C_{i;n_1,\ldots,n_N}^{(h)}\, 
\chi_i^{(n_1)}(x(p_1)) \cdots \chi_i^{(n_N)}(x(p_N))\,,
\label{kernel_exp}
\end{align}
where $\alpha_1=0$ and $\alpha_2=4\kappa/c^2$ are branch points of the disk amplitude, and 
the coefficients $C_{i;n_1,\ldots,n_N}^{(h)}$ do not depend on $x_1,\ldots,x_N$.


\section{Dynamical Triangulation (Strip Type)}
\label{sec:DTSTD}

\subsection{Fundamental properties}
\label{sec:PropertiesDiscretStripDT}

\begin{figure}[t] 
\begin{center}
\includegraphics[width=11cm,pagebox=cropbox,clip]{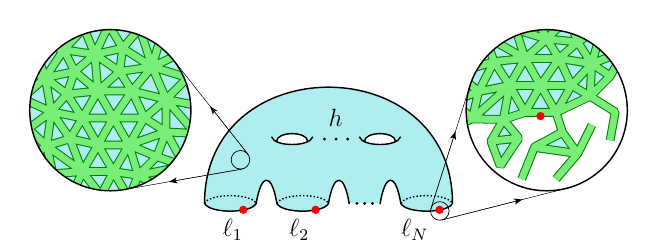}
  \caption{A (strip type) triangulated 2D surface 
with $\NN$ boundaries and $h$ handles.
Each thick green line denotes a strip. 
Red points on the boundaries denote marked points.
The insets enclosed by the two large circles show 
the detailed structures of the triangulated surface.
}
  \label{fig:DTamplitudeWithStrip}
\end{center}
\end{figure}

In this section,
we define and compute the amplitudes of
pure DT
of the strip type, 
formulated on a modified triangulation.
Unlike the original triangulation in DT (basic type), 
which contains only triangles, 
the modified triangulation also includes 
strips of a small but nonzero width, 
in addition to the triangles.
Each strip appears in one of the following three configurations:
\begin{enumerate}
\item
it connects two triangles, being sandwiched between their edges;
\item
it connects a triangle edge and a boundary edge, 
acting as a buffer between the triangle and the boundary;
\item
it connects two boundary edges and thus serves as an intermediate bridge 
between different regions of the triangulated surface.
\end{enumerate}
Not all triangles are directly adjacent to the boundary; 
some are connected through such strips. 
The boundary coincides with one or both edges of a strip.
A typical triangulated 2D surface 
is shown in Fig.~\ref{fig:DTamplitudeWithStrip}. 
We refer to this variant as ``DT (strip type)''.

\begin{figure}[t] 
\begin{center}
\includegraphics[width=15cm,pagebox=cropbox,clip]{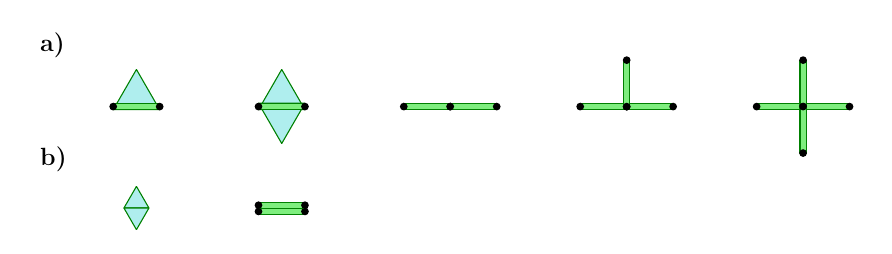}
  \caption{%
The figures in a) show various local connections between triangles and strips, 
while 
the figures in b) illustrate configurations that are not permitted. 
The black dots denote the endpoints of the strips, 
at which the strips connect to one another.
}
  \label{fig:DTstriptypeAllowedForbiddenConnection}
\end{center}
\end{figure}

In DT (basic type), triangles are connected directly along their edges.
By contrast, 
in DT (strip type), strip structures appear in the connecting regions, 
and the edge of a triangle attaches to the edge of a strip.
Consequently, triangles no longer touch one another directly.
Moreover, 
configurations can arise in which the endpoints of multiple strips 
are connected to each other
(see Fig.~\ref{fig:DTstriptypeAllowedForbiddenConnection}).
From the DT viewpoint,
the introduction of the strip structure may seem somewhat artificial.
However, 
in the dual graph of DT---which underlies 
the matrix model---the strip corresponds to 
the matrix propagator, making the construction natural.
In the next subsection, 
we explain the rationale behind the introduction of strips 
and how this modification affects the theory.

\subsubsection{Decomposition of triangulated surface}

\begin{figure}[t] 
\begin{center}
\includegraphics[width=15cm,pagebox=cropbox,clip]{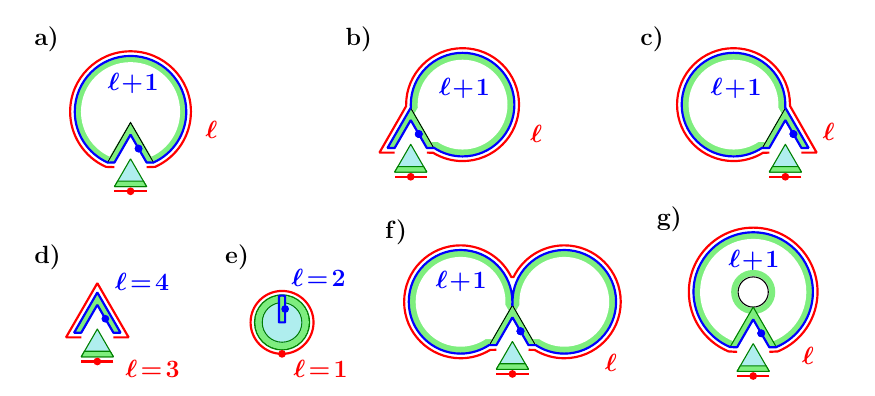}
  \caption{Decompositions by removing a triangle, leaving two strip-like edges.
The solid red line represents the initial boundary of length $\ell$, 
while the solid blue line represents the boundary 
after removing one triangle in each figure.
The strips adjacent to the boundary are shaded in light green.
Both the removed triangle and the strip-shaped region adjacent to 
its two edges are also shaded in light green.
The triangle is removed, while the strip-shaped region remains intact, 
and potentially serves as part of the new boundary of the triangulated surface.
}
  \label{fig:TriangleDecompositionLeavingStrip}
  \end{center}
\end{figure}

Let us make a slight modification to DT (basic type), as depicted in 
Fig.~\ref{fig:TriangleDecomposition}.
When a triangle is removed, 
two of its edges remain, 
as illustrated in Fig.~\ref{fig:TriangleDecompositionLeavingStrip} {\bf a)} 
for the decomposition in Fig.~\ref{fig:TriangleDecomposition} {\bf a)}.
We will refer to these edges as ``strips'' from now on.
Each remaining strip has a small but nonzero width 
and remains attached, facing the new boundary.
Under this modified decomposition, 
a strip remains in every case shown in 
Figs.~\ref{fig:TriangleDecomposition} {\bf b)}--{\bf g)}, 
yielding the corresponding modified versions depicted in 
Figs.~\ref{fig:TriangleDecompositionLeavingStrip} {\bf b)}--{\bf g)}, 
respectively.
Since a strip remains in all cases, 
all modified decompositions in 
Fig.~\ref{fig:TriangleDecompositionLeavingStrip} 
are equivalent to 
the one in  
Fig.~\ref{fig:TriangleDecompositionLeavingStrip} {\bf a)}. 
Notably, in Fig.~\ref{fig:TriangleDecompositionLeavingStrip} {\bf f)}, 
the boundary does not split, 
and in Fig.~\ref{fig:TriangleDecompositionLeavingStrip} {\bf g)}, 
the two boundaries do not merge.
From the perspective of the string field theory,
all the figures in Fig.~\ref{fig:TriangleDecompositionLeavingStrip} 
represent propagators; 
string interactions such as 
vanishing, splitting, and merging are absent.
Unlike the case of 2-gons, 
a strip is regarded as connected to another strip or a triangle 
if they share a common endpoint.

\begin{figure}[t] 
\begin{center}
\includegraphics[width=15cm,pagebox=cropbox,clip]{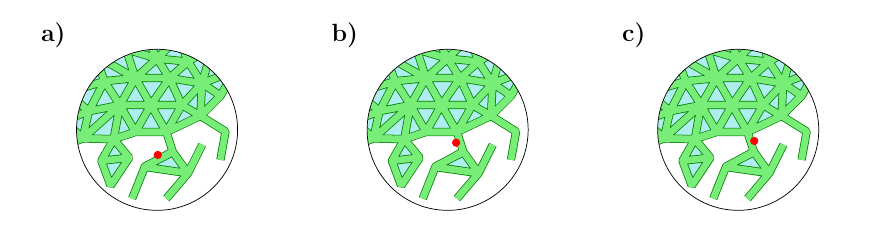}
  \caption{Several detailed structures 
of (strip type) triangulated 2D surface for a portion of a boundary.
Note that 
Figs.~{\bf b)} and {\bf c)} 
represent different configurations because the positions of the marked points differ.
The red point indicates a marked point on the boundary.
}
  \label{fig:DTamplitudeWithStripExamples}
  \end{center}
\end{figure}

\begin{figure}[t] 
\begin{center}
\includegraphics[width=15cm,pagebox=cropbox,clip]{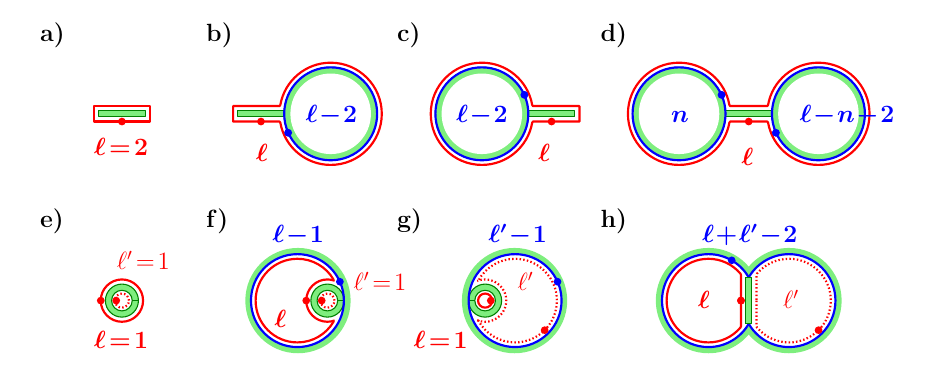}
  \caption{Decomposition by removing a strip.
The solid red line represents the initial boundary of length $\ell$, 
while the solid blue line represents the boundary after one strip is removed 
in each figure. 
The dashed red line indicates the adjacent boundary of length $\ell'$
which merges with the initial boundary. 
The strips shaded in light green indicate the ones being removed.
}
  \label{fig:StripDecomposition}
  \end{center}
\end{figure}

However, introducing strips creates a new issue: 
a single marked point now appears on all possible boundaries, 
as shown in
Fig.~\ref{fig:DTamplitudeWithStripExamples}. 
Figs.~\ref{fig:DTamplitudeWithStripExamples} {\bf b)} and {\bf c)}
display new configurations that must be addressed. 
To resolve this problem, 
we introduce strip removal via eight new decompositions 
shown in 
Figs.~\ref{fig:StripDecomposition} {\bf a)}--{\bf h)}.
Thus, 
the modified decomposition is carried out using 
Fig.~\ref{fig:TriangleDecompositionLeavingStrip} {\bf a)} 
together with 
Figs.~\ref{fig:StripDecomposition} {\bf a)}--{\bf h)}.%
\footnote{%
This decomposition is the standard one 
used in the literature; see, for example, 
\cite{preSFT:KKMW,SFT:Watabiki}. 
}

\subsubsection{String field theory of dynamical triangulation}

The decompositions in  
Fig.~\ref{fig:TriangleDecompositionLeavingStrip} {\bf a)} and 
Figs.~\ref{fig:StripDecomposition} {\bf a)}--{\bf h)} 
are
\begin{equation}\label{StripTypeWaveFun}
\Psi^\dagger(\ell)
\,\to\,
\left\{
  \begin{array}{ll}
   \displaystyle
    \kappa {\tinyspace} \Psi^\dagger(\ell{\negdbltinyspace}+{\negdbltinyspace}1)
    &
    \hbox{[\,$\ell {\negdbltinyspace}\ge{\negdbltinyspace} 1$\,]}
   \\
   \displaystyle
    1
    &
    \hbox{[\,$\ell {\negdbltinyspace}={\negdbltinyspace} 2$\,]}
   \\
   \displaystyle
    \Psi^\dagger(\ell{\negdbltinyspace}-{\negdbltinyspace}2)
    &
    \hbox{[\,$\ell {\negdbltinyspace}\ge{\negdbltinyspace} 3$\,]}
   \\
   \displaystyle
    \Psi^\dagger(\ell{\negdbltinyspace}-{\negdbltinyspace}2)
    &
    \hbox{[\,$\ell {\negdbltinyspace}\ge{\negdbltinyspace} 3$\,]}
   \\
   \displaystyle
    \sum_{n=1}^{\ell-3}{\negtrpltinyspace}
      \Psi^\dagger(n)
      \Psi^\dagger(\ell{\negdbltinyspace}-{\negdbltinyspace}n
                       {\negdbltinyspace}-{\negdbltinyspace}2)
    &
    \hbox{[\,$\ell {\negdbltinyspace}\ge{\negdbltinyspace} 4$\,]}
   \\
   \displaystyle
    \Psi(1)
    &
    \hbox{[\,$\ell {\negdbltinyspace}={\negdbltinyspace} 1$\,]}
   \\
   \displaystyle
      \Psi^\dagger(\ell{\negdbltinyspace}-{\negdbltinyspace}1)
      \Psi(1)
    \hspace{30pt}
    &
    \hbox{[\,$\ell {\negdbltinyspace}\ge{\negdbltinyspace} 2$\,]}
   \\
   \displaystyle
    \sum_{\ell'=2}^{\infty}{\negtinyspace}
      \Psi^\dagger(\ell'{\negdbltinyspace}-{\negdbltinyspace}1){\tinyspace}
      \ell'{\halftinyspace}
      \Psi(\ell')
    &
    \hbox{[\,$\ell {\negdbltinyspace}={\negdbltinyspace} 1$\,]}
   \\
   \displaystyle
    \sum_{\ell'=2}^{\infty}{\negtinyspace}
      \Psi^\dagger(\ell{\negdbltinyspace}+{\negdbltinyspace}\ell'{\negdbltinyspace}-{\negdbltinyspace}2){\tinyspace}
      \ell'{\halftinyspace}
      \Psi(\ell')
    \hspace{30pt}
    &
    \hbox{[\,$\ell {\negdbltinyspace}\ge{\negdbltinyspace} 2$\,]}
  \end{array}
\right.
\,.
\end{equation}
The decompositions in each line of \rf{StripTypeWaveFun} 
correspond, respectively, to 
Fig.~\ref{fig:TriangleDecompositionLeavingStrip} {\bf a)} and 
Figs.~\ref{fig:StripDecomposition} {\bf a)}--{\bf h)}.
Note that the last two cases in \rf{StripTypeWaveFun}, 
which correspond to Figs.~\ref{fig:StripDecomposition} {\bf g)} and {\bf h)},  
account for $\ell'$ distinct configurations 
representing the possible positions 
of the dot (marked point) on the merged boundary.
Therefore, upon removing a triangle or a strip,
the creation operator $\Psi^\dagger(\ell)$ 
is transformed as follows: 
\begin{eqnarray}
\Psi^\dagger(\ell)
\!\!&\to&\!\!
\Psi^{{\rm [Decomp]}\dagger}(\ell) =
    \kappa{\tinyspace}
    \Psi^\dagger(\ell{\negdbltinyspace}+{\negdbltinyspace}1)
+
    \delta_{\ell,2}
+
    2{\tinyspace}\theta_{\ell,3}{\tinyspace}
    \Psi^\dagger(\ell{\negdbltinyspace}-{\negdbltinyspace}2)
+
    \theta_{\ell,4}{\negtinyspace}
    \sum_{n=1}^{\ell-3}{\negtrpltinyspace}
      \Psi^\dagger(n)
      \Psi^\dagger(\ell{\negdbltinyspace}-{\negdbltinyspace}n
                       {\negdbltinyspace}-{\negdbltinyspace}2)
\nonumber\\&&\!\!\phantom{%
\Psi^{{\rm [Decomp]}\dagger}(\ell) =
}%
+
    \delta_{\ell,1}
      \Psi(1)
+
    \theta_{\ell,2}
      \Psi^\dagger(\ell{\negdbltinyspace}-{\negdbltinyspace}1)
      \Psi(1)
+
    \sum_{\ell'=2}^{\infty}{\negtinyspace}
      \Psi^\dagger(\ell{\negdbltinyspace}+{\negdbltinyspace}\ell'{\negdbltinyspace}-{\negdbltinyspace}2){\tinyspace}
      \ell'{\halftinyspace}
      \Psi(\ell')
\,.
\qquad
\end{eqnarray}
The Hamiltonian that implements the decompositions 
Fig.~\ref{fig:TriangleDecompositionLeavingStrip} {\bf a)} and 
Figs.~\ref{fig:StripDecomposition} {\bf a)}--{\bf h)} 
satisfies the same relation \rf{OneOverLstepMovement}.
Then, one finds the following Hamiltonian $\Hop$, 
which satisfies the ``no big-bang condition'' \rf{NoBigBangCondition}, 
i.e., $\Hop \!\> \cuum =0$, as
\begin{eqnarray}\label{StripTypeHamiltonianLength}
\Hop
\!\!&=&\!\!
\sum_{\ell=1}^\infty{\negtinyspace}
  \Psi^\dagger(\ell)
  {\tinyspace}\ell{\tinyspace}\Psi(\ell)
-
  \sum_{\ell=1}^\infty{\negtinyspace}
    \kappa{\tinyspace}
    \Psi^\dagger(\ell{\negdbltinyspace}+{\negdbltinyspace}1)
  {\tinyspace}\ell{\tinyspace}\Psi(\ell)
-
  2{\tinyspace}
  \Psi(2)
- 2
  \sum_{\ell=3}^\infty{\negtinyspace}
    \Psi^\dagger(\ell{\negdbltinyspace}-{\negdbltinyspace}2)
  {\tinyspace}\ell{\tinyspace}\Psi(\ell)
\nonumber\\&&\!\!
-{\dbltinyspace}
  \sum_{\ell=1}^\infty \sum_{n=1}^{\ell-3}
    \Psi^\dagger(n)
    \Psi^\dagger(\ell{\negdbltinyspace}-{\negdbltinyspace}n{\negdbltinyspace}-{\negdbltinyspace}2)
  {\tinyspace}\ell{\tinyspace}\Psi(\ell)
\nonumber\\&&\!\!
-{\dbltinyspace}
 \G \Big(
  \Psi(1) {\tinyspace} \Psi(1)
  +
  \sum_{\ell=1}^\infty \sum_{\ell'=1}^\infty
  \theta_{\ell+\ell',3}
  \Psi^\dagger(\ell{\negdbltinyspace}+{\negdbltinyspace}\ell'
                {\negdbltinyspace}-{\negdbltinyspace}2)
  {\tinyspace}\ell'{\tinyspace}\Psi(\ell')
  {\tinyspace}\ell{\tinyspace}\Psi(\ell)
 \Big)
\,,
\end{eqnarray}
where the parameter $G$ is introduced to count the number of handles 
according to \rf{DTamplitude}.

Note that in DT (strip type), 
adding $\delta_{\ell,0}$ to $\Psi^\dagger(\ell)$ 
reduces 
Figs.~\ref{fig:StripDecomposition} {\bf a)}--{\bf d)} to 
Fig.~\ref{fig:StripDecomposition} {\bf d)}, 
and 
Figs.~\ref{fig:StripDecomposition} {\bf e)}--{\bf h)} to 
Fig.~\ref{fig:StripDecomposition} {\bf h)}. 
In other words, the decomposition figures reduce to just 
three: 
Fig.~\ref{fig:TriangleDecompositionLeavingStrip} {\bf a)} and 
Figs.~\ref{fig:StripDecomposition} {\bf d)} and {\bf h)}. 
The geometric meaning of 
$\delta_{\ell,0}$ in 
$\Psi^\dagger(\ell) {\negtinyspace}+{\negtinyspace} \delta_{\ell,0}$ 
is to generate a single 0-gon 
(i.e., a 0D space with no handles, zero area, and a single boundary of length 0)
when $\ell {\negtrpltinyspace}={\negtrpltinyspace} 0$, 
and nothing 
when $\ell {\negtrpltinyspace}\neq{\negtrpltinyspace} 0$. 
When a strip with a marked edge is removed,
a 0-gon (i.e., a point) appears at each endpoint of the strip 
that lies on the boundary of the triangulated surface; 
that is, if one or both endpoints of the strip lie on the boundary,
then one or two 0-gons are created accordingly.
Consequently, the Hamiltonian \rf{StripTypeHamiltonianLength} 
reduces to the following form: 
\begin{eqnarray}\label{StripTypeAnotherHamiltonianLength}
\Hop
\!\!&=&\!\!
  \sum_{\ell=1}^\infty{\negdbltinyspace}
    \big(
      \Psi^\dagger(\ell)
      + \delta_{\ell,0}
    \big)
  {\tinyspace}\ell{\tinyspace}\Psi(\ell)
-
  \sum_{\ell=1}^\infty
    \kappa{\tinyspace}
    \big(
      \Psi^\dagger(\ell{\negdbltinyspace}+{\negdbltinyspace}1)
      + \delta_{\ell+1,0}
    \big)
  {\tinyspace}\ell{\tinyspace}\Psi(\ell)
\nonumber\\&&\!\!
-{\dbltinyspace}
  \sum_{\ell=1}^\infty \sum_{n=0}^{\ell-2}{\negdbltinyspace}
    \big(
      \Psi^\dagger(n)
      + \delta_{n,0}
    \big)
    {\negdbltinyspace}
    \big(
      \Psi^\dagger(\ell{\negdbltinyspace}-{\negdbltinyspace}n
                       {\negdbltinyspace}-{\negdbltinyspace}2)
      + \delta_{\ell-n-2,0}
    \big)
  {\tinyspace}\ell{\tinyspace}\Psi(\ell)
\nonumber\\&&\!\!
-{\dbltinyspace}
 \G
  \sum_{\ell=1}^\infty \sum_{\ell'=1}^\infty{\negdbltinyspace}
    \big(
      \Psi^\dagger(\ell{\negdbltinyspace}+{\negdbltinyspace}\ell'
                       {\negdbltinyspace}-{\negdbltinyspace}2)
      + \delta_{\ell+\ell'-2,0}
    \big)
  {\tinyspace}\ell'{\tinyspace}\Psi(\ell')
  {\tinyspace}\ell{\tinyspace}\Psi(\ell)
\,,
\end{eqnarray}
where $\Psi^\dagger(0) {\negtrpltinyspace}\define{\negtrpltinyspace} 0$. 
The $\delta_{\ell,0}$ that appears in 
$\Psi^\dagger(\ell) {\negtinyspace}+{\negtinyspace} \delta_{\ell,0}$ 
in the first and second terms of 
\rf{StripTypeAnotherHamiltonianLength} 
is identically zero and is included only as a formal expression.
The contribution $\delta$ in the third term 
corresponds to Figs.~\ref{fig:StripDecomposition} {\bf a)}--{\bf c)}, 
and matches the third and fourth terms of \rf{StripTypeHamiltonianLength}.
The final $\delta$ 
corresponds to Fig.~\ref{fig:StripDecomposition} {\bf e)}, 
and matches the $\Psi(1)\Psi(1)$ term 
on the last line of \rf{StripTypeHamiltonianLength}.

The Hamiltonians 
\rf{StripTypeHamiltonianLength} 
and 
\rf{StripTypeAnotherHamiltonianLength} 
are equivalent.
Using the Laplace-transformed operators defined in \rf{DiscreteLaplaceTransfWaveFun}, 
the Hamiltonian \rf{StripTypeAnotherHamiltonianLength} can be rewritten as follows:
\begin{eqnarray}
\Hop
\!\!&=&\!\!
\underset{z=0}{\rm Res}{\trpltinyspace}
 \frac{1}{z}
\bigg[
  \Big\{{\negqdrpltinyspace}-{\negdbltinyspace}
    1
    +
	\frac{
      V'{\negtinyspace}(z)}{z}
    {\negtinyspace}\big(
      z \tilde\Psi^\dagger(z) + 1
    {\tinyspace}\big)
    -
    \frac{1}{z^2}
    \big(
      z \tildePsidag(z) + 1
    {\tinyspace}\big)^{{\negtinyspace}2}
  {\tinyspace}\Big\}
  \Big\{{\negtinyspace}
    z \pder{z}
    \Big( \frac{1}{z} \tilde\Psi \Big(\frac{1}{z}\Big)\!\Big)
  \!\Big\}
\nonumber\\&&\!\!\phantom{%
\underset{z=0}{\rm Res}{\trpltinyspace}
 \frac{1}{z}
\bigg[
}-
  \frac{\G}{z^2}
    {\negtinyspace}\big(
      z \tilde\Psi^\dagger(z) + 1
    {\tinyspace}\big)
  \Big\{{\negtinyspace}
    z \pder{z}
    \Big( \frac{1}{z} \tilde\Psi \Big(\frac{1}{z}\Big)\!\Big)
  \!\Big\}^{{\negtinyspace}2}
{\dbltinyspace}\bigg]
\,.
\label{StripTypeHamiltonianConjLength}
\end{eqnarray}
The derivative of the potential 
$V'{\negtinyspace}(z) = z - \kappa{\halftinyspace}z^2$ 
(see \rf{MxModelPotentialDer}) 
in the cubic matrix model appears 
in the kinetic term of the Hamiltonian \rf{StripTypeHamiltonianConjLength}. 
The terms $\kappa{\halftinyspace}z$ and $1/z^2$ 
in \rf{StripTypeHamiltonianConjLength}---arising from $V'(z)/z$ 
and as 
prefactors of two three-loop interactions---represent 
the operations of removing one triangle and removing one strip, 
respectively. 
Notably, 
the creation operator $\tilde\Psi^\dagger(z)$ 
always appears in \rf{StripTypeHamiltonianConjLength} as 
$z \tilde\Psi^\dagger(z) + 1$, indicating that a single 0-gon 
is effectively added to the disk amplitude.
However, the tadpole term $-1$,
which appears as the first term in \rf{StripTypeHamiltonianConjLength},
was originally introduced to cancel the point-like state 
represented by $\Psi^\dagger(0)$.
Nevertheless, since $\Psi^\dagger(0)$ is not an operator 
but simply a constant, there is actually nothing to cancel.
Consequently, the $-1$ term has no effect on the dynamics
and is present merely as a formal expression.

The two Hamiltonians 
\rf{BasicTypeHamiltonianConjLength} and \rf{StripTypeHamiltonianConjLength} 
are structurally identical except for two differences: 
the Hamiltonian \rf{BasicTypeHamiltonianConjLength} includes 
an additional 2-gon 
and, in the three-loop interactions, removes a single triangle, 
by contrast, the 
Hamiltonian \rf{StripTypeHamiltonianConjLength} includes 
an additional 0-gon 
and, in the three-loop interactions, removes a single strip.


\subsubsection{Schwinger-Dyson equation}

Let us consider the SD equation in the case $\NN=1$. 
For this purpose, we introduce the well-known square-completion technique via the shift 
\rf{WaveFunShiftDef}.
Then, the Hamiltonian 
\rf{StripTypeHamiltonianConjLength} becomes 
\begin{eqnarray}\label{StripTypeHamiltonianConjLengthShift}
\Hop
\!\!&=&\!\!
\underset{z=0}{\rm Res}{\trpltinyspace}
 \frac{1}{z}
\bigg[
  \big\{
    \Omega(z)
    -
    \big(
      \tilde\Phi^\dagger(z)
    \big)^{{\negtrehalftinyspace}2}
  {\tinyspace}\big\}
  \Big\{{\negtinyspace}
    z \pder{z}
    \Big( \frac{1}{z} \tilde\Psi\Big(\frac{1}{z}\Big)\!\Big)
  \!\Big\}
\nonumber\\&&\!\!\phantom{%
\underset{z=0}{\rm Res}{\trpltinyspace}
 \frac{1}{z}
\bigg[
}-
  \frac{\G}{z} {\tinyspace}
  \Big(
    \tilde\Phi^\dagger(z) +
    \frac{V'{\negtinyspace}(z)}{2}
  {\tinyspace}\Big)
  \Big\{{\negtinyspace}
    z \pder{z}
    \Big( \frac{1}{z} \tilde\Psi\Big(\frac{1}{z}\Big)\!\Big)
  \!\Big\}^{{\negtinyspace}2}
{\dbltinyspace}\bigg]
\,,
\end{eqnarray}
where 
\begin{align}
\lambda(x)
&\define
  \frac{V'{\negtinyspace}(x)}{2}
-
  \frac{1}{x}
\,,
\label{StripTypeLambdaDef}
\\
\Omega(x)
&\define
 \big( \lambda(x) \big)^{{\negtrehalftinyspace}2}
 - \frac{1}{x^2}
\,.
\label{StripTypeOmegaDef}
\end{align}

For $\NN=1$, 
the commutation relation between the Hamiltonian 
and 
$\tilde\Psi^\dagger(x)$ is 
\begin{eqnarray}\label{StripTypeHamiltonianCommutationRelationN1}
&&
\commutator{\Hop}
           {\tilde\Psi^\dagger(x)}
\cuum
\,=\,
-\,
\pder{x}
  \Big(
    \big\{
      \OmegaXXX(x)
      -
      \big(
        \tilde\Phi^\dagger(x)
      \big)^{{\negtrehalftinyspace}2}
    \big\}
  {\negdbltinyspace}\Big)
\cuum
\,.
\end{eqnarray}
Then, 
the SD equation \rf{SDeqDT} together with the ``no big-bang condition'' \rf{NoBigBangCondition} becomes 
\begin{eqnarray}\label{StripTypeSDeqN1}
0
\!\!&=&\!\!
\lim_{T \to \infty}
\vac \E^{-T \Hop} {\dbltinyspace}
\Hop {\dbltinyspace} \tilde\Psi^\dagger(x) \cuum
\nonumber\\&=&\!\!
\pder{x}
  \Big(
      \tilde{F}_2^{{\tinyspace}{\rm conn}}(x,x;G)
      +
      \big( \tilde{F}_1^{{\tinyspace}{\rm conn}}(x{\halftinyspace};G) \big)^2
    -
      \Omega(x)
  {\neghalftinyspace}\Big)
\,,
\end{eqnarray}
where 
$\tilde{F}_1^{{\tinyspace}{\rm conn}}(x;G)$ 
and 
$\tilde{F}_2^{{\tinyspace}{\rm conn}}(x_1,x_2;G)$ 
are defined in 
\rf{DefFcon1} and \rf{DefFcon2}, 
respectively.

For $\NN=2$, 
the commutation relation between the Hamiltonian 
and 
$\tilde\Psi^\dagger(x_1) \tilde\Psi^\dagger(x_2)$ 
is 
\begin{eqnarray}\label{StripTypeHamiltonianCommutationRelationN2}
&&
\commutator{\Hop}
           {\tilde\Psi^\dagger(x_1) \tilde\Psi^\dagger(x_2)}
\cuum
\nonumber\\
&&=\,
-\,
\pder{x_1}
  \Big(
    \big\{
      \OmegaXXX(x_1)
      -
      \big(
        \tilde\Phi^\dagger(x_1)
      \big)^{{\negtrehalftinyspace}2}
    \big\}
  {\negdbltinyspace}\Big)
{\tinyspace}
\tilde\Psi^\dagger(x_2)
\cuum
\nonumber\\
&&\phantom{=\,}%
-
\pder{x_2}
  \Big(
    \big\{
      \OmegaXXX(x_2)
      -
      \big(
        \tilde\Phi^\dagger(x_2)
      \big)^{{\negtrehalftinyspace}2}
    \big\}
  {\negdbltinyspace}\Big)
{\tinyspace}
\tilde\Psi^\dagger(x_1)
\cuum
\nonumber\\
&&\phantom{=\,}%
-
2{\tinyspace}\G{\dbltinyspace}
\pder{x_1}
\pder{x_2}
{\tinyspace}
  \frac{
        \big(
          \tilde\Phi^\dagger(x_1) +
          \frac{V'{\negtinyspace}(x_1)}
               {2}
        \big)
        -
        \big(
          \tilde\Phi^\dagger(x_2) +
          \frac{V'{\negtinyspace}(x_2)}
               {2}
        \big)
       }
       {x_1 {\negtinyspace}-{\negtinyspace} x_2}
\,.
\end{eqnarray}
Then, 
the SD equation \rf{SDeqDT} together with the ``no big-bang condition'' \rf{NoBigBangCondition} 
becomes${}^{\ref{fn:AppConnectivity}}$
\begin{eqnarray}\label{StripTypeSDeqN2}
0
\!\!&=&\!\!
\lim_{T \to \infty}
\vac \E^{-T \Hop} {\dbltinyspace}
  \Hop {\dbltinyspace} \tilde\Psi^\dagger(x_1) \tilde\Psi^\dagger(x_2)
\cuum^{{\negtinyspace}{\rm conn}}
\nonumber\\&=&\!\!
\pder{x_1}
 \Big(
    \tilde{F}_3^{{\tinyspace}{\rm conn}}(x_1,x_1,x_2;G)
    {\tinyspace}+{\tinyspace}
    2{\tinyspace}
    \tilde{F}_1^{{\tinyspace}{\rm conn}}(x_1;G)
    \tilde{F}_2^{{\tinyspace}{\rm conn}}(x_1,x_2;G)
 {\negdbltinyspace}\Big)
\nonumber\\&&\!\!
+{\trpltinyspace}
\pder{x_2}
 \Big(
    \tilde{F}_3^{{\tinyspace}{\rm conn}}(x_1,x_2,x_2;G)
    {\tinyspace}+{\tinyspace}
    2{\tinyspace}
      \tilde{F}_1^{{\tinyspace}{\rm conn}}(x_2;G)
      \tilde{F}_2^{{\tinyspace}{\rm conn}}(x_1,x_2;G)
 {\negdbltinyspace}\Big)
\nonumber\\&&\!\!
+{\trpltinyspace}
2{\tinyspace}\G{\dbltinyspace}
\pder{x_1}
\pder{x_2}
{\tinyspace}
  \frac{
        \tilde{F}_1^{{\tinyspace}{\rm conn}}(x_1;G)
        -
        \tilde{F}_1^{{\tinyspace}{\rm conn}}(x_2;G)
       }
       {x_1 {\negtinyspace}-{\negtinyspace} x_2}
\,,
\end{eqnarray}
where 
$\tilde{F}_1^{{\tinyspace}{\rm conn}}(x;G)$, 
$\tilde{F}_2^{{\tinyspace}{\rm conn}}(x_1,x_2;G)$, 
and 
$\tilde{F}_3^{{\tinyspace}{\rm conn}}(x_1,x_2,x_3;G)$ 
are defined in 
\rf{DefFcon1}, \rf{DefFcon2}, and \rf{DefFcon3}, 
respectively. 

Note that 
the SD equation \rf{StripTypeSDeqN2} is derived 
from the ``multi-peeling decomposition''. 
Its first and second lines involve removing 
a single triangle or a single strip 
from two boundaries with $x_1$ and $x_2$, respectively, 
while the last line describes the 
merging of two boundaries through strip removal. 
By transitioning to the ``single-peeling decomposition'' 
(the right-hand figure of Fig.~\ref{fig:DTamplitudeSingleSlicingPeeling}), 
we obtain the modified SD equation 
\begin{eqnarray}\label{StripTypeSDeqN2singlePeeling}
0
\!\!&=&\!\!
\pder{x_1}
 \Big(
    \tilde{F}_3^{{\tinyspace}{\rm conn}}(x_1,x_1,x_2;G)
    {\tinyspace}+{\tinyspace}
    2{\tinyspace}
    \tilde{F}_1^{{\tinyspace}{\rm conn}}(x_1;G)
    \tilde{F}_2^{{\tinyspace}{\rm conn}}(x_1,x_2;G)
 {\negdbltinyspace}\Big)
\nonumber\\&&\!\!
+{\trpltinyspace}
\G{\dbltinyspace}
\pder{x_1}
\pder{x_2}
{\tinyspace}
  \frac{
        \tilde{F}_1^{{\tinyspace}{\rm conn}}(x_1;G)
        -
        \tilde{F}_1^{{\tinyspace}{\rm conn}}(x_2;G)
       }
       {x_1 {\negtinyspace}-{\negtinyspace} x_2}
\,.
\end{eqnarray}
The calculations for this transition proceed in the same way  
as those for DT (basic type), i.e.,
from \rf{BasicTypeSDeqN2} to \rf{BasicTypeSDeqN2singlePeeling}.

For $\NN \ge 3$, 
the SD equation \rf{SDeqDT} together with the ``no big-bang condition'' \rf{NoBigBangCondition} 
becomes${}^{\ref{fn:AppConnectivity}}$
\begin{eqnarray}\label{StripTypeSDeqGeneralNN}
0
\!\!&=&\!\!
\lim_{T \to \infty}
\vac \E^{-T \Hop} {\dbltinyspace}
\Hop {\dbltinyspace} \prod_{k=1}^\NN{\negtinyspace} \tildePsidag(x_k)
 \cuum^{{\negtinyspace}{\rm conn}}
\nonumber\\&=&\!\!
\sum_{i=1}^\NN {\negtinyspace}
  \pder{x_i}
    \bigg({\negtinyspace}
      \tilde{F}_{\NN+1}^{{\tinyspace}{\rm conn}}(x_i,x_i,\X_{I \backslash \{i\}};G)
      + 
\sum_{I_1 \cup I_2=I \backslash \{i\}}
      \tilde{F}_{|I_1|+1}^{{\tinyspace}{\rm conn}}(x_i,\bm{x}_{I_1};G)
      \tilde{F}_{|I_2|+1}^{{\tinyspace}{\rm conn}}(x_i,\bm{x}_{I_2};G)
\nonumber\\&&\!\!
\phantom{%
\sum_{i=1}^\NN {\negtinyspace}
  \pder{x_i}
    \bigg({\negtinyspace}
}{\negqdrpltinyspace}%
+
\G {\negdbltinyspace}
\sum_{{\scriptstyle j=1} \atop (j \neq i)}^{\NN}
{\negtrpltinyspace}
  \pder{x_j}
     \frac{
           \tilde{F}_{\NN-1}^{{\tinyspace}{\rm conn}}(\X_{I \backslash \{j\}};G)
           -
           \tilde{F}_{\NN-1}^{{\tinyspace}{\rm conn}}(\X_{I \backslash \{i\}};G)
          }
          {x_i - x_j}
    \bigg)
\,,
\end{eqnarray}
where $I=\{1, \ldots, \NN\}$, $\X_I=\{x_1, \ldots, x_\NN\}$, and 
$\tilde{F}_\NN^{{\tinyspace}{\rm conn}}$ 
are defined in  
\rf{DefFcon1} and \rf{DefFconN}. 

For DT (strip type) with general $\NN$: 
As in DT (basic type), 
the SD equations \rf{StripTypeSDeqGeneralNN} 
originate from the ``multi-peeling decomposition''. 
Using the same reasoning as in the transition 
from \rf{BasicTypeSDeqGeneralNN} 
to \rf{BasicTypeSDeqGeneralNNsinglePeeling}, 
we modify the decomposition to the ``single-peeling decomposition''.
The SD equations with general $\NN$ take the form
\begin{eqnarray}\label{StripTypeSDeqGeneralNNsinglePeeling}
0
\!\!&=&\!\!
  \pder{x_i}
    \bigg({\negtinyspace}
      \tilde{F}_{N+1}^{{\tinyspace}{\rm conn}}(x_i,x_i,\X_{I \backslash \{i\}};G)
      + 
\sum_{I_1 \cup I_2=I \backslash \{i\}}
      \tilde{F}_{|I_1|+1}^{{\tinyspace}{\rm conn}}(x_i,\bm{x}_{I_1};G)
      \tilde{F}_{|I_2|+1}^{{\tinyspace}{\rm conn}}(x_i,\bm{x}_{I_2};G)
\nonumber\\&&\!\!
\phantom{%
  \pder{x_i}
    \bigg({\negtinyspace}
}{\negoctpltinyspace}%
- \OmegaXXX(x_i)\, \delta_{N,1}
+
\G {\negdbltinyspace}
\sum_{{\scriptstyle j=1} \atop (j \neq i)}^{\NN}
{\negtrpltinyspace}
  \pder{x_j}
     \frac{
           \tilde{F}_{N-1}^{{\tinyspace}{\rm conn}}(\X_{I \backslash \{j\}};G)
           -
           \tilde{F}_{N-1}^{{\tinyspace}{\rm conn}}(\X_{I \backslash \{i\}};G)
          }
          {x_i - x_j}
    \bigg)
\,,
\end{eqnarray}
where the amplitudes $F_N^{{\tinyspace}{\rm conn}}(x_1,\ldots,x_\NN;G)$ are also expanded with respect to $G$ as in  \eqref{AmplitudeGexpansionN}.

\subsection{Amplitudes and topological recursion}

\subsubsection{Disk and cylinder amplitudes}

Using the expansion \rf{AmplitudeGexpansionN},
the zeroth-order term in $G$ from \rf{StripTypeSDeqN1}   
is extracted, which yields
\begin{eqnarray}\label{StripTypeSDeqN1G0}
0
\!\!&=&\!\!
\pder{x}
  \big\{{\negtinyspace}
      \big( \tilde{F}_1^{{\tinyspace}{\rm conn(0)}}(x)
      \big)^2
    -
    \Omega(x)
  {\neghalftinyspace}\big\}
\,.
\end{eqnarray}
Integrating \rf{StripTypeSDeqN1G0} with respect to $x$, one finds 
\begin{equation}\label{StripTypeDiskAmpTemp}
 \tilde{F}_1^{{\tinyspace}{\rm conn}(0)}{\negdbltinyspace}(x)
\,=\,
\sqrt{
  \Omega(x) + C_1
}
\,,
\end{equation}
where $C_1$ is an integration constant.

Here, we assume that
$\tilde{F}_1^{{\tinyspace}{\rm conn}(0)}{\negdbltinyspace}(x)$
has a single cut on the complex $x$-plane.
That is, 
the disk amplitude \rf{StripTypeDiskAmpTemp} becomes 
\begin{equation}\label{StripTypeDiskAmpAssumption}
 \tilde{F}_1^{{\tinyspace}{\rm conn}(0)}{\negdbltinyspace}(x)
\,=\,
 \frac{\kappa}{2}
 \Big({\tinyspace}
   x
   - \frac{d}{\kappa}
 {\tinyspace}\Big)
 \sqrt{
  \Big(
    x
    -
    \frac{a}{\kappa}
  {\tinyspace}\Big)
  {\negdbltinyspace}
  \Big(
    x
    -
    \frac{b}{\kappa}
  {\tinyspace}\Big)
 }
\,.
\end{equation}
This assumption is justified by 
\begin{eqnarray}\label{StripTypeEdgeCondition}
&&\!\!
d \,=\, 1 - \frac{a + b}{2}
\,,
\qquad\quad
( 1 - d {\tinyspace}) {\tinyspace} d
\,=\,
\frac{( a - b )^2}{8}
\,,
\qquad
\Big(
  \frac{( a - b )^2}{8} - \frac{1 - d}{2}
{\tinyspace}\Big) {\tinyspace} d
\,=\,
\kappa^2
\,,
\nonumber
\\
&&\!\!
C_1
\,=\,
1 + \frac{a{\halftinyspace}b {\tinyspace}
          d^{{\trehalftinyspace}2}}
         {4{\halftinyspace}\kappa^2}
\,,
\quad\qquad
\hbox{[\,$a_{\rm c} \le a \le 0 \le b \le b_{\rm c}$\,]}
\,.
\end{eqnarray}
Note that if we set 
\begin{equation}
c \,\define\, 
2{\halftinyspace}d - 1
\,,
\end{equation}
the third equation in \rf{StripTypeEdgeCondition} 
coincides with  
the second equation in \rf{BasicTypeEdgeConditionC}.

Repeating the same analysis as in Section \ref{subsubsec:Disk_Basic}, one finds that 
the critical values of $\kappa_c$ and $c_c$ are given in \rf{BasicCriticalValues},
and then we have
\begin{equation}
a_{\rm c} = -{\trpltinyspace} \frac{\sqrt{3} - 1}{2}
\,,
\qquad\quad
b_{\rm c} = \frac{3 + \sqrt{3}}{6}
\,,
\qquad\quad
x_{\rm c} = \frac{b_c}{\kappa_c} = 3^{3/4} + 3^{1/4}
\,. 
\label{crit_strip}
\end{equation}

Under the relations in \rf{StripTypeEdgeCondition}, 
the disk amplitude \rf{StripTypeDiskAmpAssumption} 
can be written as
\begin{equation}\label{StripTypeDiskAmp}
 \tilde{F}_1^{{\tinyspace}{\rm conn}(0)}{\negdbltinyspace}(x)
\,=\,
 \frac{\kappa}{2}
 \Big({\tinyspace}
   x
   - \frac{2 {\negtinyspace}-{\negtinyspace} a
             {\negtinyspace}-{\negtinyspace} b}
          {2{\halftinyspace}\kappa}
 {\tinyspace}\Big)
 \sqrt{
  \Big(
    x
    -
    \frac{a}{\kappa}
  {\tinyspace}\Big)
  {\negdbltinyspace}
  \Big(
    x
    -
    \frac{b}{\kappa}
  {\tinyspace}\Big)
 }
\,=\,
M(x) \sqrt{\sigma(x)}
\,,
\end{equation}
where
\begin{align}\label{StripTypeSpDef}
M(x):=
\frac{\kappa}{2}\left(x-\frac{2-a-b}{2\kappa}\right)\,,
\qquad
\sigma(x):=
\left(x-\frac{a}{\kappa}\right)\left(x-\frac{b}{\kappa}\right)\,.
\end{align}
As in \rf{BasicZ}, by introducing the Zhukovsky variable $p \in \mathbb{P}^1$ via
\begin{align}\label{StripZ}
x(p):=\frac{a+b}{2\kappa}
+\frac{b-a}{4\kappa} \Big( p + \frac{1}{p} \Big)
\,,
\end{align}
the square root $\sqrt{\sigma(x)}$ in the disk amplitude yields
\begin{align}
\sqrt{\sigma(x(p))}=\frac{b-a}{4\kappa}\Big( p - \frac{1}{p} \Big)
\,.
\end{align}
Here the branch points $x=a/\kappa$ and $b/\kappa$ of 
the disk amplitude are mapped to $p=-1$ and $1$, respectively.

Using the expansion \rf{AmplitudeGexpansionN},
the first-order term of \rf{StripTypeSDeqN2} with respect to $G$ 
is extracted, which yields
\begin{eqnarray}\label{StripTypeSDeqN2G1}
0
\!\!&=&\!\!
    \tilde{F}_1^{{\tinyspace}{\rm conn}(0)}{\negdbltinyspace}(x_1)
    \tilde{F}_2^{{\tinyspace}{\rm conn}(0)}{\negdbltinyspace}(x_1,x_2)
\nonumber\\&&\!\!
+{\trpltinyspace}
\pder{x_2}
{\tinyspace}
  \frac{
        \tilde{F}_1^{{\tinyspace}{\rm conn}(0)}{\negdbltinyspace}(x_1)
        -
        \tilde{F}_1^{{\tinyspace}{\rm conn}(0)}{\negdbltinyspace}(x_2)
       }
       {2{\tinyspace} ( x_1 {\negtinyspace}-{\negtinyspace} x_2 )}
{\dbltinyspace}+{\dbltinyspace}
\Const_2(x_2)
\,,
\end{eqnarray}
where $\Const_2(x_2)$ is an integration constant with respect to $x_1$. 
Here, we assume that
$\tilde{F}_2^{{\tinyspace}{\rm conn}(0)}{\negdbltinyspace}(x_1,x_2)$ has no poles at the zeros of $M(x)$ in \eqref{StripTypeSpDef}.
This assumption fixes $\Const_2(x_2)$ as \cite{Eynard:2004mh},
\begin{equation}
\Const_2(x_2)
\,=\,
-\,
\frac{
      \kappa{\tinyspace} \big( x_2 - \frac{a + b}{2\kappa} \big)
     }
     {4 \sqrt{ \big( x_2 - \frac{a}{\kappa} \big)
               \big( x_2 - \frac{b}{\kappa} \big) }
     }
\,.
\end{equation}
Then, 
using the disk amplitude \rf{StripTypeDiskAmp}, 
the cylinder amplitude is obtained as 
\begin{align}\label{StripTypeCylinderAmp}
&
\tilde{F}_2^{{\tinyspace}{\rm conn}(0)}{\negdbltinyspace}(x_1,x_2)
\nonumber\\
&
\,=\,
 \frac{1}
      {2{\tinyspace} ( x_1 {\negtinyspace}-{\negtinyspace} x_2 )^2}
  \Bigg(
    \frac{
            \big(
              x_1
              -
              \frac{a}{\kappa}
            \big)
            {\negdbltinyspace}
            \big(
              x_1
              -
              \frac{b}{\kappa}
            \big)
            +
            \big(
              x_2
              -
              \frac{a}{\kappa}
            \big)
            {\negdbltinyspace}
            \big(
              x_2
              -
              \frac{b}{\kappa}
            \big)
            -
            ( x_1 - x_2 )^2
         }
         {2
          \sqrt{
            \big(
              x_1
              -
              \frac{a}{\kappa}
            \big)
            {\negdbltinyspace}
            \big(
              x_1
              -
              \frac{b}{\kappa}
            \big)
          }
          \sqrt{
            \big(
              x_2
              -
              \frac{a}{\kappa}
            \big)
            {\negdbltinyspace}
            \big(
              x_2
              -
              \frac{b}{\kappa}
            \big)
          }
         }
    {\dbltinyspace}-{\dbltinyspace} 1
  \Bigg)
\nonumber\\
&
\,=\,
\frac{1}
      {2{\tinyspace} ( x_1 {\negtinyspace}-{\negtinyspace} x_2 )^2}
\left(
\frac{x_1x_2-\frac{a+b}{2\kappa}\left(x_1+x_2\right)+\frac{ab}{\kappa^2}}
{\sqrt{\sigma(x_1)\sigma(x_2)}}-1\right),
\end{align}
where $\sigma(x)$ is defined in \rf{StripTypeSpDef}. 
Under the map \rf{StripZ}, this cylinder amplitude takes the same simple form as \rf{BasicBergman}:
\begin{align}
\AmpFXXX_2^{{\tinyspace}{\rm conn}(0)}{\negdbltinyspace}(x(p_1),x(p_2))\, dx(p_1)dx(p_2)
\,&=\,
-B(p_{1},p_{2}^{-1})
\,=\,
\frac{dp_{1}dp_{2}}{\left(p_{1}p_{2}-1\right)^2}\,.
\label{BasicBergman_st}
\end{align}

\subsubsection{Topological recursion}

Integrating \rf{StripTypeSDeqGeneralNNsinglePeeling} 
with respect to $x_i$, one finds 
\begin{eqnarray}\label{StripTypeSDeqGeneralNNsinglePeelingInt}
0
\!\!&=&\!\!
      \tilde{F}_{\NN+1}^{{\tinyspace}{\rm conn}}(x_i,x_i,\X_{I \backslash \{i\}})
      + 
\sum_{I_1 \cup I_2=I \backslash \{i\}}
      \tilde{F}_{|I_1|+1}^{{\tinyspace}{\rm conn}}(x_i,\bm{x}_{I_1})
      \tilde{F}_{|I_2|+1}^{{\tinyspace}{\rm conn}}(x_i,\bm{x}_{I_1})
- \OmegaXXX(x_i)\, \delta_{N,1}
\nonumber\\&&\!\!
{\negqdrpltinyspace}%
+
\G {\negdbltinyspace}
\sum_{{\scriptstyle j=1} \atop (j \neq i)}^{\NN}
{\negtrpltinyspace}
  \pder{x_j}
     \frac{
           \tilde{F}_{\NN-1}^{{\tinyspace}{\rm conn}}(\X_{I \backslash \{j\}})
           -
           \tilde{F}_{\NN-1}^{{\tinyspace}{\rm conn}}(\X_{I \backslash \{i\}})
          }
          {x_i - x_j}
{\dbltinyspace}+{\dbltinyspace}
C_\NN(\X_{I \backslash \{i\}})
\,,
\end{eqnarray}
where $C_N(\X_{I \backslash \{i\}})$ is a function of $\X_{I \backslash \{i\}}$. 
We remark that the equation \eqref{StripTypeSDeqGeneralNNsinglePeelingInt} is exactly the loop equation, for multi-point resolvents amplitudes of the matrix model with potential $V(x) = x^2/2 - \kappa{\halftinyspace}x^3/3$.\footnote{%
We discuss the relation between them in Appendix \ref{app:SD_MM}.
}
The following result was shown by Eynard in \cite{Eynard:2004mh}, which we recall here for completeness.

By the expansion \rf{AmplitudeGexpansionN} with respect to $G$, 
the equation \rf{StripTypeSDeqGeneralNNsinglePeelingInt} for the case $i=1$ yields
\begin{align}
\label{StripSD}
\AmpFXXX_{N}^{{\tinyspace}{\rm conn}(h)}{\negdbltinyspace}(\X_I)
&=
\frac{(-1)}{2\AmpFXXX_{1}^{{\tinyspace}{\rm conn}(0)}{\negdbltinyspace}(x_1)}
\Biggl[\AmpFXXX_{N+1}^{{\tinyspace}{\rm conn}(h-1)}{\negdbltinyspace}(x_1,x_1,\X_{I \backslash \{1\}})
\nonumber\\
&\hspace{7em}
+
\mathop{\sum_{h_1+h_2=h}}_{I_1 \cup I_2=\{2,\ldots,N\}}^{\textrm{no $(0,1)$}}
\AmpFXXX_{|I_1|+1}^{{\tinyspace}{\rm conn}(h_1)}{\negdbltinyspace}(x_1,\bm{x}_{I_1})
\AmpFXXX_{|I_2|+1}^{{\tinyspace}{\rm conn}(h_2)}{\negdbltinyspace}(x_1,\bm{x}_{I_1})
\nonumber\\
&\hspace{7em}
+
\sum_{i=2}^{\NN}
\frac{\AmpFXXX_{N-1}^{{\tinyspace}{\rm conn}(h)}{\negdbltinyspace}(\X_{I \backslash \{i\}})}
{(x_1-x_i)^2}\Biggr]
+R(x_1;\X_{I \backslash \{1\}})\,,
\\
R(x_1;\X_{I \backslash \{1\}})&:=
\frac{(-1)}{2 \AmpFXXX_{1}^{{\tinyspace}{\rm conn}(0)}{\negdbltinyspace}(x_1)}
\Biggl[
-\sum_{i=2}^{\NN}
\pder{x_i}
\frac{\AmpFXXX_{N-1}^{{\tinyspace}{\rm conn}(h)}{\negdbltinyspace}(\X_{I \backslash \{1\}})}{x_1-x_i}
+ \ConstXXX_\NN^{(h)}(\X_{I \backslash \{1\}})
\Biggr]\,,
\end{align}
for $(h,N)\ne (0,1)$, 
where ``no $(0,1)$'' in the summation indicates that the summation does not contain 
the disk amplitude 
$\AmpFXXX_{1}^{{\tinyspace}{\rm conn}(0)}{\negdbltinyspace}(x_1)$, 
and $\ConstXXX_\NN^{(h)}(\X_{I \backslash \{1\}})$ is a function of $\X_{I \backslash \{1\}}$. 
We now follow the same strategy as before to derive the topological recursion \rf{top_rec_F_d} or \rf{top_rec_W_d}. 
Assuming that the amplitudes 
$\AmpFXXX_{N}^{{\tinyspace}{\rm conn}(h)}{\negdbltinyspace}(\X_I)$ have no poles away from the branch cut $[a/\kappa, b/\kappa]$ of the disk amplitude 
\rf{StripTypeDiskAmp}, we obtain \cite{Eynard:2004mh},
\begin{align}\label{top_rec_F_d_strip}
\AmpFXXX_{N}^{{\tinyspace}{\rm conn}(h)}{\negdbltinyspace}(\X_I)
&=
\sum_{i=1,2}
\mathop{\mathrm{Res}}_{x_0=\alpha_i}
\frac{(-1)\, dx_0\, dS_{p_0}(p_1)}{2\AmpFXXX_{1}^{{\tinyspace}{\rm conn}(0)}{\negdbltinyspace}(x_0)\, dx(p_1)}
\Biggl[
\AmpFXXX_{N+1}^{{\tinyspace}{\rm conn}(h-1)}{\negdbltinyspace}(x_0,x_0,\X_{I \backslash \{1\}})
\nonumber\\
&\hspace{8em}
+
\mathop{\sum_{h_1+h_2=h}}_{I_1 \cup I_2=\{2,\ldots,N\}}^{\textrm{no $(0,1)$}}
\mathfrak{\AmpFXXX}_{|I_1|+1}^{{\tinyspace}{\rm conn}(h_1)}{\negdbltinyspace}(x_0,\bm{x}_{I_1})\,
\mathfrak{\AmpFXXX}_{|I_2|+1}^{{\tinyspace}{\rm conn}(h_2)}{\negdbltinyspace}(x_0,\bm{x}_{I_2})
\Biggr]\,,
\end{align}
where $\alpha_1=a/\kappa$ and $\alpha_2=b/\kappa$, 
and the topological recursion as in \rf{top_rec_W_d} for the variables $p_i \in \mathbb{P}^1$ by the map $x_i=x(p_i)$ in \rf{StripZ}, where $dS_{p_0}(p_1)$ is the third-kind differential defined in \rf{differential_ds}.
The spectral curve data $(\mathbb{P}^1; x,y,B)$ of the topological recursion consists of the disk amplitude $y=\AmpFXXX_1^{{\tinyspace}{\rm conn}(0)}{\negdbltinyspace}(x)$ in \rf{StripTypeDiskAmp},
\begin{align}
&
y=M(x) \sqrt{\sigma(x)}=
\frac{\kappa}{2}\left(x-\gamma\right)\sqrt{\left(x-\alpha_1\right)\left(x-\alpha_2\right)}\,,
\nonumber\\
&
\alpha_1:=\frac{a}{\kappa}\,,\ \
\alpha_2:=\frac{b}{\kappa}\,,\ \
\gamma:=\frac{2-a-b}{2\kappa}\,,
\end{align}
and the cylinder amplitude given by the bi-differential $B=B(p_{1},p_{2})$ in \rf{BasicBergman_st}. 
Some computational results for the amplitudes are listed in Appendix \ref{app:list_dt_strip}.
Note that, as remarked in \eqref{kernel_exp}, the amplitudes $\AmpFXXX_N^{{\tinyspace}{\rm conn}(h)}{\negdbltinyspace}(\bm{x}_I)$ for $2h+N \ge 3$ are expressed in terms of kernel differentials \eqref{kernel_diff} with $\alpha_1=a/\kappa, \alpha_2=b/\kappa$.


\section{Dynamical Triangulation (Continuous Level)}
\label{sec:DTcontinuouslevel}

\subsection{Continuum limit and mode expansion}
\label{sec:PropertiesContinuousDT}

\subsubsection{Continuum limit}
\label{sec:DTcontinuumLimit}

The continuum 2D surface is realized as 
a surface composed of infinitely many triangles and strips.
Because there are infinitely many triangles and strips, their edges are also infinite in number.
Consequently,
we consider a 2D surface 
in which the length of each link and the area of each triangle 
become infinitesimally small.%
\footnote{
In Appendix \ref{app:1DQG}, the essential meaning of the continuum limit is clarified through a toy model of 1D pure quantum gravity.
} 
(Here, the edges of triangles and the boundaries of strips 
are referred to as ``links''.)
Technically, this continuum limit is realized by setting%
\footnote{
In the continuum limit, triangulations with a large number of triangles $N_2$ and boundary lengths $\ell_i$ in the discrete DT models should statistically dominate in order to yield a smooth manifold \cite{Kazakov:1988ch,Ginsparg:1991bi}. 
In the case of the disk amplitude, divergences in the expectation values of $N_2$ and $\ell_1$ appear when taking $\kappa\to\kappa_{\mathrm c}$ and  $x\to x_{\mathrm c}$.
The expectation values $\langle N_2\rangle$ and $\langle \ell_1\rangle$ are 
\begin{equation}\label{divergence_area_boundary}
\langle N_2\rangle=\kappa\frac{\partial}{\partial \kappa} \AmpFXXX_1^{{\tinyspace}{\rm conn}(0)}{\negdbltinyspace}(x)\,,
\quad 
\langle \ell_1\rangle =x\frac{\partial}{\partial x} \AmpFXXX_1^{{\tinyspace}{\rm conn}(0)}{\negdbltinyspace}(x)\,.
\nonumber
\end{equation}
By substituting $\AmpFXXX_1^{{\tinyspace}{\rm conn}(0)}{\negdbltinyspace}(x)$ from \rf{BasicTypeDiskAmpAssumption} (resp. \rf{StripTypeDiskAmpAssumption})
for the basic (resp. strip) type model into the above equations, one finds the critical values $\kappa_{\rm c}$ and $x_{\mathrm c}$ from the divergent factors in the expectation values $\langle N_2\rangle$ and $\langle \ell_1\rangle$.  
}
\begin{equation}\label{Continuum}
x_1 =
x_{\rm c}{\dbltinyspace}
\E^{{\tinyspace}\ep{\tinyspace}\l_{\xi}{\dbltinyspace}\xi_1}
\,,
\quad
\dots
\,,
\quad
x_\NN =
x_{\rm c}{\dbltinyspace}
\E^{{\tinyspace}\ep{\tinyspace}\l_{\xi}{\dbltinyspace}\xi_\NN}
\,,
\quad
\kappa =
\kappa_{\rm c}{\dbltinyspace}
\E^{-{\tinyspace}\ep^2 \l_{\mu}{\halftinyspace}\cc}
\,,
\quad
G =
\frac{\ep^{{\tinyspace}d_G} \calG}{2}
\,,
\end{equation}
in the limit 
$\ep\to +0$, 
where $\l_{\xi}$ and $\l_{\mu}$ are positive constants that normalize 
$\xi_1, \ldots, \xi_N$ and the cosmological constant $\cc$, respectively. 
The critical values $x_{\rm c}$ and $\kappa_{\rm c}$ 
are chosen so as to remain nonzero and finite in size as in \eqref{BasicCriticalValues} and \eqref{crit_strip}. 
For convenience, these values are summarized in Table \ref{crit_v_summary}.
In the continuum limit \rf{Continuum}, 
the amplitudes
\rf{DTamplitude} and 
\rf{DTamplitudeMatrixModel} 
become 
\begin{equation}\label{DTamplitudeContinuum}
F_\NN^{{\tinyspace}{\rm conn}}(\ell_1,\ldots,\ell_\NN;G)
\,=\,
\sum_{h=0}^\infty\,
\sum_{N_2=1}^\infty\,
\sum_{S \in {\cal T}_\NN^{(h)}(\ell_1,\ldots,\ell_\NN, N_2)}
( \ep^{{\tinyspace}d_G} \calG/2 )^{\;\!h+\NN-1}
      \kappa_{\rm c}^{N_2}{\tinyspace}
      \E^{-{\tinyspace}\cc A_2}
\,,
\end{equation}
and 
\begin{eqnarray}\label{DTamplitudeMatrixModelContinuum}
&&\hspace{-16pt}
\widetilde{\calF}_\NN^{{\tinyspace}{\rm conn}}(\xi_1,\ldots,\xi_\NN;\calG)
\nonumber
\\
&&\hspace{-16pt}\define
\lim_{\ep \to +0}
\big( 2 {\tinyspace} \l_{\xi}{\dbltinyspace}x_{\rm c}
 {\tinyspace}\ep^{- d_F} \big)^{{\negtinyspace}\NN}
\tilde{F}_N^{{\tinyspace}{\rm conn}}(x_1,\ldots,x_N;G)
\nonumber
\\
&&\hspace{-16pt}=
\lim_{\ep \to +0}
\sum_{\ell_1=1}^\infty \ldots \sum_{\ell_\NN=1}^\infty
\E^{- \xi_1 L_1 - \ldots - \xi_\NN L_\NN}
  \big( 2 {\tinyspace} \l_{\xi}{\dbltinyspace}x_{\rm c}
 {\tinyspace}\ep^{- d_F} \big)^{{\negtinyspace}\NN}
  x_{\rm c}^{- \ell_1 - \ldots - \ell_\NN -\NN}
  F_\NN^{{\tinyspace}{\rm conn}}(\ell_1,\ldots,\ell_\NN;G)
\,,
\qquad\quad
\end{eqnarray}
respectively, 
where 
$L_1 {\negdbltinyspace}\define{\negdbltinyspace}
 ( \ell_1 {\negdbltinyspace}+{\negdbltinyspace} 1 ){\tinyspace}
 \l_{\xi}{\dbltinyspace}\ep$, 
\ldots,
$L_\NN {\negdbltinyspace}\define{\negdbltinyspace}
 ( \ell_\NN {\negdbltinyspace}+{\negdbltinyspace} 1 ){\tinyspace}
 \l_{\xi}{\dbltinyspace}\ep$ 
represent the continuous boundary lengths, 
and 
$A_2 {\negdbltinyspace}\define{\negdbltinyspace}
 N_2 {\tinyspace} \l_{\mu}{\halftinyspace}\ep^2$ 
denotes the area of the 2D surface.
The exponents 
$d_G$ and $d_F$ are chosen so that 
both sides of 
\rf{DTamplitudeContinuum} 
and 
\rf{DTamplitudeMatrixModelContinuum} 
remain finite and nonzero 
in the limit 
$\ep\to{\negdbltinyspace}+{\negdbltinyspace}0$.
In this limit, 
the discrete Laplace transform 
becomes 
the continuous Laplace transform.

Here, we assume that relations 
similar to \rf{AmplitudeGexpansionN}, 
i.e., 
\begin{eqnarray}\label{AmplitudeGexpansionContinuumN}
\widetilde{\calF}_\NN^{{\tinyspace}{\rm conn}}(\xi_1,\ldots,\xi_\NN;\calG)
\!\!&=&\!\!
\sum_{h=0}^\infty \calG^{\;\!h+\NN-1}
  \widetilde{\calF}_\NN^{{\tinyspace}{\rm conn}(h)}{\negdbltinyspace}(\xi_1,\ldots,\xi_\NN)
\,,
\end{eqnarray}
also hold in the continuum limit.
Note that 
all amplitudes 
$\widetilde{\calF}_\NN^{{\tinyspace}{\rm conn}(h)}{\negdbltinyspace}(\xi_1,\ldots,\xi_\NN)$ 
defined by \rf{AmplitudeGexpansionContinuumN} 
are finite and nonzero.
This is the essential property of 
the double scaling limit \cite{MM:DS,MM:GM,MM:BK}.
The scaling property \rf{DTamplitudeMatrixModelContinuum} 
makes it possible to define the wave function at the continuous level.

\begin{table}[t]
\begin{center}
\begin{tabular}{c|ccc|cc}
 & $x_{\rm c}$ & $\kappa_{\rm c}$ & $c_{\rm c}$ & $\l_{\xi}$ & $\l_{\mu}$ \\
\hline
Basic type & $2 \!\cdot\! 3^{1/4}$ & $1/(2 \cdot 3^{3/4})$ & $1/\sqrt{3}$ & $1$ & $3/16$ \\
Strip type & $3^{3/4} + 3^{1/4}$ & $1/(2 \cdot 3^{3/4})$ & $1/\sqrt{3}$ & $1/(1 + \sqrt{3})$ & $3/16$
\end{tabular}
\caption{
Critical values and normalization factors for each discrete model.}
\label{crit_v_summary}
\end{center}
\end{table}

Using \rf{Continuum}, 
the second relation in \rf{BasicTypeEdgeConditionC} leads to 
\begin{equation}\label{ContinuumC}
c \,=\,
c_{\rm c} 
\Big(
  1 + 2\sqrt{\frac{\l_{\mu}{\tinyspace}\cc}{3}} {\dbltinyspace} \ep
    + {\cal O}(\ep^2)
\Big)
\,,
\end{equation}
and
the second and third relations in \rf{StripTypeEdgeCondition} lead to 
\begin{eqnarray}\label{ContinuumAB}
a \!\!&=&\!\!
a_{\rm c}
+
\frac{c_{\rm c}{\dbltinyspace}\l_{\mu}{\tinyspace}\cc}{2}
  {\dbltinyspace} \ep^2
    + {\cal O}(\ep^3)
\,,
\qquad\quad
b \,=\,
b_{\rm c}
- 2{\tinyspace}c_{\rm c}\sqrt{\frac{\l_{\mu}{\tinyspace}\cc}{3}}
  {\dbltinyspace} \ep
    + {\cal O}(\ep^2)
\,,
\end{eqnarray}
where, note that $a+b+c=1$. In the following, we 
normalize the cosmological constant $\cc$ by setting
$\l_{\mu} = 3/16.$
Taking the continuum limit \rf{Continuum} 
together with \rf{ContinuumC} or \rf{ContinuumAB}, 
the disk amplitude 
\rf{BasicTypeDiskAmp} or \rf{StripTypeDiskAmp} 
and 
the cylinder amplitude 
\rf{BasicTypeCylinderAmp} or \rf{StripTypeCylinderAmp} 
become 
\begin{eqnarray}
 \AmpFXXX_1^{{\tinyspace}{\rm conn}(0)}{\negdbltinyspace}(x)
\!\!&=&\!\!
\frac{\ep^{3/2}}
     {2 {\tinyspace} \l_{\xi}{\dbltinyspace}x_{\rm c}}
\Big(
  \xi - \frac{\sqrt{\cc}}{2}
{\tinyspace}\Big)
\sqrt{\xi + \sqrt{\cc}}
+
{\cal O}(\ep^2)
\,,
\label{BasicTypeDiskAmpContinuumRelation}
\\
 \AmpFXXX_2^{{\tinyspace}{\rm conn}(0)}{\negdbltinyspace}(x_1,x_2)
\!\!&=&\!\!
\frac{2{\tinyspace}\ep^{-2}}
     {(2 {\tinyspace} \l_{\xi}{\dbltinyspace}x_{\rm c})^2
     }{\dbltinyspace}
\frac{1}{(\xi_1 {\negdbltinyspace}-{\negdbltinyspace} \xi_2)^2}
\Big(
  \frac{\xi_1 + \xi_2 + 2\sqrt{\cc}}
       {2 \sqrt{\xi_1 + \sqrt{\cc}}
          \sqrt{\xi_2 + \sqrt{\cc}}}
  - 1
{\tinyspace}\Big)
+
{\cal O}(\ep^{-1})
\,,
\qquad
\label{BasicTypeCylinderAmpContinuumRelation}
\end{eqnarray}
respectively. 
Here we set
$\l_{\xi} {\negtrpltinyspace}={\negdbltinyspace} 1$ 
for the basic type and
$\l_{\xi} {\negtrpltinyspace}={\negdbltinyspace} 1
 / (1 {\negtrpltinyspace}+{\negtrpltinyspace} \sqrt{3} )$ 
for the strip type 
in order to simplify the expressions.
The critical values $x_{\rm c}$, $\kappa_{\rm c}$, $c_{\rm c}$, and the normalization factors $\l_{\xi}$, $\l_{\mu}$ for each model are summarized in Table \ref{crit_v_summary}.
By comparing 
\rf{AmplitudeGexpansionContinuumN} and \rf{AmplitudeGexpansionN} 
under the relations \rf{DTamplitudeMatrixModelContinuum} and 
\rf{BasicTypeDiskAmpContinuumRelation}--%
\rf{BasicTypeCylinderAmpContinuumRelation}, 
we obtain the same amplitudes for both the basic type and the strip type as%
\footnote{%
This property is a kind of universality. 
}
\begin{eqnarray}
 \widetilde{\calF}_1^{{\dbltinyspace}{\rm conn}(0)}{\negdbltinyspace}(\xi)
\!\!&=&\!\!
\lim_{\ep \to 0}
 2 {\tinyspace} \l_{\xi}{\dbltinyspace}x_{\rm c}{\tinyspace}\ep^{-3/2}
 \AmpFXXX_1^{{\tinyspace}{\rm conn}(0)}{\negdbltinyspace}(x)
\nonumber
\\
\!\!&=&\!\!
\Big(
  \xi - \frac{\sqrt{\cc}}{2}
{\tinyspace}\Big)
\sqrt{\xi + \sqrt{\cc}}
\,,
\label{DiskAmpContinuum}
\\
 \widetilde{\calF}_2^{{\dbltinyspace}{\rm conn}(0)}{\negdbltinyspace}(\xi_1,\xi_2)
\!\!&=&\!\!
\lim_{\ep \to 0}
 (
 2 {\tinyspace} \l_{\xi}{\dbltinyspace}x_{\rm c}{\tinyspace}\ep^{-3/2}
 {\tinyspace})^2
 {\dbltinyspace}
 \frac{\ep^5}{2}
 \AmpFXXX_2^{{\tinyspace}{\rm conn}(0)}{\negdbltinyspace}(x_1,x_2)
\nonumber
\\
\!\!&=&\!\!
\frac{1}{(\xi_1 {\negdbltinyspace}-{\negdbltinyspace} \xi_2)^2}
\Big(
  \frac{\xi_1 + \xi_2 + 2\sqrt{\cc}}
       {2 \sqrt{\xi_1 + \sqrt{\cc}}
          \sqrt{\xi_2 + \sqrt{\cc}}}
  - 1
{\tinyspace}\Big)
\,,
\label{CylinderAmpContinuum}
\end{eqnarray}
and 
\begin{equation}
d_F = \frac{3}{2}
\,,
\qquad
d_G = 5
\,.
\end{equation}
Note that the values of $d_F$ and $d_G$ are independent of $\NN$ and $h$ 
\cite{MM:DS,MM:GM,MM:BK}. 
This fact makes it possible to construct the string field theory 
at the continuous level \cite{SFT:IK,SFT:Watabiki}.

\subsubsection{Schwinger-Dyson equation}
\label{sec:SD_continuum}

We now consider the continuum limit of 
the SD equation \eqref{BasicTypeSDeqGeneralNNsinglePeeling} for the basic type and 
the SD equation \eqref{StripTypeSDeqGeneralNNsinglePeeling} for the strip type. 
By \eqref{Continuum} and \eqref{DTamplitudeMatrixModelContinuum}, 
the SD equations for both types yield 
\begin{align}
0&=
\pder{\xi_i}\bigg(
\widetilde{\calF}_{N+1}^{{\tinyspace}{\rm conn}}(\xi_i,\xi_i,\bm{\xi}_{I \backslash \{i\}};\calG)
+ 
\sum_{I_1 \cup I_2=I \backslash \{i\}}
\widetilde{\calF}_{|I_1|+1}^{{\tinyspace}{\rm conn}}(\xi_i,\bm{\xi}_{I_1};\calG)\,
\widetilde{\calF}_{|I_2|+1}^{{\tinyspace}{\rm conn}}(\xi_i,\bm{\xi}_{I_2};\calG)
\nonumber\\
&\hspace{4em}
- \calomega(\xi_i)\, \delta_{N,1}
+ 2\, \calG {\negdbltinyspace}
\sum_{{\scriptstyle j=1} \atop (j \neq i)}^{\NN}
\pder{\xi_j}
\frac{\widetilde{\calF}_{N-1}^{{\tinyspace}{\rm conn}}(\bm{\xi}_{I \backslash \{j\}};\calG)
-\widetilde{\calF}_{N-1}^{{\tinyspace}{\rm conn}}(\bm{\xi}_{I \backslash \{i\}};\calG)}
{\xi_i - \xi_j}\bigg)\,,
\label{cont_lim_sd}
\end{align}
where $\bm{\xi}_I=\{\xi_1, \ldots, \xi_N\}$, and 
\begin{align}
\calomega(\xi)
\,\define\,
\xi^3 - \frac{3}{4} \cc\, \xi\,.
\end{align}
Here, for the continuum limit, we used relations
\begin{align}
\frac{\partial}{\partial x} \left(x^3\, \Omega(x)\right)& =
\frac14\, \frac{\partial}{\partial \xi}\, \calomega(\xi)\,\ep^2 
+ \mathcal{O}(\ep^3)
\quad
\textrm{for basic type}\,,
\nonumber\\
\frac{\partial}{\partial x} \left(\Omega(x)\right)& =
\frac{3^{1/4}}{12}\, \frac{\partial}{\partial \xi}\, \calomega(\xi)\,\ep^2 
+ \mathcal{O}(\ep^3)
\quad
\textrm{for strip type}\,,
\end{align}
where $x=x_{\rm c}{\dbltinyspace}
\E^{{\tinyspace}\ep{\tinyspace}\l_{\xi}{\dbltinyspace}\xi}$ by \eqref{Continuum}.

We expand the disk amplitude \rf{DiskAmpContinuum} 
and the cylinder amplitude \rf{CylinderAmpContinuum} 
around the point at infinity in $\xi$:
\begin{align}
 \widetilde{\calF}_1^{{\dbltinyspace}{\rm conn}(0)}{\negdbltinyspace}(\xi)
&=
\calOmega_1{\negtinyspace}(\xi)
\,+
\sum_{\ell=1,3,5,\ldots}
  \xi^{-\ell/2 -1}
  f_1^{{\dbltinyspace}{\rm conn}(0)}{\negdbltinyspace}(\ell)
\,,
\label{DiskAmpContinuumModeExpansion}
\\
 \widetilde{\calF}_2^{{\dbltinyspace}{\rm conn}(0)}{\negdbltinyspace}(\xi_1,\xi_2)
&=
\frac{1}{2 \sqrt{(\xi_1 + \sqrt{\cc})\, (\xi_2 + \sqrt{\cc})}
\left(\sqrt{\xi_1 + \sqrt{\cc}}+\sqrt{\xi_2 + \sqrt{\cc}}\right)^2}
\nonumber\\
&=
\calOmega_2(\xi_1,\xi_2)
\,
+
\sum_{\ell_1=1,3,5,\ldots} \sum_{\ell_2=1,3,5,\ldots}
  \xi_1^{-\ell_1/2 -1} \xi_2^{-\ell_2/2 -1}
  f_2^{{\dbltinyspace}{\rm conn}(0)}{\negdbltinyspace}(\ell_1,\ell_2)
\label{CylinderAmpContinuumModeExpansion}
\,,
\end{align}
where 
\begin{align}
\calOmega_1{\negtinyspace}(\xi)
\,\define\,
\xi^{3/2} - \frac{3}{8}\cc\, \xi^{-1/2}
\,,
\qquad
\calOmega_2(\xi_1,\xi_2)
\,\define\,
\frac{1}
     {2 \sqrt{\xi_1 \xi_2}{\tinyspace}
      \big({\negtinyspace}
        \sqrt{\xi_1} +{\negtrehalftinyspace} \sqrt{\xi_2}
      {\qdrpltinyspace}\big)^{{\negtinyspace}2}}\,,
\end{align}
are non-universal parts, as they are polynomial in $\cc$ and do not contribute in the finite-area limit.
All amplitudes $\widetilde{\calF}_\NN^{{\dbltinyspace}{\rm conn}(h)}{\negdbltinyspace}(\xi_1,\ldots,\xi_\NN)$ other than 
the disk amplitude and the cylinder amplitude are also expanded as
\begin{equation}\label{GeneralAmpContinuumModeExpansion}
\widetilde{\calF}_\NN^{{\dbltinyspace}{\rm conn}(h)}{\negdbltinyspace}(\xi_1,\ldots,\xi_\NN)
\,=
\sum_{\ell_1=1,3,5,\ldots} \ldots \sum_{\ell_\NN=1,3,5,\ldots}
  \xi_1^{-\ell_1/2 -1} \ldots \xi_\NN^{-\ell_\NN/2 -1}
  f_\NN^{{\dbltinyspace}{\rm conn}(h)}{\negdbltinyspace}(\ell_1,\ldots,\ell_\NN)
\,.
\end{equation}
This follows from the fact that the amplitudes $\AmpFXXX_N^{{\tinyspace}{\rm conn}(h)}{\negdbltinyspace}(x_1, \ldots,x_N)$ of $2h+N \ge 3$ for both the basic and the strip types are expanded in terms of the kernel differentials as in \eqref{kernel_exp}, and that, in the continuum limit, the kernel differentials behave as
\begin{align}
\chi_1^{(n)}(x) &= \mathcal{O}(\ep^{-n+1/2})\, dx,
\nonumber\\
\chi_2^{(n)}(x) &= \left(
\frac{2\, x_{\rm c}^{-n+1}}{(n-1)!\sqrt{\xi + \sqrt{\cc}}}
\frac{\partial^{n-1}}{\partial \xi_0^{n-1}}\bigg|_{\xi_0=-\sqrt{\cc}}
\frac{\ep^{-n-3/2}}{\left(\xi_0-\sqrt{\cc}/2\right)\left(\xi-\xi_0\right)}
+ \mathcal{O}(\ep^{-n-1/2})\right) dx\,.
\label{kernel_diff_limit}
\end{align}
Then the amplitudes $\widetilde{\calF}_\NN^{{\dbltinyspace}{\rm conn}(h)}{\negdbltinyspace}(\xi_1,\ldots,\xi_\NN)$ of $2h+N \ge 3$ are also expanded as
\begin{align}
\widetilde{\calF}_\NN^{{\dbltinyspace}{\rm conn}(h)}{\negdbltinyspace}(\xi_1,\ldots,\xi_\NN)\,
d\xi_1 \cdots d\xi_N=
\sum_{n_1,\ldots,n_N \ge 1}
\widetilde{C}_{n_1,\ldots,n_N}^{(h)}\, 
\widetilde{\chi}^{(n_1)}(\xi_1) \cdots \widetilde{\chi}^{(n_N)}(\xi_N)\,,
\label{kernel_exp_cont}
\end{align}
where the coefficients $\widetilde{C}_{n_1,\ldots,n_N}^{(h)}$ do not depend on $\xi_1,\ldots,\xi_N$, and
\begin{align}
\widetilde{\chi}^{(n)}(\xi):=
\frac{d\xi}{(n-1)!\sqrt{\xi + \sqrt{\cc}}}
\frac{\partial^{n-1}}{\partial \xi_0^{n-1}}\bigg|_{\xi_0=-\sqrt{\cc}}
\frac{1}{\left(\xi_0-\sqrt{\cc}/2\right)\left(\xi-\xi_0\right)}\,.
\end{align}
Consequently, we obtain 
the expansion \eqref{GeneralAmpContinuumModeExpansion}.

Removing the non-universal parts in \eqref{DiskAmpContinuumModeExpansion} and 
\eqref{CylinderAmpContinuumModeExpansion}, for 
general $N\ge 1, h \ge 0$, we define
\begin{align}
\tilde{f}_{N}^{{\dbltinyspace}{\rm conn}(h)}{\negdbltinyspace}(\xi_1,\ldots,\xi_N)
:=
\widetilde{\calF}_\NN^{{\dbltinyspace}{\rm conn}(h)}{\negdbltinyspace}(\xi_1,\ldots,\xi_\NN)
-\calOmega_1{\negtinyspace}(\xi)\, \delta_{N,1}\delta_{h,0}
-\calOmega_2{\negtinyspace}(\xi_1,\xi_2)\, \delta_{N,2}\delta_{h,0}\,,
\label{conti_fF}
\end{align}
and
\begin{align}
\tilde{f}_{N}^{{\dbltinyspace}{\rm conn}}(\xi_1, \ldots, \xi_N;\calG):=
\sum_{h=0}^{\infty} \calG^{h+N-1} 
\tilde{f}_{N}^{{\dbltinyspace}{\rm conn}(h)}{\negdbltinyspace}(\xi_1, \ldots, \xi_N)\,.
\label{conti_gen_amp}
\end{align}
In the following, we rewrite the SD equation \eqref{cont_lim_sd} in terms of the disconnected amplitudes $\tilde{f}_{N}(\xi_1, \ldots, \xi_N;\calG)$ for \eqref{conti_gen_amp}.
For that purpose, it is useful to note the following identities for the non-universal parts:
\begin{align}
&
\calOmega_1{\negtinyspace}(\xi)^2 - \omega(\xi) 
= \Big(\frac{3}{8}\cc\Big)^2  \xi^{-1}\,,
\label{om1_rel}
\\
&
\calOmega_2(\xi_1,\xi_2)
=\frac{\partial}{\partial \xi_2} 
\frac{\xi_1^{-1/2}\xi_2^{1/2} - 1}{\xi_1-\xi_2}\,,
\label{om2_rel}
\\
&
\calOmega_2{\negtinyspace}(\xi_1,\xi_2)\,\calOmega_2{\negtinyspace}(\xi_1,\xi_3)
+
\frac{\partial}{\partial \xi_2}
\frac{\calOmega_2{\negtinyspace}(\xi_1,\xi_3)-\calOmega_2{\negtinyspace}(\xi_2,\xi_3)}{\xi_1-\xi_2}
+
\frac{\partial}{\partial \xi_3}
\frac{\calOmega_2{\negtinyspace}(\xi_1,\xi_2)-\calOmega_2{\negtinyspace}(\xi_3,\xi_2)}{\xi_1-\xi_3}
\nonumber\\
&= 
\frac14 \xi_1^{-1}\xi_2^{-3/2}\xi_3^{-3/2}\,.
\label{om22_rel}
\end{align}
Using \eqref{om1_rel}, 
the SD equation \eqref{cont_lim_sd} for $N=1$ yields
\begin{align}
0&=
\frac{\partial}{\partial \xi} 
\tilde{f}_{2}(\xi,\xi;\calG)
+\frac{\partial}{\partial \xi} 
\Big(2 \calOmega_1{\negtinyspace}(\xi)\,\tilde{f}_{1}(\xi;\calG)\Big)
-\left(
\Big(\frac{3}{8}\cc\Big)^2  \xi^{-2} 
+\frac14 \calG \xi^{-3}
\right).
\label{conti_sd_n1}
\end{align}
From this equation and \eqref{om2_rel}, the SD equation \eqref{cont_lim_sd} for $N=2$ yields
\begin{align}
0&=
\frac{\partial}{\partial \xi_i} 
\tilde{f}_{3}(\xi_i,\xi_i, \xi_j;\calG)
+\frac{\partial}{\partial \xi_i} 
\Big(2 \calOmega_1{\negtinyspace}(\xi_i)\,\tilde{f}_{2}(\xi_i,\xi_j;\calG)\Big)
-\left(
\Big(\frac{3}{8}\cc\Big)^2  \xi_i^{-2} 
+\frac14 \calG \xi_i^{-3}
\right) \tilde{f}_{1}(\xi_j;\calG)
\nonumber\\
&\ \
+ \frac{3}{8}\cc\, \calG \xi_i^{-2}\xi_j^{-3/2}
+ 2\calG 
\frac{\partial^2}{\partial \xi_i \partial \xi_j}
\frac{\xi_i^{-1/2}\xi_j^{1/2} \tilde{f}_{1}(\xi_i;\calG)
-\tilde{f}_{1}(\xi_j;\calG)}
{\xi_i-\xi_j}\,,
\label{conti_sd_n2}
\end{align}
where $\{i,j\}=\{1,2\}$. 
Using \eqref{conti_sd_n1} and \eqref{conti_sd_n2}  together with \eqref{om22_rel}, the SD equation \eqref{cont_lim_sd} for $N=3$ yields
\begin{align}
0&=
\frac{\partial}{\partial \xi_i} 
\tilde{f}_{4}(\xi_i,\xi_i, \xi_j, \xi_k;\calG)
+\frac{\partial}{\partial \xi_i} 
\Big(2 \calOmega_1{\negtinyspace}(\xi_i)\,\tilde{f}_{3}(\xi_i,\xi_j,\xi_k;\calG)\Big)
\nonumber\\
&\ \
-\left(
\Big(\frac{3}{8}\cc\Big)^2  \xi_i^{-2} 
+\frac14 \calG \xi_i^{-3}
\right) \tilde{f}_{2}(\xi_j,\xi_k;\calG)
\nonumber\\
&\ \
+\frac{3}{8}\cc\, \calG 
\xi_i^{-2}\xi_j^{-3/2}\, \tilde{f}_{1}(\xi_k;\calG)
+\frac{3}{8}\cc\, \calG 
\xi_i^{-2}\xi_k^{-3/2}\, \tilde{f}_{1}(\xi_j;\calG)
-\frac12 \calG^2 
\xi_i^{-2}\xi_j^{-3/2}\xi_k^{-3/2}
\nonumber\\
&\ \
+2\calG 
\frac{\partial^2}{\partial \xi_i \partial \xi_j}
\frac{\xi_i^{-1/2}\xi_j^{1/2} \tilde{f}_{2}(\xi_i,\xi_k;\calG)
-\tilde{f}_{2}(\xi_j,\xi_k;\calG)}
{\xi_i-\xi_j}
\nonumber\\
&\ \
+2\calG 
\frac{\partial^2}{\partial \xi_i \partial \xi_k}
\frac{\xi_i^{-1/2}\xi_k^{1/2} \tilde{f}_{2}(\xi_i,\xi_j;\calG)
-\tilde{f}_{2}(\xi_k,\xi_j;\calG)}
{\xi_i-\xi_k}\,,
\label{conti_sd_n3}
\end{align}
where $\{i,j,k\}=\{1,2,3\}$. 
In the same way, we find that the SD equation \eqref{cont_lim_sd} for general $N$ yields
\begin{align}
0&=
\frac{\partial}{\partial \xi_i} 
\tilde{f}_{N+1}(\xi_i,\xi_i, \bm{\xi}_{I \backslash \{i\}};\calG)
+\frac{\partial}{\partial \xi_i} 
\Big(2 \calOmega_1{\negtinyspace}(\xi_i)\,\tilde{f}_{N}(\bm{\xi}_I;\calG)\Big)
\nonumber\\
&\ \
-\left(
\Big(\frac{3}{8}\cc\Big)^2  \xi_i^{-2} 
+\frac14 \calG \xi_i^{-3}
\right) \tilde{f}_{N-1}(\bm{\xi}_{I \backslash \{i\}};\calG)
+\frac{3}{8}\cc\, \calG \sum_{j=1\,(j \neq i)}^{N}
\xi_i^{-2}\xi_j^{-3/2}\, \tilde{f}_{N-2}(\bm{\xi}_{I \backslash \{i,j\}};\calG)
\nonumber\\
&\ \
-\frac12 \calG^2 \mathop{\sum_{1 \le j< k \le N}}_{(j,k \neq i)}
\xi_i^{-2}\xi_j^{-3/2}\xi_k^{-3/2}\, \tilde{f}_{N-3}(\bm{\xi}_{I \backslash \{i,j,k\}};\calG)
\nonumber\\
&\ \
+2\calG \sum_{j=1\,(j \neq i)}^{N}
\frac{\partial^2}{\partial \xi_i \partial \xi_j}
\frac{\xi_i^{-1/2}\xi_j^{1/2} \tilde{f}_{N-1}(\bm{\xi}_{I \backslash \{i,j\}};\calG)
-\tilde{f}_{N-1}(\bm{\xi}_{I \backslash \{i\}};\calG)}
{\xi_i-\xi_j}\,.
\label{conti_sd}
\end{align}
By summing over $1\le i \le N$, we obtain the symmetrized SD equation for the disconnected amplitudes:
\begin{align}
0&=
\sum_{i=1}^N \frac{\partial}{\partial \xi_i}
\tilde{f}_{N+1}(\xi_i,\xi_i, \bm{\xi}_{I \backslash \{i\}};\calG)
+\sum_{i=1}^N \frac{\partial}{\partial \xi_i} 
\Big(2 \calOmega_1{\negtinyspace}(\xi_i)\,\tilde{f}_{N}(\bm{\xi}_I;\calG)\Big)
\nonumber\\
&\ \
-\sum_{i=1}^N \left(
\Big(\frac{3}{8}\cc\Big)^2  \xi_i^{-2} 
+\frac14 \calG \xi_i^{-3}
\right) \tilde{f}_{N-1}(\bm{\xi}_{I \backslash \{i\}};\calG)
\nonumber\\
&\ \
+\frac38 \mu\, \calG \sum_{1\le i < j \le N}
\left(\xi_i^{-3/2}\xi_j^{-2}+\xi_i^{-2}\xi_j^{-3/2}\right)
\tilde{f}_{N-2}(\bm{\xi}_{I \backslash \{i,j\}};\calG)
\nonumber\\
&\ \
-\frac12 \calG^2 \sum_{1 \le i< j< k \le N}
\left(
\xi_i^{-3/2}\xi_j^{-3/2}\xi_k^{-2}+\xi_i^{-3/2}\xi_j^{-2}\xi_k^{-3/2}
+\xi_i^{-2}\xi_j^{-3/2}\xi_k^{-3/2}
\right)
\tilde{f}_{N-3}(\bm{\xi}_{I \backslash \{i,j,k\}};\calG)
\nonumber\\
&\ \
+2\calG \sum_{1\le i < j \le N} \frac{\partial^2}{\partial \xi_i \partial \xi_j}
\frac{\xi_i^{-1/2} \tilde{f}_{N-1}(\bm{\xi}_{I \backslash \{j\}};\calG)
-\xi_j^{-1/2} \tilde{f}_{N-1}(\bm{\xi}_{I \backslash \{i\}};\calG)}
{\xi_i^{1/2}-\xi_j^{1/2}}\,.
\label{conti_sd_sum}
\end{align}

\subsubsection{Hamiltonian}
\label{sec:Hamiltonian_continuum}

The purpose of this section is to search for a Hamiltonian $\cH$ such that the disconnected amplitudes defined by
\begin{align}
\tilde{f}_{N}(\xi_1, \ldots, \xi_N;\calG)=
\sum_{\ell_1,\ldots,\ell_N = 1}^{\infty}
\xi_1^{-\ell_1/2-1} \cdots \xi_N^{-\ell_N/2-1}
\lim_{\T\rightarrow\infty}
  \vac {\tinyspace} \E^{- T{\tinyspace}\cH} {\dbltinyspace}
    \phi^\dagger_{\ell_1} \ldots \phi^\dagger_{\ell_\NN}
  \cuum
\,,
\label{conti_amp_exp}
\end{align}
obey \eqref{conti_sd_sum}, 
where $\phi_n$ and $\phi_n^\dagger$ satisfy the commutation relations
\begin{equation}\label{CommutationRelationConjMode}
\commutator{\phi_m}{\phi_n^\dagger}
=
\delta_{m,n}
\,,
\qquad
\commutator{\phi_m^\dagger}{\phi_n^\dagger}
=
0
\,,
\qquad
\commutator{\phi_m}{\phi_n}
=
0
\,,
\end{equation}
and the vacuum state satisfies the conditions 
$\vac \phi_n^\dagger =\phi_n \cuum = 0$.
In the following, we show that such a Hamiltonian is the one given in \cite{SFT:AW}:
\begin{eqnarray}\label{pureDT_HamiltonianModeExpansion}
\cH \!&=&\!
   -\, \frac{1}{4}
       \Big(
         \frac{3{\halftinyspace}\cc}{4}
         -
         \calG {\halftinyspace} \phi_1
       {\negdbltinyspace}\Big)^{{\negfemhalftinyspace}2}
       \phi_2
 \,-\, \frac{\calG}{4}{\tinyspace} \phi_4
 \,-\, \sum_{\ell=1}^\infty \phi_{\ell+1}^\dagger \ell \phi_{\ell}
 \,+\, \frac{3{\halftinyspace}\cc}{8}
       \sum_{\ell=4}^\infty \phi_{\ell-3}^\dagger \ell \phi_{\ell}
\nonumber\\&&\!
   -\> \frac{1}{2}
       \sum_{\ell=6}^\infty \sum_{n=1}^{l-5}
         \phi_n^\dagger \phi_{\ell-n-4}^\dagger
         \ell \phi_{\ell}
 \,-\, \frac{\calG}{4}
       \sum_{\ell=1}^\infty \sum_{n=\max(5-\ell,1)}^\infty\!\!
         \phi_{n+\ell-4}^\dagger
         n \phi_n \ell \phi_{\ell}
\,.
\end{eqnarray}
The amplitudes \eqref{conti_amp_exp} are evaluated using the commutation relations with the Hamiltonian via the SD equation:
\begin{align}
0&=
\lim_{T \to \infty} \pder{{\tinyspace}T}
\vac \E^{-T{\tinyspace}\calH} {\dbltinyspace}
\phi^\dagger_{\ell_1} \ldots \phi^\dagger_{\ell_\NN}
\cuum
=
\lim_{T \to \infty} \pder{{\tinyspace}T}
\vac \E^{-T{\tinyspace}\calH}\,
[-\mathcal{H}, \phi_{\ell_1}^{\dagger}\cdots \phi_{\ell_N}^{\dagger}]
\cuum
\nonumber\\
&=
\sum_{i=1}^N
\lim_{T \to \infty} \pder{{\tinyspace}T}
\vac \E^{-T{\tinyspace}\calH} {\dbltinyspace}
\phi_{\ell_1}^{\dagger}\cdots \phi_{\ell_{i-1}}^{\dagger}
\,[-\mathcal{H}, \phi_{\ell_i}^{\dagger}]\,
\phi_{\ell_{i+1}}^{\dagger}\cdots \phi_{\ell_N}^{\dagger}
\cuum\,.
\label{conti_sd_eq}
\end{align}
The commutation relation
\begin{align}
[-\mathcal{H}, \phi_{\ell}^{\dagger}]&=
-\frac12 \delta_{\ell,1} 
\Big(\frac{3\cc}{4}-\calG \phi_1\Big) \calG \phi_2
+ \frac14 \delta_{\ell,2} 
\Big(\frac{3\cc}{4}-\calG \phi_1\Big)^{2} 
+ \frac14 \delta_{\ell, 4}\, \calG
+ \ell\, \phi_{\ell+1}^{\dagger}
\nonumber\\
&\ \ \
- \frac{3\cc}{8} \ell\, \theta_{\ell-3,1}\, \phi_{\ell-3}^{\dagger}
+\frac12 \ell\, \sum_{n=1}^{\ell-5} \phi_n^{\dagger} \phi_{\ell-n-4}^{\dagger}
+\frac12 \ell\, \calG \sum_{n= \mathrm{max}(5-\ell,1)}^{\infty} n\, \phi_{n+\ell-4}^{\dagger} \phi_n\,,
\label{comut_H_1}
\end{align}
leads to
\begin{align}
&
\phi_{\ell_1}^{\dagger}\cdots \phi_{\ell_{i-1}}^{\dagger}
\,[-\mathcal{H}, \phi_{\ell_i}^{\dagger}]\,
\phi_{\ell_{i+1}}^{\dagger}\cdots \phi_{\ell_N}^{\dagger}
\nonumber\\
&=
\phi_{\ell_1}^{\dagger}\cdots \breve{\phi}_{\ell_i}^{\dagger}\cdots
\phi_{\ell_N}^{\dagger}
\left(
- \frac12 \delta_{\ell_i,1}
\Big(\frac{3\cc}{4}-\calG \phi_1\Big) \calG \phi_2
+ \frac14 \delta_{\ell_i,2} 
\Big(\frac{3\cc}{4}-\calG \phi_1\Big)^{2} 
+ \frac14 \delta_{\ell_i, 4}\, \calG
\right)
\nonumber\\
&\ \
+ \ell_i\,
\phi_{\ell_1}^{\dagger}\cdots \phi_{\ell_i+1}^{\dagger}\cdots \phi_{\ell_N}^{\dagger}
- \frac{3\cc}{8} \ell_i\, \theta_{\ell_i-3,1}\, 
\phi_{\ell_1}^{\dagger}\cdots \phi_{\ell_i-3}^{\dagger}\cdots \phi_{\ell_N}^{\dagger}
\nonumber\\
&\ \
+\phi_{\ell_1}^{\dagger}\cdots \breve{\phi}_{\ell_i}^{\dagger}\cdots
\phi_{\ell_N}^{\dagger}\left(
\frac12 \ell_i \sum_{n=1}^{\ell_i-5} \phi_n^{\dagger} \phi_{\ell_i-n-4}^{\dagger}
+ \frac12 \calG\, \ell_i 
\sum_{n = \mathrm{max}(5-\ell_i,1)}^{\infty} n\, \phi_{n+\ell_i-4}^{\dagger} \phi_n
\right)
\nonumber\\
&\ \
+\frac12 \sum_{j=i+1}^N
\phi_{\ell_1}^{\dagger}\cdots \breve{\phi}_{\ell_i}^{\dagger} \cdots
\breve{\phi}_{\ell_j}^{\dagger}\cdots
\phi_{\ell_N}^{\dagger}
\left(
\delta_{\ell_i+\ell_j,2}\, \calG^2 \phi_2 
- \delta_{\ell_i+\ell_j,3}\, \Big(\frac{3\cc}{4}-\calG \phi_1\Big) \calG
\right)
\nonumber\\
&\ \
+\frac12 \calG^2 \sum_{i+1 \le j<k \le N}
\delta_{\ell_i+\ell_j+\ell_k, 4}\,
\phi_{\ell_1}^{\dagger}\cdots \breve{\phi}_{\ell_i}^{\dagger}
\cdots \breve{\phi}_{\ell_j}^{\dagger}\cdots
\breve{\phi}_{\ell_k}^{\dagger}\cdots \phi_{\ell_N}^{\dagger}
\nonumber\\
&\ \
+\frac12 \calG \sum_{j=i+1}^N
\ell_i\, \ell_j\, \theta_{\ell_i+\ell_j-4,1}\,
\phi_{\ell_i+\ell_j-4}^{\dagger}\,
\phi_{\ell_1}^{\dagger}\cdots \breve{\phi}_{\ell_i}^{\dagger}
\cdots \breve{\phi}_{\ell_j}^{\dagger} \cdots \phi_{\ell_N}^{\dagger},
\label{comut_H_phi}
\end{align}
where $\breve{\phi}_{\ell}^{\dagger}$ indicates that $\phi_{\ell}^{\dagger}$ is excluded. 
Then the SD equation \eqref{conti_sd_eq} for the amplitudes \eqref{conti_amp_exp} yields \eqref{conti_sd_sum}, thereby
confirming that the Hamiltonian \eqref{pureDT_HamiltonianModeExpansion} describes pure DT at the continuous level.

Note that the Hamiltonian \eqref{pureDT_HamiltonianModeExpansion} is realized in the string field theory. 
Using the string fields defined by
\begin{equation}\label{StringFieldModeExpansion}
\tilde{\calPhi}^\dagger{\negdbltinyspace}(\xi)
\,=\,
\calOmega_1{\negtinyspace}(\xi)
+ \sum_{\ell=1}^\infty \xi^{{\tinyspace}-\ell/2 -1}
   \phi^\dagger_\ell
\,,
\qquad
\tilde{\calPsi}(-\eta)
\,=\,
  \sum_{\ell=1}^\infty \eta^{{\tinyspace}\ell/2}
   \phi_\ell
\,,
\end{equation}
the Hamiltonian \rf{pureDT_HamiltonianModeExpansion}
can be rewritten as 
\begin{eqnarray}\label{pureDT_HamiltonianModeExpansionField}
\cH \!&=&\!
-\>
\underset{\zeta=0}{\rm Res}{\trpltinyspace}
\zeta
\bigg[
    \big(
      \tilde{\calPhi}^\dagger{\negdbltinyspace}(\zeta^2)
    \big)^{{\negdbltinyspace}2}
    {\tinyspace}
    \frac{\partial \tilde{\calPsi}(-\zeta^2)}
         {\partial \zeta^2}
  +
  \calG{\tinyspace}
    \tilde{\calPhi}^\dagger{\negdbltinyspace}(\zeta^2)
    {\negtinyspace}
    \Big(
      \frac{\partial \tilde{\calPsi}(-\zeta^2)}
           {\partial \zeta^2}
    \Big)^{{\negdbltinyspace}2}
\nonumber\\&&\!\phantom{%
-\>
\underset{\zeta=0}{\rm Res}{\trpltinyspace}
\zeta
\bigg[
}{\negtrpltinyspace}
  +
  \frac{\calG^{{\halftinyspace}2}}{3}
    \Big(
      \frac{\partial \tilde{\calPsi}(-\zeta^2)}
           {\partial \zeta^2}
    \Big)^{{\negtrpltinyspace}3}
  +
  \frac{\calG}{8}{\tinyspace} \zeta^{{\tinyspace}-4}
    \frac{\partial \tilde{\calPsi}(-\zeta^2)}
         {\partial \zeta^2}
\bigg]
\,.
\end{eqnarray}
Since the amplitudes with even indices vanish, 
one may freely add 
$\phi_{2n}^\dagger$
[\,$n {\negtrpltinyspace}\in{\negtrpltinyspace} \Dbl{N}$\,]. 
Then, 
the Hamiltonian \rf{pureDT_HamiltonianModeExpansionField} 
can be rewritten as 
\begin{equation}\label{pureDT_HamiltonianModeExpansionWop}
\cH
 \,=\,
-\>\sqrt{\calG}{\qdrpltinyspace}
\underset{\zeta=0}{\rm Res}{\trpltinyspace}
\zeta^{{\tinyspace}-5}
\Big[{\dbltinyspace}
  \frac{1}{3}
  \big(
    \calJ(\zeta^2)
  \big)^{{\negtinyspace}3}
  +
  \frac{1}{8}
    \calJ(\zeta^2)
\Big]
\,,
\end{equation}
where 
\begin{equation}
\calJ(\xi)
\,\define\,
    \xi
    \bigg({\negtinyspace}
      \frac{1}{\sqrt{\calG}}{\tinyspace}
      \tilde{\calPhi}^\dagger{\negdbltinyspace}(\xi)
      +
      \sqrt{\calG}{\trpltinyspace}
      \frac{\partial \tilde{\calPsi}(-\xi)}
           {\partial \xi}
    \bigg)
\,.
\end{equation}
Note that 
the Hamiltonian \rf{pureDT_HamiltonianModeExpansionWop} 
is proportional to the two-reduced $W^{(3)}$ operator \cite{MM:FKN,SFT:AW,FMW2,FMW3} (see, e.g., \cite{SFT:WatabikiReview} for a review).

\subsection{Amplitudes and topological recursion}

\subsubsection{Disk and cylinder amplitudes}

We integrate \eqref{cont_lim_sd} with respect to $\xi_i$ and obtain
\begin{align}
0&=
\widetilde{\calF}_{N+1}^{{\tinyspace}{\rm conn}}(\xi_i,\xi_i,\bm{\xi}_{I \backslash \{i\}};\calG)
+ 
\sum_{I_1 \cup I_2=I \backslash \{i\}}
\widetilde{\calF}_{|I_1|+1}^{{\tinyspace}{\rm conn}}(\xi_i,\bm{\xi}_{I_1};\calG)\,
\widetilde{\calF}_{|I_2|+1}^{{\tinyspace}{\rm conn}}(\xi_i,\bm{\xi}_{I_2};\calG)
- \calomega(\xi_i)\, \delta_{N,1}
\nonumber\\
& \ \ \ \
+ 2\, \calG {\negdbltinyspace}
\sum_{{\scriptstyle j=1} \atop (j \neq i)}^{\NN}
\pder{\xi_j}
\frac{\widetilde{\calF}_{N-1}^{{\tinyspace}{\rm conn}}(\bm{\xi}_{I \backslash \{j\}};\calG)
-\widetilde{\calF}_{N-1}^{{\tinyspace}{\rm conn}}(\bm{\xi}_{I \backslash \{i\}};\calG)}
{\xi_i - \xi_j}
+C_N(\bm{\xi}_{I \backslash \{i\}})\,,
\label{cont_lim_sd_int}
\end{align}
where $C_N(\bm{\xi}_{I \backslash \{i\}})$ is a function of $\bm{\xi}_{I \backslash \{i\}}$.

For $N=1$, using the expansion \eqref{AmplitudeGexpansionContinuumN},
the zeroth-order term in $\calG$ from \eqref{cont_lim_sd_int} is
\begin{align}
0=
\widetilde{\calF}_1^{{\dbltinyspace}{\rm conn(0)}}{\negdbltinyspace}(\xi)^2
- \omega(\xi) +C_1
=
\widetilde{\calF}_1^{{\dbltinyspace}{\rm conn(0)}}{\negdbltinyspace}(\xi)^2
-\left(\xi^3 - \frac{3}{4} \cc\, \xi\right) +C_1^{(0)}
\,.
\label{ContinuousSDeqN1G0}
\end{align}
Here $C_1^{(0)}=-2f_1^{{\dbltinyspace}{\rm conn}(0)}{\negdbltinyspace}(1)$, obtained from the expansion \eqref{DiskAmpContinuumModeExpansion}, 
and one finds the disk amplitude \eqref{DiskAmpContinuum}, i.e.,  
\begin{equation}
 \widetilde{\calF}_1^{{\dbltinyspace}{\rm conn}(0)}{\negdbltinyspace}(\xi)
\,=\,
\Big(
  \xi - \frac{\sqrt{\cc}}{2}
{\dbltinyspace}\Big)
\sqrt{
  \xi + \sqrt{\cc}
}
\,=\,
M(\xi) \sqrt{\sigma(\xi)}
\,,
\label{ContinuousDiskAmp}
\end{equation}
where
\begin{align}
M(\xi) :=
\xi - \frac{\sqrt{\cc}}{2}\,,
\qquad
\sigma(\xi) :=
\xi + \sqrt{\cc}\,.
\label{conti_SpDef}
\end{align}
Here, we introduce a variable $\eta \in \mathbb{P}^1$ defined by
\begin{align}
\xi(\eta) = \eta^2 - \sqrt{\cc}\,,
\label{local_conti}
\end{align}
so that $\sqrt{\sigma(\xi(\eta))}= \eta$.

For $N=2$, using the expansion \eqref{AmplitudeGexpansionContinuumN},
the zeroth-order term in $\calG$ from \eqref{cont_lim_sd_int} is
\begin{align}
0&=
\widetilde{\calF}_{1}^{{\dbltinyspace}{\rm conn(0)}}{\negdbltinyspace}(\xi_1)\,
\widetilde{\calF}_{2}^{{\dbltinyspace}{\rm conn(0)}}{\negdbltinyspace}(\xi_1,\xi_2)
+ \pder{\xi_2}
\frac{\widetilde{\calF}_{1}^{{\dbltinyspace}{\rm conn(0)}}{\negdbltinyspace}(\xi_1)
-\widetilde{\calF}_{1}^{{\dbltinyspace}{\rm conn(0)}}{\negdbltinyspace}(\xi_2)}
{\xi_1 - \xi_2}
+C_2^{(0)}(\xi_2)\,,
\label{ContinuousSDeqN2G0}
\end{align}
where $C_2^{(0)}(\xi_2)$ is a function of $\xi_2$.
From the disk amplitude \eqref{ContinuousDiskAmp}, the equation \eqref{ContinuousSDeqN2G0} follows as
\begin{align}
0=
\sqrt{\sigma(\xi_1)}\,
\widetilde{\calF}_{2}^{{\dbltinyspace}{\rm conn(0)}}{\negdbltinyspace}(\xi_1,\xi_2)
+\pder{\xi_2}
\frac{\sqrt{\sigma(\xi_1)}-\sqrt{\sigma(\xi_2)}}{\xi_1-\xi_2}
+\frac{\pder{\xi_2}\sqrt{\sigma(\xi_2)}+C_2^{(0)}(\xi_2)}{\xi_1-\sqrt{\cc}/2}\,.
\label{ContinuousSDeqN2G0_2}
\end{align}
Assuming that $\widetilde{\calF}_{2}^{{\dbltinyspace}{\rm conn(0)}}{\negdbltinyspace}(\xi_1,\xi_2)$ has no poles at $\xi_1=\sqrt{\cc}/2$, the last term disappears, and $C_2^{(0)}(\xi_2)$ is fixed as in \cite{Eynard:2004mh}. Then, we obtain the cylinder amplitude \eqref{CylinderAmpContinuum}:
\begin{equation}\label{ContinuousCylinderAmp}
 \widetilde{\calF}_2^{{\dbltinyspace}{\rm conn}(0)}{\negdbltinyspace}(\xi_1,\xi_2)
\,=\,
\frac{1}{( \xi_1 - \xi_2 )^2}
\bigg(
  \frac{
        \xi_1 + \xi_2 + 2 \sqrt{\cc}
       }
       {2
        \sqrt{\xi_1 + \sqrt{\cc}}
        \sqrt{\xi_2 + \sqrt{\cc}}
       }
  - 1
{\tinyspace}\bigg)\,.
\end{equation}
With the map \eqref{local_conti}, the cylinder amplitude is also written as
\begin{align}\label{BasicBergman_cont}
\widetilde{\calF}_2^{{\dbltinyspace}{\rm conn}(0)}{\negdbltinyspace}(\xi(\eta_1),\xi(\eta_2))\, d\xi(\eta_1)d\xi(\eta_2)
\,&=\,
\frac{2\, d\eta_{1}d\eta_{2}}{\left(\eta_{1}-\eta_{2}\right)^2}
-\frac{2\, d\xi(\eta_1)d\xi(\eta_2)}{\left(\xi(\eta_1)-\xi(\eta_2)\right)^2}
\nonumber\\
\,&=\, 2\, B(\eta_{1},\eta_{2})
- \frac{2\,d\xi(\eta_1)d\xi(\eta_2)}{\left(\xi(\eta_1)-\xi(\eta_2)\right)^2}
\nonumber\\
\,&=\, \frac{2\,d\eta_{1}d\eta_{2}}{\left(\eta_{1}+\eta_{2}\right)^2}
\,=\, -2\, B(\eta_{1},-\eta_{2})\,.
\end{align}

\subsubsection{Topological recursion}

From the expansion \eqref{AmplitudeGexpansionContinuumN} with respect to $\calG$, 
the equation \eqref{cont_lim_sd_int} with $i=1$ gives
\begin{align}
\label{sp_sd_N_con_pert}
\widetilde{\calF}_{N}^{{\tinyspace}{\rm conn}(h)}{\negdbltinyspace}(\xi_1,\bm{\xi}_{I \backslash \{1\}})
&=
\frac{(-1)}{2\widetilde{\calF}_1^{{\tinyspace}{\rm conn}(0)}{\negdbltinyspace}(\xi_1)}
\Biggl[\widetilde{\calF}_{N+1}^{{\tinyspace}{\rm conn}(h-1)}{\negdbltinyspace}(\xi_1,\xi_1,\bm{\xi}_{I \backslash \{1\}})
\nonumber\\
&\hspace{7em}
+ 
\mathop{\sum_{h_1+h_2=h}}_{I_1 \cup I_2=\{2,\ldots,N\}}^{\textrm{no $(0,1)$}}
\widetilde{\calF}_{|I_1|+1}^{{\tinyspace}{\rm conn}(h_1)}{\negdbltinyspace}(\xi_1,\bm{\xi}_{I_1})
\widetilde{\calF}_{|I_2|+1}^{{\tinyspace}{\rm conn}(h_2)}{\negdbltinyspace}(\xi_1,\bm{\xi}_{I_2})
\nonumber\\
&\hspace{7em}
+2 \sum_{i=2}^{N} 
\frac{\widetilde{\calF}_{N-1}^{{\tinyspace}{\rm conn}(h)}{\negdbltinyspace}(\bm{\xi}_{I \backslash \{i\}})}{(\xi_1-\xi_i)^2}\Biggr]
+R(\xi_1;\bm{\xi}_{I \backslash \{1\}})\,,
\\
R(\xi_1;\bm{\xi}_{I \backslash \{1\}})&:=
\frac{(-1)}{2\widetilde{\calF}_{1}^{{\tinyspace}{\rm conn}(0)}{\negdbltinyspace}(\xi_1)}
\Biggl[
-2\sum_{i=2}^{N} \pder{\xi_i} 
\frac{\widetilde{\calF}_{N-1}^{{\tinyspace}{\rm conn}(h)}{\negdbltinyspace}(\bm{\xi}_{I \backslash \{1\}})}{\xi_1-\xi_i}
+C_N^{(h)}(\bm{\xi}_{I \backslash \{1\}})\Biggr]
\,,
\label{r_cont}
\end{align}
for $(h,N)\ne (0,1)$,
where $C^{(h)}(\bm{\xi}_{I \backslash \{1\}})$ is a function of $\bm{\xi}_{I \backslash \{1\}}$. 
The equation \eqref{sp_sd_N_con_pert} has the same form as \eqref{BasicSD} and \eqref{StripSD}. 
Assuming that the amplitudes 
$\widetilde{\calF}_{N}^{{\tinyspace}{\rm conn}(h)}{\negdbltinyspace}(\bm{\xi}_I)$ have no poles away from the branch cut $[-\sqrt{\cc}, \infty)$ of the disk amplitude \eqref{ContinuousDiskAmp}, we obtain \cite{Eynard:2004mh},
\begin{align}
\widetilde{\calF}_{N}^{{\tinyspace}{\rm conn}(h)}{\negdbltinyspace}(\bm{\xi}_I)
&=
\frac{1}{2\pi \mathrm{i}}\oint_{\xi_0=\xi_1}
\frac{d\xi_0}{\xi_0-\xi_1}\,
\sqrt{\frac{\sigma(\xi_0)}{\sigma(\xi_1)}}\,
\widetilde{\calF}_{N}^{{\tinyspace}{\rm conn}(h)}{\negdbltinyspace}(\xi_0,\bm{\xi}_{I \backslash \{1\}})
\nonumber\\
&=
\frac{(-1)}{2\pi \mathrm{i}}\oint_{[-\sqrt{\cc}, \infty)}
\frac{d\xi_0\, dS_{\eta_0}(\eta_1)}{2\widetilde{\calF}_{1}^{{\tinyspace}{\rm conn}(0)}{\negdbltinyspace}(\xi_0)\, d\xi(\eta_1)}
\Biggl[
\widetilde{\calF}_{N+1}^{{\tinyspace}{\rm conn}(h-1)}{\negdbltinyspace}(\xi_0,\xi_0,\bm{\xi}_{I \backslash \{1\}})
\nonumber\\
&\hspace{8em}
+ 
\mathop{\sum_{h_1+h_2=h}}_{I_1 \cup I_2=\{2,\ldots,N\}}^{\textrm{no $(0,1)$}}
\widetilde{\mathfrak{F}}_{|I_1|+1}^{{\tinyspace}{\rm conn}(h_1)}{\negdbltinyspace}(\xi_0,\bm{\xi}_{I_1})
\widetilde{\mathfrak{F}}_{|I_2|+1}^{{\tinyspace}{\rm conn}(h_2)}{\negdbltinyspace}(\xi_0,\bm{\xi}_{I_2})
\Biggr],
\label{sd_tr_cont}
\end{align}
where
\begin{align}
dS_{\eta_0}(\eta_1)&:=
\frac{d\xi(\eta_1)}{\xi(\eta_1)-\xi(\eta_0)}\,
\sqrt{\frac{\sigma(\xi(\eta_0))}{\sigma(\xi(\eta_1))}}
=
\frac{2\eta_0\, d\eta_1}{\eta_1^2-\eta_0^2}
\left(=
\int^{\eta_0}_{-\eta_0} B(\cdot, \eta_1)\right),
\\
\widetilde{\mathfrak{F}}_{|I|+1}^{{\tinyspace}{\rm conn}(h)}{\negdbltinyspace}(\xi_0, \bm{\xi}_I)&:=
\widetilde{\calF}_{|I|+1}^{{\tinyspace}{\rm conn}(h)}{\negdbltinyspace}(\xi_0, \bm{\xi}_I)+
\frac{\delta_{|I|,1}\, \delta_{h,0}}{(\xi_0-\xi_i)^2},\qquad 
i \in I\,.
\end{align}
Here the variables $\eta_i \in \mathbb{P}^1$ are introduced 
via the map \eqref{local_conti}, 
and then,
$B(\cdot, \eta_1)$ denotes the bi-differential in \eqref{BasicBergman_cont}.
The integrand of the equation \eqref{sd_tr_cont} has no branch cuts because  
$\widetilde{\mathfrak{F}}_{|I|+1}^{{\tinyspace}{\rm conn}(h)}{\negdbltinyspace}(\xi(-\eta_0), \bm{\xi}_I)
=-\widetilde{\mathfrak{F}}_{|I|+1}^{{\tinyspace}{\rm conn}(h)}{\negdbltinyspace}(\xi(\eta_0), \bm{\xi}_I)$.
Therefore, 
the equation \eqref{sd_tr_cont} can be rewritten as
\begin{align}
\widetilde{\calF}_{N}^{{\tinyspace}{\rm conn}(h)}{\negdbltinyspace}(\bm{\xi}_I)&=
\mathop{\mathrm{Res}}_{\xi=-\sqrt{\cc}}\,
\frac{(-1)\, d\xi_0\, dS_{\eta_0}(\eta_1)}
{2\widetilde{\calF}_{1}^{{\tinyspace}{\rm conn}(0)}{\negdbltinyspace}(\xi_0)\, d\xi(\eta_1)}
\Biggl[
\widetilde{\calF}_{N+1}^{{\tinyspace}{\rm conn}(h-1)}{\negdbltinyspace}(\xi_0,\xi_0,\bm{\xi}_{I \backslash \{1\}})
\nonumber
\\
&\hspace{8em}
+ 
\mathop{\sum_{h_1+h_2=h}}_{I_1 \cup I_2=\{2,\ldots,N\}}^{\textrm{no $(0,1)$}}
\widetilde{\mathfrak{F}}_{|I_1|+1}^{{\tinyspace}{\rm conn}(h_1)}{\negdbltinyspace}(\xi_0,\bm{\xi}_{I_1})
\widetilde{\mathfrak{F}}_{|I_2|+1}^{{\tinyspace}{\rm conn}(h_2)}{\negdbltinyspace}(\xi_0,\bm{\xi}_{I_2})\Biggr]\,.
\label{top_rec_conti_F}
\end{align}
By introducing multi-differentials%
\footnote{%
Compared with the (standard) normalization of the multi-differentials \eqref{F_mdiffW_d}, 
the overall factor $2$ appears in $\omega_{2}^{(0)}(\eta_{1}, \eta_{2})$. 
This is due to the normalization of $\calG$ in \eqref{Continuum}.
}
\begin{align}
&
\omega_{2}^{(0)}(\eta_{1}, \eta_{2})=2B(\eta_{1}, \eta_{2})
=\frac{2d\eta_{1}d\eta_{2}}{\left(\eta_{1}-\eta_{2}\right)^2}\,,
\nonumber\\
&
\omega_{N}^{(h)}(\eta_1,\ldots,\eta_N)=
\widetilde{\calF}_{N}^{{\tinyspace}{\rm conn}(h)}{\negdbltinyspace}(\xi(\eta_1),\ldots,\xi(\eta_N))\, 
d\xi(\eta_{1}) \cdots d\xi(\eta_{N})
\ \ \textrm{for}\ \ (h,N)\ne (0,2)\,,
\label{F_mdiffW_conti}
\end{align}
the equation \eqref{top_rec_conti_F} is rewritten in the form of the topological recursion in terms of the variables $\eta_i$,
\begin{align}
\omega_{N}^{(h)}(\bm{\eta}_I)&=
\mathop{\mathrm{Res}}_{\eta=0}\,
K_{\eta_0}(\eta_{1})
\Biggl[
\omega_{N+1}^{(h-1)}(\eta_{0},-\eta_{0},\bm{\eta}_{I \backslash \{1\}})
\nonumber\\
&\hspace{8em}
+
\mathop{\sum_{h_1+h_2=h}}_{I_1 \cup I_2=\{2,\ldots,N\}}^{\textrm{no $(0,1)$}}
\omega_{|I_1|+1}^{(h_1)}(\eta_{0}, \bm{\eta}_{I_1})\,
\omega_{|I_2|+1}^{(h_2)}(-\eta_{0}, \bm{\eta}_{I_2})\Biggr]\,,
\label{top_rec_W_conti}
\end{align}
where $\bm{\eta}_I=\{\eta_1, \ldots, \eta_N\}$, and the recursion kernel $K_{\eta_0}(\eta_{1})$ is given by
\begin{align}
K_{\eta_0}(\eta_{1})=
\frac{dS_{\eta_{0}}(\eta_{1})}{4\omega_1^{(0)}(\eta_0)}\,.
\end{align}
Here the spectral curve data $(\mathbb{P}^1; \xi,y,B)$ consists of the disk amplitude $y=\widetilde{\calF}_{1}^{{\tinyspace}{\rm conn}(0)}{\negdbltinyspace}(\xi)$ as given in \eqref{ContinuousDiskAmp},
\begin{align}
y=M(\xi) \sqrt{\sigma(\xi)}
=
\left(\xi - \frac{\sqrt{\cc}}{2}\right)
\sqrt{\xi + \sqrt{\cc}}\,,
\end{align}
and the cylinder amplitude, which is 
the bi-differential $B=B(\eta_{1},\eta_{2})$ given in \eqref{F_mdiffW_conti}.
Some computational results of the amplitudes are summarized 
in Appendix \ref{app:list_dt_conti}.


\section{Conclusions}\label{sec:conclusions}

In this paper we have established explicitly that the Schwinger-Dyson equations of the string field theory for noncritical strings based on dynamical triangulations (DT) of two-dimensional pure gravity, developed in the 1990s, can be reformulated as the Chekhov-Eynard-Orantin (CEO) topological recursion associated with a specific spectral curve.

Previous studies have shown, primarily through the scaling limits of matrix models \cite{Eynard:2016yaa} and analyses involving FZZT and ZZ branes \cite{Seiberg:2003nm}, that correlation functions in noncritical string theory obey the CEO topological recursion. These results have provided the conceptual foundation for numerous developments, including resurgence analyses \cite{Eynard:2023qdr} and recent progress in models such as JT gravity \cite{Saad:2019lba}. In contrast, the relation between the CEO topological recursion and another formulation of noncritical string theory—namely the string field theory description—had long been conjectured on the basis of structures such as the Virasoro constraints, but a direct and explicit proof had not been available.

In the present work, we revisit the string field theory of pure DT, originally formulated by Ishibashi and Kawai \cite{SFT:IK} and later reformulated by Watabiki and Ambj{\o}rn without introducing additional regularization, and analyze it in a fully unambiguous manner \cite{SFT:Watabiki,SFT:AW}.
On this basis we derived the Schwinger-Dyson equations for amplitudes with an arbitrary number $N$ of boundaries for two discrete models as well as for the corresponding continuum model. We then demonstrated that all of these equations can be rewritten in the form of the CEO topological recursion. In particular, the DT (basic type) model is defined purely combinatorially and does not admit a matrix-model representation. It is therefore highly nontrivial that, even for such a model, the spectral curve can be extracted from the disk amplitude and that the CEO topological recursion with this curve as input follows from the Schwinger-Dyson equations of the string field theory.

The results obtained here also suggest a natural direction for future work. One may attempt to address the inverse problem \cite{FMW3}: constructing a dictionary that associates a given CEO topological recursion with the Hamiltonian of a corresponding string field theory. Such a perspective would open the possibility of exploring new aspects of the CEO formalism from the viewpoint of string field theory. In recent years the CEO topological recursion has been actively reinterpreted and reformulated in various frameworks, including quantum Airy structure \cite{Kontsevich:2017vdc,Andersen:2017vyk} and geometric recursion \cite{GR}. A formulation based on string field theory, as initiated in this work, may therefore provide new insights into the structural foundations of the CEO recursion and potentially lead to further developments.

\vspace{1cm}
\noindent{\textbf{\Large Acknowledgements}}\\

The authors would like to thank Jan Ambj\o rn and Yasuhiko Yamada for their valuable comments.
This work was supported by JSPS KAKENHI Grant Numbers JP23K22388 and JP25K07278.

\renewcommand{\theequation}{\thesection.\arabic{equation}}
\renewcommand{\thefigure}{\thesection.\arabic{figure}}
\appendix
\allowdisplaybreaks


\section{Formulas for Computing Schwinger-Dyson Equations}
\label{app:CalculationConnectivity}

The quadratic term 
$\Phi^\dagger(z)^{2}$ appears in the Hamiltonians \rf{BasicTypeHamiltonianConjLengthShift}
and \rf{StripTypeHamiltonianConjLengthShift}.
In the computation of the Schwinger-Dyson equations, we will repeatedly use the following identities involving this term:%
\footnote{
The formulas \rf{Connectivity0} and \rf{Connectivity1}
are employed in the derivations of \rf{BasicTypeSDeqN1} and \rf{StripTypeSDeqN1}, 
\rf{BasicTypeSDeqN2} and \rf{StripTypeSDeqN2}, respectively, 
while the formula \rf{ConnectivityN} is employed in the derivations of 
\rf{BasicTypeSDeqGeneralNN} and \rf{StripTypeSDeqGeneralNN}.
}
\begin{align}
&
\lim_{T \to \infty}
\vac \E^{-T \Hop} {\tinyspace}
\Phi^\dagger(z)^{2}
\cuum^{{\negtinyspace}{\rm conn}}
\nonumber\\
&=
\tilde{F}_2^{{\tinyspace}{\rm conn}}(z,z)
+
\tilde{F}_1^{{\tinyspace}{\rm conn}}(z)^2
\,,
\label{Connectivity0}
\\
&
\lim_{T \to \infty}
\vac \E^{-T \Hop} {\tinyspace}
\Phi^\dagger(z)^{2}
\Phi^\dagger(x_1)
\cuum^{{\negtinyspace}{\rm conn}}
\nonumber\\
&=
\tilde{F}_3^{{\tinyspace}{\rm conn}}(z,z,x_1)
+
2{\tinyspace}
\tilde{F}_1^{{\tinyspace}{\rm conn}}(z)
\tilde{F}_2^{{\tinyspace}{\rm conn}}(z,x_1)
\,,
\label{Connectivity1}
\\
&
\lim_{T \to \infty}
\vac \E^{-T \Hop} {\tinyspace}
\Phi^\dagger(z)^{2}
\Phi^\dagger(x_1)\Phi^\dagger(x_2)
\cuum^{{\negtinyspace}{\rm conn}}
\nonumber\\
&=
\tilde{F}_4^{{\tinyspace}{\rm conn}}(z,z,x_1,x_2)
+
2
\tilde{F}_2^{{\tinyspace}{\rm conn}}(z,x_1)
\tilde{F}_2^{{\tinyspace}{\rm conn}}(z,x_2)
+
2
\tilde{F}_1^{{\tinyspace}{\rm conn}}(z)
\tilde{F}_3^{{\tinyspace}{\rm conn}}(z,x_1,x_2)
\,,
\label{Connectivity2}
\\
&
\lim_{T \to \infty}
\vac \E^{-T \Hop} {\tinyspace}
\Phi^\dagger(z)^{2}
\prod_{k=1}^\NN{\negtinyspace} \tildePhidag(x_k)
\cuum^{{\negtinyspace}{\rm conn}}
\nonumber\\
&=
\tilde{F}_{\NN+2}^{{\tinyspace}{\rm conn}}(z,z,x_1,\ldots,x_\NN)
\,+
\sum_{I_1 \cup I_2=\{1,\ldots,N\}}
  \tilde{F}_{|I_1|+1}^{{\tinyspace}{\rm conn}}(z,\bm{x}_{I_1})
  \tilde{F}_{|I_2|+1}^{{\tinyspace}{\rm conn}}(z,\bm{x}_{I_2})\,.
\label{ConnectivityN}
\end{align}

\section{List of Amplitudes}
\label{app:list_amplitudes}

\subsection{Dynamical triangulation (basic type)}
\label{app:list_dt_basic}

\begin{figure}[t]
\begin{center}
\vspace{-0cm}
\hspace*{0cm}
  \includegraphics[bb=0 0 595 842, width=125mm]{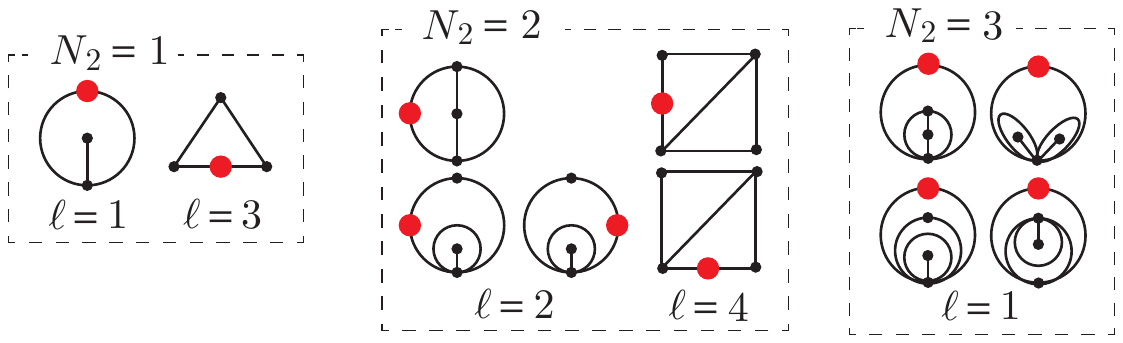}
 \vspace{-13cm}
\caption{Examples of triangulated disks with $N_2$ triangles and $\ell$ boundaries, as enumerated by \eqref{basic_disk_ex}; the red circles indicate marked points on the boundaries.}
\label{fig:disk_basic}
\end{center}
\end{figure}

The disk amplitude \eqref{BasicTypeDiskAmp} can be expanded around $\kappa=0$ as
\begingroup\makeatletter\def\f@size{10}\check@mathfonts
\def\maketag@@@#1{\hbox{\m@th\normalsize\normalfont#1}}%
\begin{align}
&
\AmpFXXX_1^{{\tinyspace}{\rm conn}(0)}{\negdbltinyspace}(x)
+ \lambda(x)
=
\frac{1}{2x^3} \!\left(x-\frac{c}{\kappa}\right)
\sqrt{x \!\left(x-\frac{4\kappa}{c^2}\right)}
+ \frac{x-\kappa x^2}{2 \kappa x^{3}}-\frac{1}{x^{3}}
\nonumber\\
&=
\left(\frac{1}{x^{2}}+\frac{1}{x^{4}}\right) \kappa +\left(\frac{3}{x^{3}}+\frac{2}{x^{5}}\right) \kappa^{2}
+\left(\frac{4}{x^{2}}+\frac{10}{x^{4}}+\frac{5}{x^{6}}\right)\kappa^{3}
+\left(\frac{24}{x^{3}}+\frac{35}{x^{5}}+\frac{14}{x^{7}}\right) \kappa^{4}
\nonumber\\
&\ \
+\left(\frac{32}{x^{2}}+\frac{120}{x^{4}}+\frac{126}{x^{6}}+\frac{42}{x^{8}}\right) \kappa^{5}
+\left(\frac{256}{x^{3}}+\frac{560}{x^{5}}+\frac{462}{x^{7}}+\frac{132}{x^{9}}\right) \kappa^{6}
\nonumber\\
&\ \
+\left(\frac{336}{x^{2}}+\frac{1600}{x^{4}}+\frac{2520}{x^{6}}+\frac{1716}{x^{8}}+\frac{429}{x^{10}}\right) \kappa^{7}
+\left(\frac{3168}{x^{3}}+\frac{8960}{x^{5}}+\frac{11088}{x^{7}}+\frac{6435}{x^{9}}+\frac{1430}{x^{11}}\right) \kappa^{8}
+ \mathcal{O}(\kappa^{9})\,,
\label{basic_disk_ex}
\end{align}
\endgroup
where the first few terms are illustrated in Fig.~\ref{fig:disk_basic}.
The cylinder amplitude \eqref{BasicTypeCylinderAmp} is expanded around $\kappa=0$ as
\begingroup\makeatletter\def\f@size{10}\check@mathfonts
\def\maketag@@@#1{\hbox{\m@th\normalsize\normalfont#1}}%
\begin{align}
&
\AmpFXXX_2^{{\tinyspace}{\rm conn}(0)}{\negdbltinyspace}(x_1,x_2)
=
\frac{1}{2{\tinyspace} ( x_1 {\negtinyspace}-{\negtinyspace} x_2 )^2}
\left(
\frac{x_1x_2-\frac{2\kappa}{c^2}\left(x_1+x_2\right)}
{\sqrt{x_1 \!\left(x_1-\frac{4\kappa}{c^2}\right)}
\sqrt{x_2 \!\left(x_2-\frac{4\kappa}{c^2}\right)}}-1\right)
\nonumber\\
&=
\frac{1}{x_{1}^{2} x_{2}^{2}} \kappa^{2}
+\left(\frac{4}{x_{1}^{2}x_{2}^{3}}+\frac{4}{x_{1}^{3}x_{2}^{2}}\right) \kappa^{3}
+\left(\frac{16}{x_{1}^{2} x_{2}^{2}}+\frac{18}{x_{2}^{3} x_{1}^{3}}+\frac{15}{x_{1}^{2} x_{2}^{4}}+\frac{15}{x_{1}^{4} x_{2}^{2}}\right) \kappa^{4}
\nonumber\\
&\ \
+\left(\frac{96}{x_{1}^{2} x_{2}^{3}}+\frac{96}{x_{1}^{3} x_{2}^{2}}+\frac{56}{x_{1}^{2} x_{2}^{5}}+\frac{56}{x_{1}^{5} x_{2}^{2}}+\frac{72}{x_{1}^{3} x_{2}^{4}}+\frac{72}{x_{1}^{4} x_{2}^{3}}\right) \kappa^{5}
\nonumber\\
&\ \
+\left(\frac{256}{x_{1}^{2} x_{2}^{2}}+\frac{576}{x_{1}^{3} x_{2}^{3}}+\frac{480}{x_{1}^{2} x_{2}^{4}}+\frac{480}{x_{1}^{4} x_{2}^{2}}+\frac{300}{x_{1}^{4} x_{2}^{4}}+\frac{210}{x_{1}^{2} x_{2}^{6}}+\frac{210}{x_{1}^{6} x_{2}^{2}}+\frac{280}{x_{1}^{3} x_{2}^{5}}+\frac{280}{x_{1}^{5} x_{2}^{3}}\right) \kappa^{6}
+\mathcal{O}(\kappa^{7})\,.
\end{align}
\endgroup
By the topological recursion, the higher amplitudes in DT (basic type), expanded around $\kappa=0$, are obtained, e.g., as 
$[\alpha=4\kappa/c^2,\ \gamma=c/\kappa]$,
\begin{figure}[t] 
\begin{center}
\vspace{-5cm}
\hspace*{3cm}
  \includegraphics[bb=0 0 595 842, width=125mm]{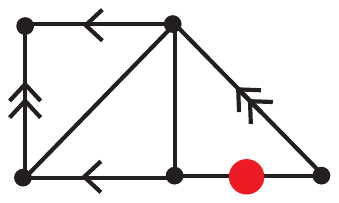}
 \vspace{-10cm}
\caption{An example of a triangulated torus with one boundary, as enumerated by the first term of \eqref{basic_torus_ex}; the red circle indicates a marked point on the boundary, and edges with the matching arrow types are identified.}
\label{fig:torus_basic}
\end{center}
\end{figure}
\begingroup\makeatletter\def\f@size{10}\check@mathfonts
\def\maketag@@@#1{\hbox{\m@th\normalsize\normalfont#1}}%
\begin{align}
&
\AmpFXXX_{1}^{{\tinyspace}{\rm conn}(1)}{\negdbltinyspace}(x)
=
\frac{\alpha^{2}
\left(\gamma  x-\alpha^{2}\right)}{16
\left(\alpha-\gamma\right)^{2} x^{1/2} \left(x-\alpha\right)^{5/2}}
\nonumber\\
&=
\frac{1}{x^{2}}\kappa^{3} + \frac{10}{x^{3}} \kappa^{4}+\left(\frac{28}{x^{2}}+\frac{70}{x^{4}}\right) \kappa^{5}+\left(\frac{344}{x^{3}}+\frac{420}{x^{5}}\right) \kappa^{6}
+\left(\frac{664}{x^{2}}+\frac{2920}{x^{4}}+\frac{2310}{x^{6}}\right) \kappa^{7}
\nonumber\\
&\ \
+ \left(\frac{9072}{x^{3}}+\frac{20720}{x^{5}}+\frac{12012}{x^{7}}\right) \kappa^{8}
+\left(\frac{14912}{x^{2}}+\frac{131880}{x^{6}}+\frac{60060}{x^{8}}+\frac{86800}{x^{4}}\right) \kappa^{9}
+ \mathcal{O}(\kappa^{10})\,,
\label{basic_torus_ex}
\\
&
\AmpFXXX_{3}^{{\tinyspace}{\rm conn}(0)}{\negdbltinyspace}(x_1,x_2,x_3)
=
\frac{\alpha^{4}}
{8 \left(\gamma -\alpha \right) \left(x_{1} x_{2} x_{3}\right)^{1/2}
\left(x_{1}-\alpha\right)^{3/2}\left(x_{2}-\alpha\right)^{3/2}\left(x_{3}-\alpha\right)^{3/2}}
\nonumber\\
&=
\frac{32}{x_{1}^{2} x_{2}^{2} x_{3}^{2}} \kappa^{5}+\left(\frac{192}{x_{1}^{2} x_{2}^{2} x_{3}^{3}}+\frac{192}{x_{1}^{2} x_{2}^{3} x_{3}^{2}}+\frac{192}{x_{1}^{3} x_{2}^{2} x_{3}^{2}}\right) \kappa^{6}
\nonumber\\
&\ \
+\left(\frac{1280}{x_{1}^{2} x_{2}^{2} x_{3}^{2}}+\frac{960}{x_{3}^{4} x_{1}^{2} x_{2}^{2}}+\frac{1152}{x_{3}^{3} x_{1}^{2} x_{2}^{3}}+\frac{960}{x_{3}^{2} x_{1}^{2} x_{2}^{4}}+\frac{1152}{x_{3}^{3} x_{1}^{3} x_{2}^{2}}+\frac{1152}{x_{3}^{2} x_{1}^{3} x_{2}^{3}}+\frac{960}{x_{3}^{2} x_{1}^{4} x_{2}^{2}}\right) \kappa^{7}
+ \mathcal{O}(\kappa^{8})\,,
\\
&
\AmpFXXX_{2}^{{\tinyspace}{\rm conn}(1)}{\negdbltinyspace}(x_1,x_2)
\nonumber\\
&=
\frac{104}{x_{1}^{2} x_{2}^{2}} \kappa^{6}+\left(\frac{1040}{x_{1}^{2} x_{2}^{3}}+\frac{1040}{x_{1}^{3} x_{2}^{2}}\right) \kappa^{7}+\left(\frac{5536}{x_{1}^{2} x_{2}^{2}}+\frac{7920}{x_{2}^{4} x_{1}^{2}}+\frac{9120}{x_{1}^{3} x_{2}^{3}}+\frac{7920}{x_{2}^{2} x_{1}^{4}}\right) \kappa^{8}
\nonumber\\
&\ \
+\left(\frac{62272}{x_{1}^{2} x_{2}^{3}}+\frac{62272}{x_{1}^{3} x_{2}^{2}}+\frac{52640}{x_{2}^{5} x_{1}^{2}}+\frac{63840}{x_{2}^{4} x_{1}^{3}}+\frac{63840}{x_{2}^{3} x_{1}^{4}}+\frac{52640}{x_{2}^{2} x_{1}^{5}}\right) \kappa^{9}
+ \mathcal{O}(\kappa^{10})\,,
\\
&
\AmpFXXX_{1}^{{\tinyspace}{\rm conn}(2)}{\negdbltinyspace}(x)
\nonumber\\
&=
\frac{105}{x^{2}} \kappa^{7}
+ \frac{2310}{x^{3}} \kappa^{8}
+\left(\frac{8112}{x^{2}}+\frac{30030}{x^{4}}\right) \kappa^{9}
+\left(\frac{177296}{x^{3}}+\frac{300300}{x^{5}}\right) \kappa^{10}
\nonumber\\
&\ \
+ \left(\frac{396792}{x^{2}}+\frac{2438016}{x^{4}}+\frac{2552550}{x^{6}}\right) \kappa^{11}
+\left(\frac{8592016}{x^{3}}+\frac{26188512}{x^{5}}+\frac{19399380}{x^{7}}\right) \kappa^{12}
\nonumber\\
&\ \
+\left(\frac{15663360}{x^{2}}+\frac{122687760}{x^{4}}+\frac{239793120}{x^{6}}+\frac{135795660}{x^{8}}\right) \kappa^{13}
+ \mathcal{O}(\kappa^{14})\,,
\\
&
\AmpFXXX_{1}^{{\tinyspace}{\rm conn}(3)}{\negdbltinyspace}(x)
\nonumber\\
&=
\frac{50050}{x^{2}} \kappa^{11}+ \frac{1701700}{x^{3}} \kappa^{12}+\left(\frac{6722816}{x^{2}}+\frac{32332300}{x^{4}}\right) \kappa^{13}
+ \left(\frac{212442240}{x^{3}}+\frac{452652200}{x^{5}}\right) \kappa^{14}
\nonumber\\
&\ \
+\left(\frac{518329776}{x^{2}}+\frac{4052554880}{x^{4}}+\frac{5205500300}{x^{6}}\right) \kappa^{15}
\nonumber\\
&\ \
+\left(\frac{15476799328}{x^{3}}+\frac{58423636480}{x^{5}}+\frac{52055003000}{x^{7}}\right) \kappa^{16}
+ \mathcal{O}(\kappa^{17})\,,
\end{align}
\endgroup
where the first term of $\AmpFXXX_{1}^{{\tinyspace}{\rm conn}(1)}{\negdbltinyspace}(x)$ is illustrated in Fig.~\ref{fig:torus_basic}.

\subsection{Dynamical triangulation (strip type)}
\label{app:list_dt_strip}

\begin{figure}[t] 
\begin{center}
\vspace{-4cm}
\hspace*{0cm}
  \includegraphics[bb=0 0 595 842, width=125mm]{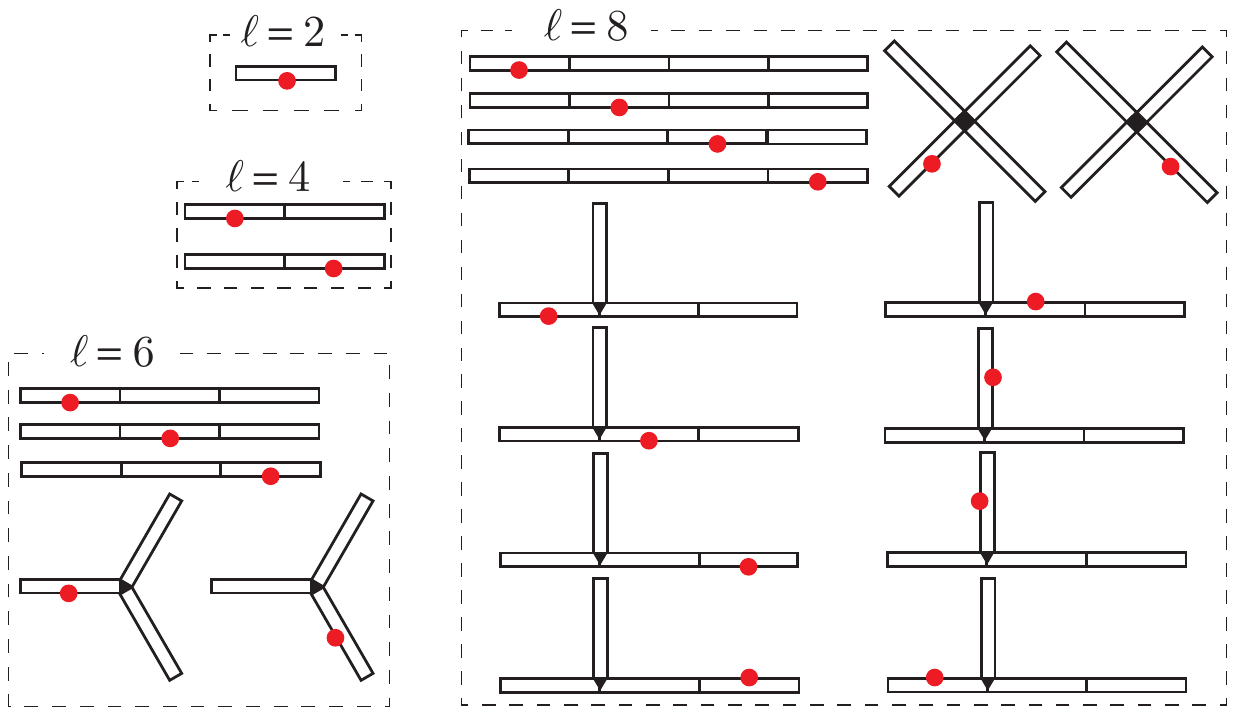}
 \vspace{-6cm}
\caption{Examples of strips of length $\ell$, as enumerated by the first line in the second equality of \eqref{strip_disk_ex}; the red circles indicate marked points on the boundaries.}
\label{fig:catalan}
\end{center}
\end{figure}

The disk amplitude \eqref{StripTypeDiskAmp} is expanded around $\kappa=0$ and 
$x=\infty$ as
\begingroup\makeatletter\def\f@size{10}\check@mathfonts
\def\maketag@@@#1{\hbox{\m@th\normalsize\normalfont#1}}%
\begin{align}
&
\tilde{F}_1^{{\tinyspace}{\rm conn}(0)}{\negdbltinyspace}(x)+\lambda(x)
=
\frac{\kappa}{2}\left(x-\frac{2-a-b}{2\kappa}\right)
\sqrt{\left(x-\frac{a}{\kappa}\right)\left(x-\frac{b}{\kappa}\right)}
+\frac{x-\kappa x^2}{2}-\frac{1}{x}
\nonumber\\
&=
\frac{x-\kappa x^2}{2}-\frac{1}{x}
-\frac{\sqrt{x^{2}-4}}{2}
+\frac{x \left(x^{2}-2\right)}{2 \sqrt{x^{2}-4}} \kappa 
+\frac{4\sqrt{x^{2}-4}}{\left(x+2\right)^{2} \left(x-2\right)^{2}} \kappa^{2}
+\frac{4x \sqrt{x^{2}-4}\, \left(x^{2}-2\right)}{\left(x+2\right)^{3} \left(x-2\right)^{3}} \kappa^{3}
\nonumber\\
&\ \
+\frac{\left(32 x^{4}-176 x^{2}+272\right) \sqrt{x^{2}-4}}{\left(x+2\right)^{4} \left(x-2\right)^{4}} \kappa^{5}
+\mathcal{O}(\kappa^{7})
\nonumber\\
&=
\left(\frac{1}{x^{3}}+\frac{2}{x^{5}}+\frac{5}{x^{7}}+\frac{14}{x^{9}}+\frac{42}{x^{11}}+\frac{132}{x^{13}}+\frac{429}{x^{15}}+\frac{1430}{x^{17}}
+\cdots\right)
\nonumber\\
&\ \
+\left(\frac{1}{x^{2}}+\frac{4}{x^{4}}+\frac{15}{x^{6}}+\frac{56}{x^{8}}+\frac{210}{x^{10}}+\frac{792}{x^{12}}+\frac{3003}{x^{14}}+\frac{11440}{x^{16}}
+\cdots\right)\kappa
\nonumber\\
&\ \
+\left(
\frac{4}{x^{3}}+\frac{24}{x^{5}}+\frac{120}{x^{7}}+\frac{560}{x^{9}}+\frac{2520}{x^{11}}+\frac{11088}{x^{13}}+\frac{48048}{x^{15}}+\frac{205920}{x^{17}}
+\cdots\right)\kappa^2
\nonumber\\
&\ \
+\left(\frac{4}{x^{2}}+\frac{32}{x^{4}}+\frac{200}{x^{6}}+\frac{1120}{x^{8}}+\frac{5880}{x^{10}}+\frac{29568}{x^{12}}+\frac{144144}{x^{14}}+\frac{686400}{x^{16}}
+\cdots\right)\kappa^3
+\mathcal{O}(\kappa^{4})\,,
\label{strip_disk_ex}
\end{align}
\endgroup
where the first line of the second equality is illustrated only by strips as in Fig.~\ref{fig:catalan} and yields the Catalan numbers $(2n)!/((n+1)!n!)$, $n \ge 1$.
The cylinder amplitude \eqref{StripTypeCylinderAmp} is expanded around $\kappa=0$, 
$x_1=\infty$ and $x_2=\infty$ as
\begingroup\makeatletter\def\f@size{10}\check@mathfonts
\def\maketag@@@#1{\hbox{\m@th\normalsize\normalfont#1}}%
\begin{align}
&
\AmpFXXX_2^{{\tinyspace}{\rm conn}(0)}{\negdbltinyspace}(x_1,x_2)
=
\frac{1}{2{\tinyspace} ( x_1 {\negtinyspace}-{\negtinyspace} x_2 )^2}
\left(
\frac{x_1x_2-\frac{a+b}{2\kappa}\left(x_1+x_2\right)+\frac{ab}{\kappa^2}}
{\sqrt{\left(x_1-\frac{a}{\kappa}\right)\left(x_1-\frac{b}{\kappa}\right)}
\sqrt{\left(x_2-\frac{a}{\kappa}\right)\left(x_2-\frac{b}{\kappa}\right)}}-1\right)
\nonumber\\
&=
\left(
\frac{1}{x_{1}^{2} x_{2}^{2}}+\frac{3}{x_{1}^{2} x_{2}^{4}}+\frac{10}{x_{1}^{2} x_{2}^{6}}
+\frac{2}{x_{1}^{3} x_{2}^{3}}+\frac{8}{x_{1}^{3} x_{2}^{5}}+\frac{30}{x_{1}^{3} x_{2}^{7}}
+\frac{12}{x_{1}^{4} x_{2}^{4}}+\frac{45}{x_{1}^{4} x_{2}^{6}}
+\cdots\right)
\nonumber\\
&\ \
+\left(
\frac{4}{x_{1}^{2} x_{2}^{3}}+\frac{24}{x_{1}^{2} x_{2}^{5}}+\frac{120}{x_{1}^{2} x_{2}^{7}}
+\frac{24}{x_{1}^{3} x_{2}^{4}}+\frac{120}{x_{1}^{3} x_{2}^{6}}
+\frac{144}{x_{1}^{4} x_{2}^{5}}+\frac{720}{x_{1}^{4} x_{2}^{7}}
+\frac{720}{x_{1}^{5} x_{2}^{6}}
+\cdots\right)\kappa
\nonumber\\
&\ \
+\left(\frac{4}{x_{1}^{2} x_{2}^{2}}+\frac{36}{x_{1}^{2} x_{2}^{4}}+\frac{240}{x_{1}^{2} x_{2}^{6}}
+\frac{32}{x_{1}^{3} x_{2}^{3}}+\frac{240}{x_{1}^{3} x_{2}^{5}}+\frac{1440}{x_{1}^{3} x_{2}^{7}}
+\frac{288}{x_{1}^{4} x_{2}^{4}}+\frac{1800}{x_{1}^{4} x_{2}^{6}}
+\cdots\right)\kappa^2
\nonumber\\
&\ \
+\left(\frac{40}{x_{1}^{2} x_{2}^{3}}+\frac{368}{x_{1}^{2} x_{2}^{5}}+\frac{2640}{x_{1}^{2} x_{2}^{7}}
+\frac{384}{x_{1}^{3} x_{2}^{4}}+\frac{2800}{x_{1}^{3} x_{2}^{6}}
+\cdots\right)\kappa^3
+\mathcal{O}(\kappa^{4})\,.
\end{align}
\endgroup
By the topological recursion, the higher amplitudes in DT (strip type) are obtained e.g., as 
$[\alpha_1=a/\kappa,\
\alpha_2=b/\kappa,\
\gamma=(2-a-b)/(2\kappa)]$,
\begingroup\makeatletter\def\f@size{10}\check@mathfonts
\def\maketag@@@#1{\hbox{\m@th\normalsize\normalfont#1}}%
\begin{align}
&
\AmpFXXX_{1}^{{\tinyspace}{\rm conn}(1)}{\negdbltinyspace}(x)
=
\frac{\left(x^{2} - \left(\alpha_1+\alpha_2\right)\left(x+\gamma\right) + 2\alpha_1\alpha_2
+\gamma^{2}\right)\left(\alpha_1 -\alpha_2 \right)^{2} \left(x-\alpha_1-\alpha_2+\gamma \right)}{16 \kappa \left(\alpha_1 -\gamma \right)^{2} \left(\alpha_2 -\gamma \right)^{2} \left(x-\alpha_1\right)^{5/2} \left(x-\alpha_2\right)^{5/2}}
\nonumber\\
&=
\left(\frac{1}{x^{5}}+\frac{10}{x^{7}}+\frac{70}{x^{9}}+\frac{420}{x^{11}}+\frac{2310}{x^{13}}+\cdots\right)
+
\left(\frac{1}{x^{4}}+\frac{20}{x^{6}}+\frac{210}{x^{8}}+\frac{1680}{x^{10}}
+\frac{11550}{x^{12}}+\frac{72072}{x^{14}}+\cdots\right) \kappa 
\nonumber\\
&\ \
+\left(\frac{1}{x^{3}}+\frac{26}{x^{5}}+\frac{370}{x^{7}}+\frac{3780}{x^{9}}+\frac{31710}{x^{11}}+\cdots\right) \kappa^{2}
+\left(\frac{1}{x^{2}}+\frac{28}{x^{4}}+\frac{510}{x^{6}}+\frac{6440}{x^{8}}+\frac{64470}{x^{10}}+\cdots\right) \kappa^{3}
\nonumber\\
&\ \
+\left(\frac{28}{x^{3}}+\frac{608}{x^{5}}+\frac{9200}{x^{7}}+\frac{108080}{x^{9}}+\frac{1064280}{x^{11}}+\frac{9225216}{x^{13}}+\cdots\right) \kappa^{4}
+\mathcal{O}(\kappa^5)\,,
\\
&
\AmpFXXX_{3}^{{\tinyspace}{\rm conn}(0)}{\negdbltinyspace}(x_1,x_2,x_3)
\nonumber\\
&=
\left(\frac{2}{x_{1}^{2} x_{2}^{2} x_{3}^{3}}+\frac{12}{x_{1}^{2} x_{2}^{2} x_{3}^{5}}+\frac{12}{x_{1}^{2} x_{2}^{3} x_{3}^{4}}+\frac{8}{x_{1}^{3} x_{2}^{3} x_{3}^{3}}+\frac{72}{x_{1}^{2} x_{2}^{4} x_{3}^{5}}+\frac{48}{x_{1}^{3} x_{2}^{3} x_{3}^{5}}+\frac{72}{x_{1}^{3} x_{2}^{4} x_{3}^{4}}+\frac{288}{x_{1}^{3} x_{2}^{5} x_{3}^{5}}+\frac{432}{x_{1}^{4} x_{2}^{4} x_{3}^{5}}
+\cdots\right)
\nonumber\\
&\ \
+\left(\frac{2}{x_{1}^{2} x_{2}^{2} x_{3}^{2}}+\frac{24}{x_{1}^{2} x_{2}^{2} x_{3}^{4}}+\frac{24}{x_{1}^{2} x_{2}^{3} x_{3}^{3}}+\frac{192}{x_{1}^{2} x_{2}^{3} x_{3}^{5}}+\frac{216}{x_{1}^{2} x_{2}^{4} x_{3}^{4}}+\frac{192}{x_{1}^{3} x_{2}^{3} x_{3}^{4}}+\frac{1440}{x_{1}^{2} x_{2}^{5} x_{3}^{5}}+\frac{1440}{x_{1}^{3} x_{2}^{4} x_{3}^{5}}
+\cdots\right)\kappa
\nonumber\\
&\ \
+\left(\frac{32}{x_{1}^{2} x_{2}^{2} x_{3}^{3}}+\frac{336}{x_{1}^{2} x_{2}^{2} x_{3}^{5}}+\frac{360}{x_{1}^{2} x_{2}^{3} x_{3}^{4}}+\frac{320}{x_{1}^{3} x_{2}^{3} x_{3}^{3}}+\frac{3312}{x_{1}^{2} x_{2}^{4} x_{3}^{5}}+\frac{2880}{x_{1}^{3} x_{2}^{3} x_{3}^{5}}
+\cdots\right)\kappa^2
+\mathcal{O}(\kappa^3)\,,
\\
&
\AmpFXXX_{2}^{{\tinyspace}{\rm conn}(1)}{\negdbltinyspace}(x_1,x_2)
\nonumber\\
&=
\left(\frac{5}{x_{1}^{2} x_{2}^{6}}+\frac{4}{x_{1}^{3} x_{2}^{5}}+\frac{3}{x_{1}^{4} x_{2}^{4}}+\frac{60}{x_{1}^{4} x_{2}^{6}}+\frac{60}{x_{1}^{5} x_{2}^{5}}
+\cdots\right)
+
\left(\frac{8}{x_{1}^{2} x_{2}^{5}}+\frac{6}{x_{1}^{3} x_{2}^{4}}+\frac{160}{x_{1}^{3} x_{2}^{6}}+\frac{156}{x_{1}^{4} x_{2}^{5}}
+\cdots\right) \kappa
\nonumber\\
&\ \
+\left(\frac{9}{x_{1}^{2} x_{2}^{4}}+\frac{8}{x_{1}^{3} x_{2}^{3}}+\frac{290}{x_{1}^{2} x_{2}^{6}}+\frac{260}{x_{1}^{3} x_{2}^{5}}+\frac{252}{x_{1}^{4} x_{2}^{4}}+\frac{4590}{x_{1}^{4} x_{2}^{6}}+\frac{4512}{x_{1}^{5} x_{2}^{5}}
+\cdots\right) \kappa^{2}
+\mathcal{O}(\kappa^3)\,,
\\
&
\AmpFXXX_{1}^{{\tinyspace}{\rm conn}(2)}{\negdbltinyspace}(x)
\nonumber\\
&=
\left(\frac{21}{x^{9}}+\frac{483}{x^{11}}+\frac{6468}{x^{13}}+\frac{66066}{x^{15}}+\frac{570570}{x^{17}}+\cdots\right)
+\left(\frac{49}{x^{8}}+\frac{1575}{x^{10}}+\frac{27258}{x^{12}}+\frac{342342}{x^{14}}+\cdots\right) \kappa 
\nonumber\\
&\ \
+\left(\frac{75}{x^{7}}+\frac{3080}{x^{9}}+\frac{65730}{x^{11}}+\frac{986832}{x^{13}}+\cdots\right) \kappa^{2}
+\left(\frac{95}{x^{6}}+\frac{4690}{x^{8}}+\frac{119490}{x^{10}}+\frac{2100252}{x^{12}}+\cdots\right) \kappa^{3}
\nonumber\\
&\ \
+\left(\frac{105}{x^{5}}+\frac{6135}{x^{7}}+\frac{182140}{x^{9}}+\frac{3683610}{x^{11}}+\frac{57520386}{x^{13}}+\cdots\right) \kappa^{4}
+\mathcal{O}(\kappa^5)\,,
\\
&
\AmpFXXX_{1}^{{\tinyspace}{\rm conn}(3)}{\negdbltinyspace}(x)
\nonumber\\
&=
\left(\frac{1485}{x^{13}}+\frac{56628}{x^{15}}+\frac{1169740}{x^{17}}+\frac{17454580}{x^{19}}+\cdots\right)
+\left(\frac{5445}{x^{12}}+\frac{258258}{x^{14}}+\frac{6414980}{x^{16}}+\cdots\right) \kappa 
\nonumber\\
&\ \
+\left(\frac{11865}{x^{11}}+\frac{674058}{x^{13}}+\frac{19627608}{x^{15}}+\cdots\right) \kappa^{2}
+\left(\frac{19985}{x^{10}}+\frac{1321628}{x^{12}}+\frac{44248204}{x^{14}}+\cdots\right) \kappa^{3}
\nonumber\\
&\ \
+\left(\frac{28630}{x^{9}}+\frac{2162720}{x^{11}}+\frac{82002074}{x^{13}}+\frac{2100862764}{x^{15}}+\cdots\right) \kappa^{4}
+\mathcal{O}(\kappa^5)\,.
\end{align}
\endgroup

\subsection{Dynamical triangulation (continuous level)}
\label{app:list_dt_conti}

The disk amplitude \eqref{ContinuousDiskAmp} and the cylinder amplitude \eqref{ContinuousCylinderAmp} are
\begingroup\makeatletter\def\f@size{10}\check@mathfonts
\def\maketag@@@#1{\hbox{\m@th\normalsize\normalfont#1}}%
\begin{align}
&
\widetilde{\calF}_1^{{\dbltinyspace}{\rm conn}(0)}{\negdbltinyspace}(\xi)
=
\left(\xi - \frac{\sqrt{\cc}}{2}\right)
\sqrt{\xi + \sqrt{\cc}}\,,
\\
&
\widetilde{\calF}_2^{{\dbltinyspace}{\rm conn}(0)}{\negdbltinyspace}(\xi_1,\xi_2)
=
\frac{1}{( \xi_1 - \xi_2 )^2}
\bigg(
  \frac{
        \xi_1 + \xi_2 + 2 \sqrt{\cc}
       }
       {2
        \sqrt{\xi_1 + \sqrt{\cc}}
        \sqrt{\xi_2 + \sqrt{\cc}}
       }
  - 1
{\tinyspace}\bigg)\,.
\end{align}
\endgroup
The topological recursion gives the higher amplitudes in DT (continuous level) as
\begingroup\makeatletter\def\f@size{10}\check@mathfonts
\def\maketag@@@#1{\hbox{\m@th\normalsize\normalfont#1}}%
\begin{align}
&
\widetilde{\calF}_{1}^{{\tinyspace}{\rm conn}(1)}{\negdbltinyspace}(\xi)
=
\frac{2 \xi +5 \sqrt{\cc}}{72\cc \left(\xi + \sqrt{\cc}\right)^{5/2}}
\,,
\\
&
\widetilde{\calF}_{3}^{{\tinyspace}{\rm conn}(0)}{\negdbltinyspace}(\xi_1,\xi_2,\xi_3)
=
\frac{1}{6\sqrt{\cc} \left(\xi_{1}+\sqrt{\cc}\right)^{3/2}
\left(\xi_{2}+\sqrt{\cc}\right)^{3/2}\left(\xi_{3}+\sqrt{\cc}\right)^{3/2}}
\,,
\\
&
\widetilde{\calF}_{2}^{{\tinyspace}{\rm conn}(1)}{\negdbltinyspace}(\xi_1,\xi_2)
=
\frac{71 \cc^{2}+91 \left(\xi_{1}+\xi_{2}\right) \cc^{3/2}+\left(35 \xi_{1}^{2}+89 \xi_{1} \xi_{2}+35 \xi_{2}^{2}\right) \cc+28 \xi_{1} \xi_{2} \left(\xi_{1}+\xi_{2}\right) \sqrt{\cc}+8 \xi_{1}^{2} \xi_{2}^{2}}
{216\cc^2 \left(\xi_{1}+\sqrt{\cc}\right)^{7/2} \left(\xi_{2}+\sqrt{\cc}\right)^{7/2}}
\,,
\\
&
\widetilde{\calF}_{1}^{{\tinyspace}{\rm conn}(2)}{\negdbltinyspace}(\xi)
=
\frac{7 \left(613 \cc^{2}+1006 \cc^{3/2} \xi +816 \cc \xi^{2}+352 \cc^{1/2} \xi^{3}+64 \xi^{4}\right)}{15552 \cc^{7/2} \left(\xi +\sqrt{\cc}\right)^{11/2}}
\,,
\\
&
\widetilde{\calF}_{1}^{{\tinyspace}{\rm conn}(3)}{\negdbltinyspace}(\xi)
=
\frac{7}{5038848 \cc^6 \left(\xi +\sqrt{\cc}\right)^{17/2}}
\left(3705145 \cc^{7/2}+11215906 \cc^{3} \xi +17949936 \cc^{5/2} \xi^{2}
\right.
\nonumber\\
&\hspace{4.5em}
\left.
+18590240 \cc^{2} \xi^{3}+12875840 \cc^{3/2} \xi^{4}+5779200 \cc \xi^{5}+1523200 \cc^{1/2} \xi^{6}+179200 \xi^{7}\right).
\end{align}
\endgroup

\section{Continuum Limit in 1D Pure Quantum Gravity}
\label{app:1DQG}

In this appendix, 
we illustrate the continuum limit in the simplest setting: 
one-dimensional pure quantum gravity (1D pure QG). 
While dynamically trivial, 
this model provides an instructive example of 
how a discrete formulation connects to its continuum counterpart.

\subsection{Continuum partition function and discretized formulation}

Consider 1D pure QG in which the 1D space has neither branches nor boundaries.
In this case, the ``space'' has $S^1$ topology and 
is represented by a closed loop of length $L$. 
At the continuous level, 
the partition function with cosmological constant $\xi$ is given by
\begin{equation}
\tilde{\mathcal{Z}}(\xi)
{\tinyspace}={\tinyspace}
\int_{0}^{\infty} dL\, e^{-\xi L}\,\mathcal{Z}(L)\,,
\end{equation}
where, in one spatial dimension, 
the intrinsic contribution of the geometry is trivial and unique; 
therefore we set $\mathcal{Z}(L){\negdbltinyspace}={\negdbltinyspace}1$.
Hence
\begin{equation}
\tilde{\mathcal{Z}}(\xi)
{\tinyspace} ={\tinyspace}
\int_{0}^{\infty} dL\, e^{-\xi L}
{\tinyspace}={\tinyspace} \frac{1}{\xi}\,.
\end{equation}

\begin{figure}[t] 
\begin{center}
\vspace{-4.5cm}
\hspace*{1.5cm}
  \includegraphics[bb=0 0 595 842, width=0.6\textwidth]{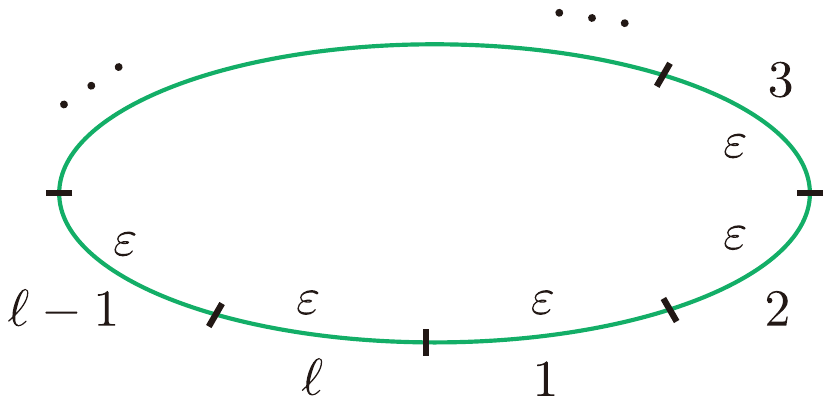}
 \vspace{-5.5cm}
\caption{\label{fig:1dqg}
A discretized loop with $\ell$ links of uniform lattice spacing $\varepsilon$; 
its physical length $L$ of this loop is $\varepsilon\ell$.}
\end{center}
\end{figure}

We now discretize the loop by dividing it into 
$\ell$ links of uniform lattice spacing $\varepsilon$, 
so that the total length is $L {\negdbltinyspace}={\negdbltinyspace} \varepsilon\ell$, as illustrated in Fig.~\ref{fig:1dqg}. 
Assign a weight (fugacity) $x$ to each link, 
and denote by $Z(\ell)$ the contribution from configurations 
with exactly $\ell$ links. 
For the present simple model 
we set $Z(\ell) {\negdbltinyspace}={\negdbltinyspace} 1$ and obtain the generating function
\begin{equation}
\tilde{Z}(x)
{\tinyspace}={\tinyspace} \sum_{\ell=1}^{\infty} x^{\ell}Z(\ell)
{\tinyspace}={\tinyspace} \sum_{\ell=1}^{\infty} x^{\ell}
{\tinyspace}={\tinyspace} x + x^2 + x^3 + \ldots
{\tinyspace}={\tinyspace} \frac{x}{1-x}\,,\qquad |x|<1\,.
\label{eq:discreteZ}
\end{equation}

\subsection{Continuum limit and scaling}

The continuum limit is obtained by tuning the discretization parameter $x$ 
to its critical value $x_{\rm c} {\negdbltinyspace}={\negdbltinyspace} 1$ 
while simultaneously sending the lattice spacing $\varepsilon {\negdbltinyspace}\to{\negdbltinyspace} 0$ 
so that the physical length $L {\negdbltinyspace}={\negdbltinyspace} \varepsilon\>\!\ell$ remains finite. 
Concretely, we introduce the continuum cosmological constant $\xi$ through
\begin{equation}
x {\tinyspace}={\tinyspace} x_{\rm c}\;\!e^{-\varepsilon\xi} 
{\tinyspace}={\tinyspace} e^{-\varepsilon\xi}\,,
\qquad 
\ell {\tinyspace}={\tinyspace} \frac{L}{\varepsilon}\,.
\label{eq:scaling}
\end{equation}
Under this identification,
\[
x^{\ell}{\tinyspace}={\tinyspace}
\bigl(x_{\rm c}\;\!e^{-\varepsilon\xi}\>\!\bigr)^{L/\varepsilon} 
{\tinyspace}={\tinyspace}\bigl(e^{-\varepsilon\xi}\>\!\bigr)^{L/\varepsilon}
\ \
        \longrightarrow 
\ \
e^{-\xi L}\,,
\]
and the discrete weight for $\ell$ links reproduces 
the continuum Boltzmann factor for a loop of length $L$.

Using the expansion 
$x {\negdbltinyspace}={\negdbltinyspace} e^{-\varepsilon\xi}
{\negdbltinyspace}={\negdbltinyspace} 1-\varepsilon\>\!\xi+O(\varepsilon^{2})$ 
for small $\varepsilon$, 
the geometric series~\eqref{eq:discreteZ} behaves as
\begin{equation}
\tilde{Z}(x)
{\tinyspace}={\tinyspace}
\frac{x}{1-x} {\tinyspace}\simeq{\tinyspace}
\frac{1}{\varepsilon\>\!\xi}+O(1)\,.
\end{equation}
Thus, multiplying the discrete generating function 
by the lattice spacing $\varepsilon$ gives the continuum partition function,
\begin{equation}
\varepsilon\;\!\tilde{Z}(x)
\ \
 \longrightarrow
\ \
\frac{1}{\xi}
 {\tinyspace}={\tinyspace}
\tilde{\mathcal{Z}}(\xi)\,,
\end{equation}
in agreement with the direct continuum computation.

\subsection{Radius of convergence and physical interpretation}

For \emph{small} $x$, 
the first few terms in the sum $\sum_{\ell=1}^{\infty}x^{\ell}$ 
dominate the discrete partition function: 
configurations with a small number of links give the leading contribution. 
However, as $x$ approaches the radius of convergence 
$x_{\rm c} {\negdbltinyspace}={\negdbltinyspace} 1$, 
terms with large $\ell$ become significant, 
and arbitrarily large link numbers must be included. 
This is precisely the mechanism by which the discrete model 
approaches a continuum description: 
tuning $x {\negdbltinyspace}\to{\negdbltinyspace} x_{\rm c}$ makes the typical number of links 
(and thus the correlation length measured in lattice units) diverge, 
so that the discrete sum approximates the continuum integral over loop lengths.

This observation---the continuum limit is reached 
by approaching the radius of convergence of the discrete generating function, 
where contributions from large $\ell$ dominate---is central 
to the construction and must be taken into account.

Although 1D pure QG is a trivial theory, 
the construction above illustrates, in the simplest possible setting, 
how a meaningful continuum limit can be extracted 
from a discretized formulation. 
The procedure naturally generalizes to higher-dimensional models, 
where a critical point in the discrete parameters (here 
$x_{\rm c} {\negdbltinyspace}={\negdbltinyspace} 1$) 
signals the emergence of continuum degrees of freedom, 
and appropriate scaling relations (such as \eqref{eq:scaling}) 
connect discrete weights to continuum couplings. 
In this sense, 
the one-dimensional case serves as a useful warm-up example 
for understanding the continuum limit in dynamical triangulations 
and related approaches to quantum gravity.

\section{Schwinger-Dyson Equations for the One-Matrix Model}\label{app:SD_MM}

In this appendix, we briefly review the Schwinger-Dyson equations and Virasoro constraints for the Hermitian one-matrix model and discuss their relation to our Hamiltonian formalism.

We start from the partition function $Z(\{t_k\})$ of the Hermitian one-matrix model:
\begin{align}
Z(\{t_k\})=\int_{\mathcal{H}_{\overline{N}}}\mathrm{e}^{-\overline{N}\,\mathrm{Tr}\,V(M)}\,dM,
\label{eq:partition_function_matrix}
\end{align}
where $\mathcal{H}_{\overline{N}}$ denotes the set of Hermitian matrices of size $\overline{N}$, and the potential $V(M)$ is given by
\begin{align}
V(M)=-\sum_{k\ge 0}t_kM^k.
\label{eq:potential}
\end{align}
We consider the correlation function given by 
\begin{align}
\langle \mathcal{O}(M)\rangle=\frac{1}{Z(\{t_k\})}\int_{\mathcal{H}_{\overline{N}}}\mathcal{O}(M)\,\mathrm{e}^{-\overline{N}\,\mathrm{Tr}\,V(M)}\,dM,
\end{align}
and denote in particular for the connected part of the correlation functions of resolvents by
\begin{align}
W_N(x_1,\ldots,x_N)
=
\left\langle
\prod_{i=1}^N \mathrm{Tr}\frac{1}{x_i-M}
\right\rangle^{\mathrm{conn}}.
\label{eq:correlator_Wbar}
\end{align}

The invariance of the matrix integral \eqref{eq:partition_function_matrix} under the change of variables $M\to M+f(M)$, with an arbitrary regular function $f$, leads to the Schwinger-Dyson equations for $W_N$ \cite{Eynard:2004mh}. The Schwinger-Dyson equations take the following form:
\begin{align}
&W_{N+1}(x_i,x_i,\bm{x}_{I\setminus\{i\}})
+\sum_{I_1\cup I_2=I\setminus\{i\}}W_{|I_1|+1}(x_i,\bm{x}_{I_1})W_{|I_2|+1}(x_i,\bm{x}_{I_2})
\nonumber \\
&\qquad
+\sum_{j\in I\setminus\{i\}}\frac{\partial}{\partial x_j}\frac{W_{N-1}(\bm{x}_{I\setminus \{j\}})-W_{N-1}(\bm{x}_{I\setminus\{i\}})}{x_i-x_j}
\nonumber \\
=\, &\overline{N}V'(x_i)W_N(\bm{x}_I)-\overline{N}U_N(x_i;\bm{x}_{I\setminus\{i\}}),
\label{eq:SD_MM}
\end{align}
where
\begin{align}
U_N(x_i;\bm{x}_{I\setminus\{i\}})
=
\left\langle
\mathrm{Tr}\frac{V'(x_i)-V'(M)}{x_i-M}
\prod_{j\in I\setminus\{i\}}\mathrm{Tr}\frac{1}{x_j-M}
\right\rangle^{\mathrm{conn}}.
\label{eq:U_MM}
\end{align}
In the limit $\overline{N}\to\infty$, the WKB expansion yields the following asymptotic expansions in powers of $\overline{N}^{-1}$:
\begin{align}
W_{N}(x_1,\ldots,x_N)
&=
\sum_{h\ge 0}\overline{N}^{2-2h-N}W^{(h)}_{N}(x_1,\ldots,x_N),
\label{eq:asymptotic_MM} \\
U_{N}(x_i;\bm{x}_{I\setminus\{i\}})
&=
\sum_{h\ge 0}\overline{N}^{2-2h-N}U^{(h)}_{N}(x_i;\bm{x}_{I\setminus\{i\}}).
\label{eq:asymptotic_U}
\end{align}

\subsection{Schwinger-Dyson equations for the cubic matrix model and DT (strip type)}\label{sec:SD_MM_strip}

To compare the above Schwinger-Dyson equations with those of DT (strip type), we specialize the potential to
\begin{align}
V(x)=\frac{x^2}{2}-\kappa \frac{x^3}{3}.
\label{eq:strip_potential}
\end{align}

For the potential \eqref{eq:strip_potential}, $U_N(x_i;\bm{x}_{I\setminus\{i\}})$ can be rewritten as
\begin{align}
U_N(x_i;\bm{x}_{I\setminus\{i\}})
&=
\left\langle
\mathrm{Tr}(1-\kappa (x_i+M))
\prod_{j\in I\setminus\{i\}}\mathrm{Tr}\frac{1}{x_j-M}
\right\rangle^{\mathrm{conn}},
\label{eq:U_specialization}
\end{align}
and $U_N(x_i;\bm{x}_{I\setminus\{i\}})$ is independent of $x_i$ for $N\ge 2$, since we take the connected part of the correlation function.

Comparing the two asymptotic expansions \eqref{AmplitudeGexpansionN} and \eqref{eq:asymptotic_MM}, we find the following relation between the expansion parameters:
\begin{align}
G=\overline{N}^{-2}.
\label{eq:parameters}
\end{align}
Under this correspondence, the connected amplitude $\tilde{F}^{\mathrm{conn}}_N(x_1,\ldots,x_N;G)$ for DT (strip type) and $W_N(x_1,\ldots,x_N)$ are related by
\begin{align}
\tilde{F}^{\mathrm{conn}}_N(x_1,\ldots,x_N;\overline{N}^{-2})
=
\overline{N}^{-N}W_N(x_1,\ldots,x_N)-\delta_{N,1}\frac{V'(x)}{2},
\end{align}
and the Schwinger-Dyson equation \eqref{eq:SD_MM} can be rewritten as
\begin{align}
&\tilde{F}^{\mathrm{conn}}_{N+1}(x_i,x_i,\bm{x}_{I\setminus \{i\}};\overline{N}^{-2})
+\sum_{I_1\cup I_2=I\setminus\{i\}}
\tilde{F}^{\mathrm{conn}}_{|I_1|+1}(x_i,\bm{x}_{I_1};\overline{N}^{-2})
\tilde{F}^{\mathrm{conn}}_{|I_2|+1}(x_i,\bm{x}_{I_2};\overline{N}^{-2})
\nonumber \\
&\qquad
+\overline{N}^{-2}\frac{\partial}{\partial x_i}
\sum_{\substack{j=1\\(j\ne i)}}^N
\frac{
\tilde{F}^{\mathrm{conn}}_{N-1}(\bm{x}_{I\setminus \{j\}};\overline{N}^{-2})
-
\tilde{F}^{\mathrm{conn}}_{N-1}(\bm{x}_{I\setminus \{i\}};\overline{N}^{-2})
}{x_i-x_j}
\nonumber \\
=
&-\overline{N}^{-N}U_N(x_i;\bm{x}_{I\setminus\{i\}})
-\delta_{N,2}\overline{N}^{-2}\frac{1}{2}\frac{\partial}{\partial x_i}
\sum_{\substack{j=1\\(j\ne i)}}^{N}\frac{V'(x_i)-V'(x_j)}{x_i-x_j}
+\delta_{N,1}\frac{1}{4}V'(x_i)^2.
\label{eq:SD_inter}
\end{align}

For $N=1$, we have
\begin{align}
U_1(x)=-\overline{N}(\kappa x-1)-\kappa\langle \mathrm{Tr}M\rangle^{\mathrm{conn}},
\end{align}
and hence obtain the Schwinger-Dyson equation
\begin{align}
\tilde{F}^{\mathrm{conn}}_{2}(x,x)+\left(\tilde{F}^{\mathrm{conn}}_{1}(x)\right)^2
=
\Omega(x)+\overline{N}^{-1}\kappa\langle \mathrm{Tr}M\rangle^{\mathrm{conn}},
\qquad
\Omega(x)=\frac{1}{4}V'(x)^2+\kappa x-1.
\end{align}
Focusing on the leading-order term in the $\overline{N}^{-1}$ expansion of this equation, we recover
\eqref{StripTypeSDeqN1G0} and \eqref{StripTypeDiskAmpTemp}.

Since the second term on the right-hand side satisfies
\begin{align}
\delta_{N,2}\overline{N}^{-2}\frac{1}{2}\frac{\partial}{\partial x_i}
\sum_{\substack{j=1\\(j\ne i)}}^{N}\frac{V'(x_i)-V'(x_j)}{x_i-x_j}
=
\delta_{N,2}\overline{N}^{-2}\left(-\frac{\kappa}{2}\right),
\end{align}
and since $U_N(x_i;\bm{x}_{I\setminus\{i\}})$ does not depend on $x_i$ for the potential \eqref{eq:strip_potential} in the case of $N\ge 2$, the right-hand side of \eqref{eq:SD_inter} can be rewritten in the form
\begin{align}
\delta_{N,1}\Omega(x_i)+C_N(\bm{x}_{I\setminus \{i\}}).
\end{align}
Substituting this expression into \eqref{eq:SD_inter}, we obtain the Schwinger-Dyson equation \eqref{StripTypeSDeqGeneralNNsinglePeelingInt} for DT (strip type).

From our analysis of the Schwinger-Dyson equations for the cubic matrix model and DT (strip type), 
we find the following correspondence between the Laplace-transformed loop creation operator in \eqref{DiscreteLaplaceTransfWaveFun} and the insertion of the shifted resolvent:
\begin{align}
\tilde{\Psi}^{\dagger}(x)=\sum_{\ell\ge 1}x^{-\ell-1}\Psi^{\dagger}(\ell)
\;\longleftrightarrow\;
\mathrm{Tr}\frac{1}{x-M}-\overline{N}x^{-1}=\sum_{n\ge 1}x^{-n-1}\mathrm{Tr}M^n\,.
\end{align}
To realize the action of the Laplace-transformed loop creation and annihilation operators explicitly in the one-matrix model, 
we consider the formal potential \eqref{eq:potential}. 
For this formal-potential model, we find
\begin{align}
\left[\overline{N}x^{-1}+\overline{N}^{-1}\sum_{n\ge 1}x^{-n-1}\frac{\partial}{\partial t_n}\right]Z(\{t_k\})=\left\langle \mathrm{Tr}\frac{1}{x-M}\right\rangle \cdot Z(\{t_k\}),
\end{align}
and the Laplace-transformed loop creation and annihilation operators in \eqref{DiscreteLaplaceTransfWaveFun} are identified with the following differential/multiplicative operators acting on $Z(\{t_k\})$ in the one-matrix model:\footnote{%
In \cite{SFT:AW}, this correspondence is referred to as the star operation, and the action of the operators $\phi^{\dagger}_n$ and $\phi_n$, which are the mode expansions of the loop creation and annihilation operators in DT (continuous level), is represented on $Z(\{j_k\})$ as
\begin{align}
&\phi^{\dagger}_n \;\longleftrightarrow\; \frac{\partial}{\partial j_n},
\qquad
\phi_n \;\longleftrightarrow\; j_n,
\nonumber \\
&Z(\{j_k\})=\lim_{T\to\infty}\langle\mathrm{vac}|\mathrm{e}^{-T\mathcal{H}}
\exp\left(\sum_{n\ge 1}\phi_n^{\dagger}j_n\right)|\mathrm{vac}\rangle.
\nonumber
\end{align}
}
\begin{align}
\tilde{\Psi}^{\dagger}(x) \;\longleftrightarrow\; \overline{N}^{-1}\sum_{n\ge 1}x^{-n-1}\frac{\partial}{\partial t_n},
\qquad
\tilde{\Psi}(x)\;\longleftrightarrow\; -\overline{N}\sum_{n\ge 1}x^{-n-1}t_n.
\end{align}
These differential/multiplicative operators also satisfy the commutation relations \eqref{CommutationRelationConj}.
After applying the differential/multiplicative operators and then specializing the couplings $t_n$ to those of the cubic potential \eqref{eq:strip_potential}, 
we recover the action of the Laplace-transformed loop creation and annihilation operators for DT (strip type).


\begin{thebibliography}{99}

\bibitem{LG:Polyakov}
  A.~M.~Polyakov, 
  \emph{Quantum Geometry of Bosonic Strings}, 
  Phys. Lett. B \textbf{103} (1981), 207-210.

\bibitem{LG:KPZ}
  V.~G.~Knizhnik, A.~M.~Polyakov, A.~B.~Zamolodchikov, 
  \emph{Fractal Structure of 2D Quantum Gravity}, 
  Mod. Phys. Lett. A \textbf{3} (1988), 819.

\bibitem{MM:BIPZ}
  E.~Br\'{e}zin, C.~Itzykson, G.~Parisi and J.~B.~Zuber, 
  \emph{Planar Diagrams},
  Commun. Math. Phys. \textbf{59} (1978), 35.

\bibitem{MM:DS}
  M.~R.~Douglas, S.~H.~Shenker,
  \emph{Strings in Less Than One-Dimension},
  Nucl. Phys. B \textbf{335} (1990), 635.

\bibitem{MM:GM}
  D.~J.~Gross, A.~A.~Migdal,
  \emph{Nonperturbative Two-Dimensional Quantum Gravity},
  Phys. Rev. Lett. \textbf{64} (1990), 127.


\bibitem{MM:BK}
  E.\ Br\'{e}zin, V.A.\ Kazakov, 
  \emph{Exactly Solvable Field Theories of Closed Strings}, 
  Phys. Lett. B \textbf{236} (1990), 144-150.

\bibitem{MM:DVV}
R.~Dijkgraaf, H.~L.~Verlinde and E.~P.~Verlinde,
\emph{Loop equations and Virasoro constraints in nonperturbative 2-D quantum gravity},
Nucl. Phys. B \textbf{348} (1991), 435-456.

\bibitem{MM:FKN}
  M.~Fukuma, H.~Kawai and R.~Nakayama,
  \emph{Continuum Schwinger-dyson Equations and Universal Structures
        in Two-dimensional Quantum Gravity},
  Int. J. Mod. Phys. A \textbf{6} (1991), 1385-1406,
  \emph{Infinite dimensional Grassmannian structure of
        two-dimensional quantum gravity},
  Commun. Math. Phys. \textbf{143} (1992), 371-404.

\bibitem{DT:David}
  F.\ David, 
  \emph{Planar Diagrams, Two-Dimensional Lattice Gravity and Surface Models},
  Nucl. Phys. B \textbf{257} (1985), 45.

\bibitem{DT:ADF}
  J.~Ambj\o rn, B.~Durhuus, J.~Fr\"{o}hlich,
  \emph{Diseases of Triangulated Random Surface Models, and Possible Cures},
  Nucl. Phys. B \textbf{257} (1985), 433-449.

\bibitem{DT:KKM}
  V.~A.~Kazakov, I.~K.~Kostov, A.~A.~Migdal,
  \emph{Critical Properties of Randomly Triangulated Planar Random Surfaces},
  Phys. Lett. B \textbf{157} (1985), 295-300.

\bibitem{preSFT:AM}
  M.~E.~Agishtein, A.A.~Migdal,
  \emph{Recursive Sampling of Planar Graphs and Fractal Properties
        of a Two-dimensional Quantum Gravity},
  Int. J. Mod. Phys. C \textbf{1} (1990), 165-179.

\bibitem{preSFT:KKSW}
  N.~Kawamoto, V.~A.~Kazakov, Y.~Saeki and Y.~Watabiki,
  \emph{Fractal structure of two-dimensional gravity coupled to D = -2 matter},
  Phys. Rev. Lett. \textbf{68} (1992), 2113-2116.

\bibitem{preSFT:KKMW}
  H.~Kawai, N.~Kawamoto, T.~Mogami and Y.~Watabiki,
  \emph{Transfer matrix formalism for two-dimensional quantum gravity
        and fractal structures of space-time},
  Phys. Lett. B \textbf{306} (1993), 19-26,
  [arXiv:hep-th/9302133 [hep-th]].

\bibitem{SFT:IK}
  N.~Ishibashi and H.~Kawai,
  \emph{String field theory of noncritical strings},
  Phys. Lett. B \textbf{314} (1993), 190-196,
  [arXiv:hep-th/9307045 [hep-th]].

\bibitem{SFT:JR}
  A.~Jevicki and J.~P.~Rodrigues,
  \emph{Loop space Hamiltonians and field theory of noncritical strings},
  Nucl. Phys. B \textbf{421} (1994), 278-292,
  [arXiv:hep-th/9312118 [hep-th]].

\bibitem{SFT:Watabiki}
  Y.~Watabiki, 
  \emph{Construction of noncritical string field theory
        by transfer matrix formalism in dynamical triangulation},
  Nucl.\ Phys.\ B\textbf{441} (1995), 119-166,
  [arXiv:hep-th/9401096 [hep-th]].

\bibitem{SFT:AW}
  J.\ Ambj\o rn and Y.\ Watabiki, 
  \emph{Noncritical string field theory for two-D quantum gravity
        coupled to $(p,q)$-conformal fields},
  Int. J. Mod. Phys. A \textbf{12} (1997), 4257-4289,
  [arXiv:hep-th/9604067 [hep-th]].


\bibitem{Eynard:2007kz}
B.~Eynard and N.~Orantin,
\emph{Invariants of algebraic curves and topological expansion},
Commun. Num. Theor. Phys. \textbf{1} (2007), 347-452,
[arXiv:math-ph/0702045 [math-ph]].

\bibitem{Alexandrov:2003pj}
A.~S.~Alexandrov, A.~Mironov and A.~Morozov,
\emph{Partition functions of matrix models as the first special functions of string theory. 1. Finite size Hermitean one matrix model},
Int. J. Mod. Phys. A \textbf{19} (2004), 4127-4165,
[arXiv:hep-th/0310113 [hep-th]].

\bibitem{Eynard:2004mh}
B.~Eynard,
\emph{Topological expansion for the 1-Hermitian matrix model correlation functions},
JHEP \textbf{11} (2004), 031,
[arXiv:hep-th/0407261 [hep-th]].

\bibitem{Chekhov:2006vd}
L.~Chekhov, B.~Eynard and N.~Orantin,
\emph{Free energy topological expansion for the 2-matrix model},
JHEP \textbf{12} (2006), 053,
[arXiv:math-ph/0603003 [math-ph]].

\bibitem{Eynard:2016yaa}
B.~Eynard, 
{\it Counting Surfaces},
Progress in Mathematical Physics,
Vol. 70, Springer (2016).

\bibitem{Bouchard:2024fih}
V.~Bouchard,
\emph{Les Houches lecture notes on topological recursion},
[arXiv:2409.06657 [math-ph]].

\bibitem{FMW2}
H.~Fuji, M.~Manabe and Y.~Watabiki,
\emph{Multicritical Dynamical Triangulations and Topological Recursion},
[arXiv:2512.10519 [hep-th]].

\bibitem{FMW3}
H.~Fuji, M.~Manabe and Y.~Watabiki,
\emph{A Hamiltonian Formalism for Topological Recursion},
[arXiv:2512.14059 [math-ph]].






\bibitem{SFT:WatabikiReview}
  Y.~Watabiki, 
  \emph{The Causality Road from Dynamical Triangulations
        to Quantum Gravity that Describes Our Universe},
  In *Handbook of Quantum Gravity*,
  edited by Cosimo Bambi, Leonardo Modesto, Ilya Shapiro,
  Springer, 2025,
  [arXiv:2212.13109v2 [gr-qc]].

\bibitem{Brini:2010fc}
A.~Brini, M.~Marino and S.~Stevan,
\emph{The Uses of the refined matrix model recursion},
J. Math. Phys. \textbf{52} (2011), 052305,
[arXiv:1010.1210 [hep-th]].

\bibitem{Bouchard:2007ys}
V.~Bouchard, A.~Klemm, M.~Marino and S.~Pasquetti,
\emph{Remodeling the B-model},
Commun. Math. Phys. \textbf{287} (2009), 117-178,
[arXiv:0709.1453 [hep-th]].

\bibitem{Kazakov:1988ch}
V.~A.~Kazakov and A.~A.~Migdal,
\emph{Recent Progress in the Theory of Noncritical Strings},
Nucl. Phys. B \textbf{311} (1988), 171.

\bibitem{Ginsparg:1991bi}
P.~H.~Ginsparg,
\emph{Matrix models of 2-d gravity},
[arXiv:hep-th/9112013 [hep-th]].


\bibitem{Seiberg:2003nm}
N.~Seiberg and D.~Shih,
\emph{Branes, rings and matrix models in minimal (super)string theory},
JHEP \textbf{02} (2004), 021
[arXiv:hep-th/0312170 [hep-th]].

\bibitem{Eynard:2023qdr}
B.~Eynard, E.~Garcia-Failde, P.~Gregori, D.~Lewanski and R.~Schiappa,
\emph{Resurgent Asymptotics of Jackiw{\textendash}Teitelboim Gravity and the Nonperturbative Topological Recursion},
Annales Henri Poincare \textbf{25} (2024) no.9, 4121-4193
[arXiv:2305.16940 [hep-th]].

\bibitem{Saad:2019lba}
P.~Saad, S.~H.~Shenker and D.~Stanford,
\emph{JT gravity as a matrix integral},
[arXiv:1903.11115 [hep-th]].

\bibitem{Kontsevich:2017vdc}
M.~Kontsevich and Y.~Soibelman,
\emph{Airy structures and symplectic geometry of topological recursion},
Contribution to: 2016 AMS von Neumann Symposium Topological Recursion and its Influence in Analysis, Geometry, and Topology
[arXiv:1701.09137 [math.AG]].

\bibitem{Andersen:2017vyk}
J.~E.~Andersen, G.~Borot, L.~O.~Chekhov and N.~Orantin,
\emph{The ABCD of topological recursion}s,
Adv. Math. \textbf{439} (2024), 109473
[arXiv:1703.03307 [math-ph]].

\bibitem{GR}
J.~E.~Andersen, G.~Borot, and N.~Orantin,
\emph{Geometric recursion},
[arXiv:1711.04729 [math.GT]].






\end{thebibliography}
\end{document}